\documentclass[aps,pra,superscriptaddress,twocolumn]{revtex4-2}
\usepackage{xcolor}
\usepackage{natbib}
\usepackage{amsmath}
\usepackage{amssymb}
\usepackage{amsthm}
\usepackage{nicefrac}
\usepackage{graphicx}
\usepackage{bm}
\usepackage{url}
\usepackage{physics}

\newcommand{\G}{\mathbb{G}}

\newcommand{\Id}{\mathbb{I}}
\newcommand\numthis{\stepcounter{equation}\tag{\theequation}}

\begin{document}

\title{Suppressing electromagnetic local density of states via \\ slow light in lossy quasi-1d gratings}

\author{Benjamin Strekha}
\thanks{These authors contributed equally to this work.}
\affiliation{Department of Electrical and Computer Engineering, Princeton University, Princeton, New Jersey 08544, USA}

\author{Pengning Chao}
\thanks{These authors contributed equally to this work.}
\affiliation{Department of Mathematics, Massachusetts Institute of Technology, Cambridge, Massachusetts 02139, USA}

\author{Rodrick Kuate Defo}
\affiliation{Department of Electrical and Computer Engineering, Princeton University, Princeton, New Jersey 08544, USA}

\author{Sean Molesky}
\affiliation{Department of Engineering Physics, Polytechnique Montréal, Montréal, Québec H3T 1J4, Canada}

\author{Alejandro W. Rodriguez}
\affiliation{Department of Electrical and Computer Engineering, Princeton University, Princeton, New Jersey 08544, USA}

\begin{abstract}
We propose a spectral-averaging procedure that enables computation of bandwidth-integrated local density of states (LDOS) from a single scattering calculation, and exploit it to investigate the minimum extinction achievable from dipolar sources over nonzero bandwidths in structured media. 
Structure-agnostic extinction bounds are derived, providing analytical insights into scaling laws and fundamental design tradeoffs with implications to bandwidth and material selection.
We find that perfect LDOS suppression over a nonzero bandwidth $\Delta\omega$ is impossible.
Inspired by limits which predict nontrivial $\sqrt{\Delta\omega}$ scaling in systems with material dissipation, we show that pseudogap edge states of quasi-1d bullseye gratings can---by simultaneously minimizing material absorption and radiation---yield arbitrarily close to perfect LDOS suppression in the limit of vanishing bandwidth. 
\end{abstract}

\maketitle

%\section{Introduction}

The electromagnetic local density of states (LDOS)
plays a central role in many optical phenomena, including spontaneous~\cite{purcell_absorption_1969} and stimulated emission~\cite{francs_optical_2010}, radiating antennas, surface enhanced Raman scattering~\cite{michon_limits_2019,gaponenko_strong_2021}, photovoltaics~\cite{wang_optimization_2014}, radiative heat transfer~\cite{cuevas_radiative_2018}, and frequency conversion~\cite{lin_cavity-enhanced_2016}.
Canonically, changing the electromagnetic environment of a quantum emitter alters its rate of spontaneous emission~\cite{purcell1995spontaneous}. Furthermore, suppressing radiative processes can reduce losses in semiconductor lasers~\cite{yablonovitch1987inhibited} and increase parametric nonlinearities~\cite{lin_cavity-enhanced_2016}.

In this article, we exploit quadratic optimization techniques to explore the minimum bandwidth-integrated LDOS possible in structured media.
Calculations are vastly simplified by the fact that LDOS, proportional to the power extracted from a subwavelength emitter, is a causal linear-response function, allowing one to relate the bandwidth-integrated response to evaluation of the integrand at a single, complex frequency (plus a nonnegative electrostatic contribution); this in turn makes techniques for computing single-frequency bounds~\cite{molesky_global_2020,shim_fundamental_2019,chao_physical_2022,chao_maximum_2022} applicable to finite-bandwidth objectives.
Prior applications of such contour-deformation techniques employed a Lorentzian spectral window that leads to ultraviolet divergences in the vacuum response~\cite{liang_formulation_2013,shim_fundamental_2019,chao_maximum_2022} and is therefore only appropriate for studying scattered power; it is unsuitable for quantifying Purcell enhancement and particularly LDOS suppression, which relies on sensitive cancellations between contributions from the radiating source and scatterer.
To remedy this issue, we introduce an averaging function that behaves similar to a Lorentzian for small bandwidths, exhibiting a single pole in the upper-half plane while decaying sufficiently fast so as to avoid such ultraviolet divergences.

Our bounds show that no structures can completely suppress LDOS for nonzero bandwidths.
Intuitively, photonic crystals (PhC) which support complete photonic band gaps and thus inhibit wave propagation at frequencies within the gap~\cite{yablonovitch1994photonic} are a natural guess for optimal suppression of average LDOS.
For lossless materials, PhCs and Bragg gratings are optimal geometries for any bandwidth but can only achieve perfect suppression in the limit of vanishing bandwidth (a single frequency), as they are unable to suppress radiation at frequencies outside the band gap. Surprisingly, we find that near-perfect suppression at a single frequency is also feasible in the presence of material loss.
For a PhC in 2d, a minimal index contrast and hence minimum feature size is needed to open a gap~\cite{joannopoulos_photonic_2008,rechtsman_method_2009,ho1994photonic}, which limits the degree to which absorption can be mitigated, leading to LDOS saturation.
Quasi-1d structures, on the other hand, such as bullseye gratings supporting pseudogap ``slow light'' resonances with low-field concentrations in the dielectric medium, approach zero extinction power in the limit of vanishing material thicknesses by enabling the gap to close more slowly than absorption. 
% These and other interesting optimal scaling properties relating to LDOS modification through structuring are presented below.

%\section{Formulation}

\textit{Formulation---} Working in dimensionless units of $\epsilon_0=\mu_0=1$, and
considering only nonmagnetic materials, the partial LDOS at frequency
$\omega$ and position $\vb{r}'$ along the direction $\hat{\vb{e}}$ is directly proportional to the
time-averaged power $\rho(\omega)$ emitted by a harmonic dipole source
$\vb{J}(\vb{r};\omega)e^{-i\omega t} = \delta(\vb{r}-\vb{r}')
e^{-i\omega t}\hat{\vb{e}}$~\cite{novotny_principles_2012,joulain2003definition},
\begin{equation}
    \rho(\omega) \equiv -\frac{1}{2} \Re{\int \vb{J}^*(\vb{r};\omega) \cdot \vb{E}(\vb{r};\omega) \,\dd\vb{r} }
    \label{eq:Jpower}
\end{equation}
where the electric field generated by the current $\vb{J}$ satisfies Maxwell's equations, 
$ \curl \curl \vb{E}(\vb{r}) - \omega^2 (1 + \chi(\vb{r})) \vb{E}(\vb{r}) = i\omega\vb{J}(\vb{r})$.
We can decompose the total field into the field emitted by the
source in vacuum and that emitted by the induced polarization $\vb{P}$ within the scatterer:
\begin{align*}
    &\vb{E}(\vb{r}) = \vb{E}_{\textrm{vac}}(\vb{r}) + \vb{E}_{\textrm{sca}}(\vb{r}), \\
    &= \frac{i}{\omega} \int \G_{0}(\vb{r},\vb{r}') \cdot \vb{J}(\vb{r}') \,\dd\vb{r}' + \int \G_{0}(\vb{r},\vb{r}') \cdot \vb{P}(\vb{r}') \,\dd\vb{r}' \numthis
\end{align*}
% \begin{align*}
%     &\vb{E}(\vb{r}) = \vb{E}_{\textrm{vac}}(\vb{r}) + \vb{E}_{\textrm{sca}}(\vb{r})
%     = \frac{i}{\omega}\mathbb{G}_{0}\mathbf{J} + \mathbb{G}_{0}\mathbf{P}
%     % \int \G_{0}(\vb{r},\vb{r}') \cdot \vb{J}(\vb{r}') \,\dd\vb{r}' + \int \G_{0}(\vb{r},\vb{r}') \cdot \vb{P}(\vb{r}') \,\dd\vb{r}' 
%     \numthis
% \end{align*}
where $\G_{0}(\vb{r},\vb{r}')$ is the vacuum dyadic Green's function, satisfying $\curl \curl \G_{0} - \omega^2 \G_{0} = \omega^2 \Id$. This in turn leads to a decomposition of $\rho$ into vacuum and scatterer contributions
\begin{equation}
    \rho(\omega) = \rho_{\textrm{vac}}(\omega) + \rho_{\textrm{sca}}(\omega) ,
    \label{eq:rho_scat}
\end{equation}
where $\rho_{\textrm{vac}}$ and $\rho_{\textrm{sca}}$ are defined by substituting $\vb{E}_{\textrm{vac}}$ and $\vb{E}_{\textrm{sca}}$ into Eq.~\eqref{eq:Jpower}, respectively. In particular,
%\begin{align*}
%    \rho_{\textrm{sca}}(\omega) &= -\frac{1}{2}\Re{\iint \vb{J}^*(\vb{r}) \cdot \G_0(\vb{r},\vb{r}') \cdot \vb{P}(\vb{r}') \,\dd\vb{r}'\dd\vb{r} }, \\
%    &= -\frac{1}{2} \Im{\omega \int \vb{E}_{\textrm{vac}}(\vb{r}')\cdot \vb{P}(\vb{r}') \,\dd\vb{r}'}, \\
%    &\equiv \Im{s(\omega)} \numthis \label{eq:rho_sca}
%\end{align*}
\begin{align*}
    \rho_{\textrm{sca}}(\omega) &= -\frac{1}{2} \Im{\omega \int \vb{E}_{\textrm{vac}}(\vb{r}')\cdot \vb{P}(\vb{r}') \,\dd\vb{r}'} \equiv \Im{s(\omega)} \numthis \label{eq:rho_sca}
\end{align*}
where we use the irrelevancy of the global phase of the source to set $\vb{J}^{*}(\mathbf{r}; \omega) = \vb{J}(\mathbf{r}; \omega)$, and reciprocity $\mathbb{G}_{0} = \mathbb{G}_{0}^{T}.$
%$\G_{0}(\vb{r}, \vb{r}') = \G^{T}_{0}(\vb{r}, \vb{r}')$.
% where in the second line we used reciprocity $\G_{0}(\vb{r},\vb{r}')=\G_{0}^T(\vb{r},\vb{r}')$, and the fact that the global phase of the dipole source is irrelevant to take $\vb{J}^* = \vb{J}$.

Since real sources emit light over a nonzero bandwidth, a key figure of merit is the bandwidth average of $\rho(\omega)$.
Prior works focused on LDOS enhancement have investigated an averaging of the form $\int_{-\infty}^\infty \rho(\omega) L(\omega) \dd \omega$, where
\begin{align}
 L(\omega) \equiv \frac{\Delta\omega/\pi}{(\omega-\omega_0)^2 + \Delta\omega^2}
 \label{eq:Lorentzian}
\end{align}
is a Lorentzian window function centered at $\omega_0$ with bandwidth $\Delta\omega$~\cite{shim_fundamental_2019,chao_maximum_2022,liang_formulation_2013}. $L(\omega)$ not only captures the spectral lineshape of many practical sources~\cite{liang_formulation_2013,hindmarsh_atomic_2014} but also offers a great computational advantage: complex contour integration simplifies the frequency integral to a single evaluation $\rho(\tilde{\omega})$ at the complex pole $\tilde{\omega}$ of $L(\omega)$ in the upper half plane, plus an electrostatic contribution. However, there is one conceptual challenge: in the ultraviolet limit $\omega \rightarrow \infty$, $\rho_{\textrm{vac}}(\omega)$ grows too quickly ($\propto\lvert\omega\rvert$ in 2d and $\propto\omega^2$ in 3d) causing the Lorentzian spectral average to diverge~\cite{liang_formulation_2013}. When investigating LDOS maximization, $\rho_{\textrm{vac}}$ and $\rho_{\textrm{sca}}$ have the same sign, and this vacuum divergence can be ignored as a constant background (precluding, however, determination of Purcell factor and related LDOS enhancement figures of merit~\cite{purcell1995spontaneous,novotny_principles_2012}).
Minimizing extinction, however, requires engineering $\rho_{\textrm{sca}}$ to cancel $\rho_{\textrm{vac}}$, making this procedure prohibitive.
 
To remedy this issue, we define an alternate bandwidth average over the positive frequencies as follows:
\begin{align}
    \langle Q\rangle
    &\equiv
    \lim_{\epsilon\to 0^{+}}\int_{\epsilon}^{\infty} Q(\omega)W(\omega) \dd\omega,
    \label{eq:Wfreq_avg0toinf}
    \\ &= \lim_{\epsilon\to 0^{+}}\int_{\epsilon}^{\infty} Q(\omega) \left[\frac{L(\omega) - L(-\omega)}{\omega\mathcal{N}}\right] \dd\omega
    \label{eq:freq_avg0toinf}
\end{align}
with the normalization factor $\mathcal{N} \equiv \frac{\omega_{0}}{\omega_{0}^{2} +\Delta\omega^{2}}$ chosen so that $\int_{0}^{\infty} W(\omega) \dd\omega = 1$, with $W(\omega)$ nonnegative and finite for $\omega\in(0,\infty)$.
This window function in most practical settings with $\Delta\omega \ll \omega_0$ resembles a Lorentzian distribution peaked around $\omega_0$; for large bandwidths $\Delta\omega \gtrsim \omega_0$ the spectrum becomes increasingly asymmetric about $\omega_{0}$
and gives greater weight to quasistatic contributions. Since $W(\omega) \sim \omega^{-4}$ as $\omega \rightarrow \infty$, $\langle \rho_{\textrm{vac}} \rangle$ is a convergent quantity that can be evaluated directly. 
%For instance, a 2d scalar (TM) source has $\langle \rho_{\textrm{vac}} \rangle = \frac{1}{4\pi \mathcal{N}}\atan(\frac{\omega_{0}}{\Delta\omega})$.

% \begin{figure}
%     \centering
%         \includegraphics[width=\linewidth]{windowfunctionexamplesFULL.pdf}
%         \caption{
%         Phenomena associated with LDOS: (a) spontaneous emission, (b) radiation from a dipole antenna, (c) optical coupling to solid state defects.
%         (d) Spectral window function $W(\omega)$ used in calculating average LDOS around a center frequency $\omega_0$, Eq.~\eqref{eq:Wfreq_avg0toinf}, for various values of the bandwidth parameter $\Delta\omega = \{1,\frac{1}{5},\frac{1}{10},\frac{1}{20},\frac{1}{50}\} \omega_0$ (top to bottom). For ease of visualization, $W(\omega)$ is normalized by $W(\omega_{0}).$
%         }  
% \label{fig:windowfunctionexamples}
% \end{figure}

\begin{figure}
    \centering
        \includegraphics[width=\linewidth]{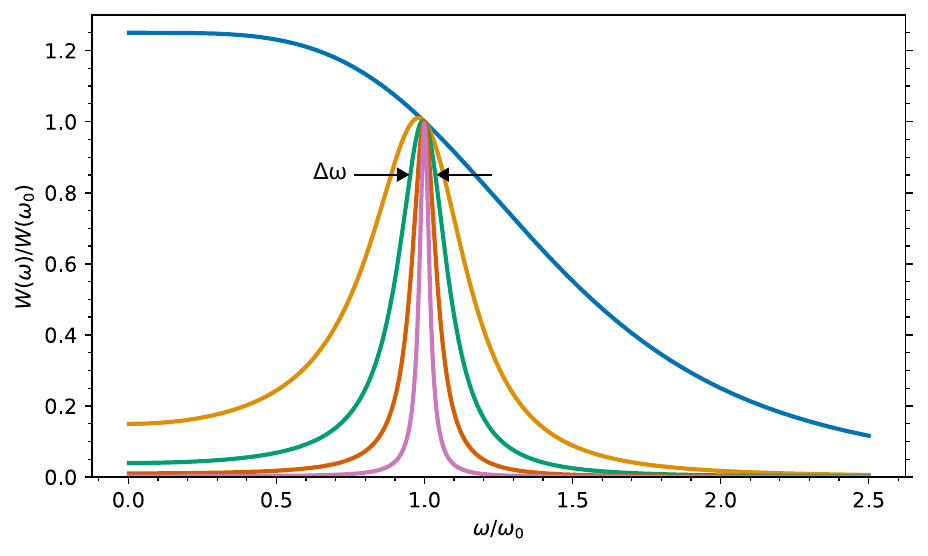}
        \caption{
        Spectral window function $W(\omega)$ used in calculating average LDOS around a ``center'' frequency $\omega_0$, Eq.~\eqref{eq:Wfreq_avg0toinf}, for various values of the bandwidth parameter $\Delta\omega = \{1,\frac{1}{5},\frac{1}{10},\frac{1}{20},\frac{1}{50}\} \omega_0$ (top to bottom). For ease of visualization, $W(\omega)$ is normalized by $W(\omega_{0}).$
        }  
\label{fig:windowfunctionexamples}
\end{figure}

Since the Fourier components of real fields at negative frequencies are conjugates of the counterparts at positive frequencies~\cite{landau_electrodynamics_2009}, $\rho_{\textrm{sca}}(\omega) = \rho_{\textrm{sca}}(-\omega)$ and Eq.~\eqref{eq:freq_avg0toinf} can be unfolded into a principal value integral over all positive and negative frequencies:
\begin{align*}
    \langle \rho_{\textrm{sca}} \rangle%_{\omega_0, \Delta\omega} 
    % &=
    % \lim_{\epsilon\to 0^{+}}\bigg(\int_{-\infty}^{-\epsilon} + \int_{\epsilon}^{\infty}\bigg) \frac{L(\omega)}{\omega\mathcal{N}} \Im{s(\omega)} \dd\omega, \\
    &= \Im{ P.V. \int_{-\infty}^\infty \frac{L(\omega)}{\omega\mathcal{N}} s(\omega) \,\dd\omega }.
    \numthis \label{eq:freq_avg}
\end{align*}
This can be evaluated using contour integration, yielding
\begin{equation}
    \langle \rho_{\textrm{sca}} \rangle%_{\omega_0, \Delta\omega} 
    = \Im\left[\frac{s(\tilde{\omega})}{\tilde{\omega}\mathcal{N}}\right]
    + \frac{2\omega_{0}\Delta\omega}{|\tilde{\omega}|^4\mathcal{N}}\alpha
    \label{eq:cplx_freq_rho_sca}
\end{equation}
where the first term is from the residue at $\tilde{\omega}\equiv \omega_0 + i\Delta\omega$, the pole of $L(\omega)$ in the upper-half plane; the second term is a nonnegative electrostatic contribution due to the singularity at $\omega = 0$: here $\alpha=\frac{1}{2} \Re{ \vb{p}_0 \cdot \vb{E}_0}$, where $\vb{p}_0$ is a unit amplitude electrostatic dipole and $\vb{E}_0$ the scattered field it generates, previously seen in Refs.~\cite{shim_fundamental_2019,chao_maximum_2022} but with a different prefactor due to our use of $W(\omega)$ instead of $L(\omega)$.

The average of allowed modifications in the emission rate around the ``center'' frequency $\omega_0$ (e.g., the spontaneous emission of an atom of corresponding transition energy) is constrained in all cases by this simple ``weighted-sum rule."
In the limit of zero bandwidth, only the first term in Eq.~\eqref{eq:cplx_freq_rho_sca} is in general nonzero, which represents single-frequency LDOS: $\langle\rho\rangle = \rho_{\textrm{vac}}(\omega_{0}) + \rho_{\textrm{sca}}(\omega_{0})$. 
As the bandwidth goes to infinity, $\Im\left[\frac{s(\tilde{\omega})}{\tilde{\omega}\mathcal{N}}\right]$ decays rapidly and the second term can be shown (under proper normalization of the window function) to yield an all-frequency sum rule $\int_{0}^{\infty}\rho_{\textrm{sca}}(\omega)\dd\omega = \pi\alpha/2$~\cite{shim_fundamental_2019}.

Since we are seeking lower bounds on $\langle \rho \rangle$, the nonnegative electrostatic term $\alpha$ can be relaxed to zero; for TM sources, it is exactly zero~\cite{chew1999waves,tsang_scattering_2004}.
We then adapt the method laid out in Ref.~\cite{chao_maximum_2022} for computing upper bounds on LDOS maximization to instead obtain lower bounds on the remaining term $\Im[\frac{s(\tilde{\omega})}{\tilde{\omega}\mathcal{N}}]$: given a prespecified design region $V$ and an isotropic material susceptibility $\chi(\tilde{\omega})$, the bounds enforce conservation of power (optical theorem~\cite{jackson_classical_1999,novotny_principles_2012}) constraints and apply to any structure that fits within $V$.

Note that the averaging procedure requires $\chi(\tilde{\omega})$ to be evaluated at a complex frequency, with causality and loss requiring that $\text{Im}[\chi(\tilde{\omega})] > 0$ be positive~\cite{hashemi_diameter-bandwidth_2012}.
While a complex susceptibility at real frequencies indicates material loss/gain~\cite{novotny_principles_2012}, at complex frequencies dispersion and loss both contribute to $\text{Im}[\chi]$, with $\text{Im}[\chi(\tilde{\omega})] \to \text{Im}[\chi(\omega_{0})]$ as $\Delta\omega \to 0$. Thus, $\text{Im}[\chi(\tilde{\omega})] = 0$ may be interpreted as a transparent and nondispersive medium.

 \begin{figure*}
    \centering
        \includegraphics[width=\linewidth]{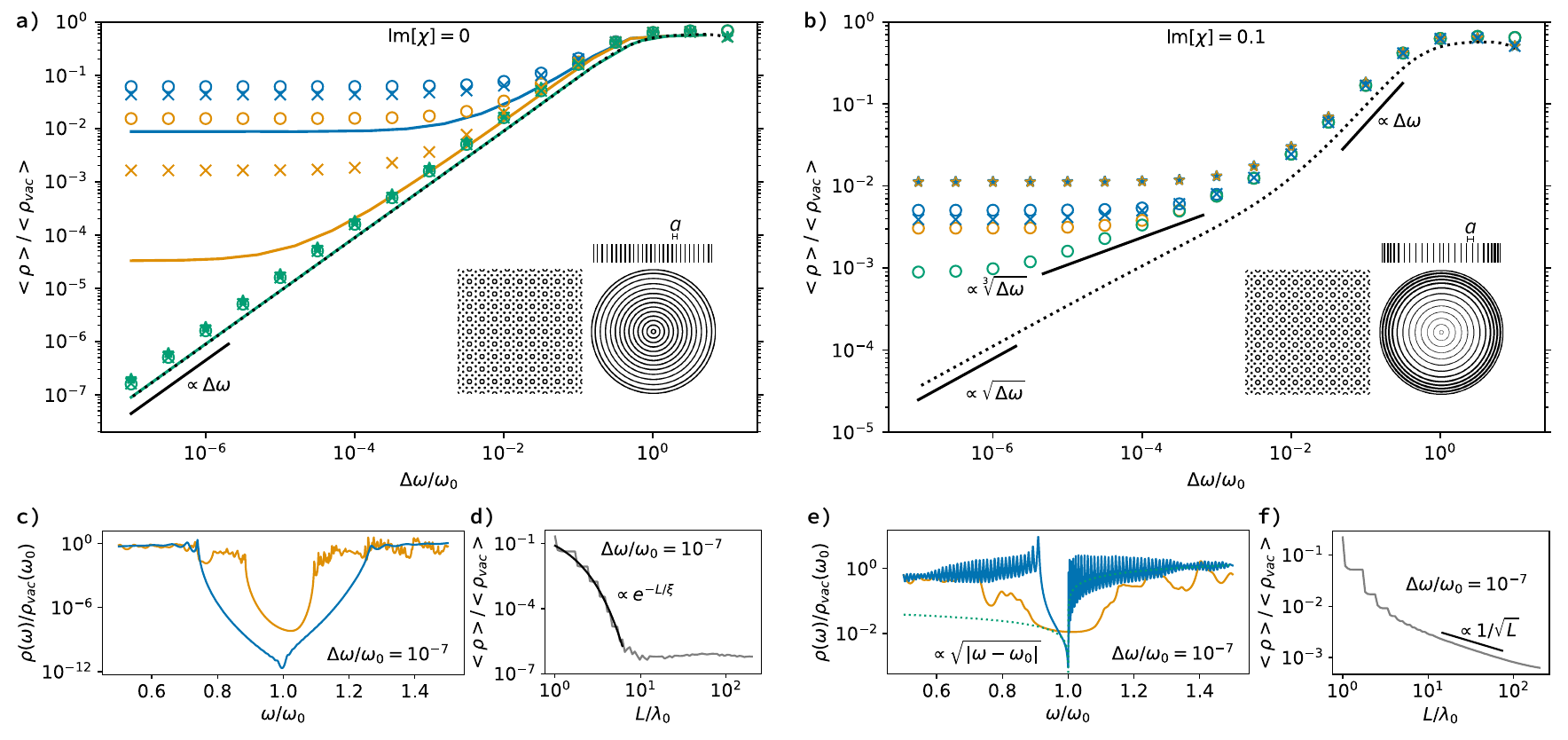}
        \caption{
        a) and b): Lower bounds on bandwidth-integrated LDOS.
        All curves and markers refer to TM line sources located at the center of $L\times L$ design domains with a) $\chi = 4$ for $L/\lambda_{0} = \{1, 1.59, 10\}$ (blue, orange, green) and b) $\chi = 4 + 0.1i$ for 
        $L/\lambda_{0} = \{5, 10, 80\}$
        (blue, orange, green) where $\lambda_0 = 2\pi c/\omega_0$.
         %$R/\lambda_{0} = \{1, 5, 10, 20, 40, \infty\}$
        Solid lines are lower bounds for finite $L\times L$ design domains,
        while dotted lines pertain to infinite systems, $L/\lambda_{0} \to \infty$, computed via Eq.~\eqref{eq:rhoscaboundexplicit}.
        All results are normalized by the corresponding bandwidth-integrated LDOS in vacuum. 
        Markers correspond to structures discovered via topology optimization, where circles $\circ$ are over rotationally symmetric structures, stars $\star$ are for truncated 2d square-lattice photonic crystals of period $\lambda_{0}$, and crosses $\times$ allow for arbitrary structuring.
        Insets of a) and b): Representative inverse designs corresponding to $\Delta\omega/\omega_{0} = 10^{-7}$ for a $10\lambda_{0} \times 10\lambda_{0}$ design region.
        c) and e): Representative spectra for $L/\lambda_{0} \gg 1$ (orange for $10\lambda_{0} \times 10\lambda_{0}$ crystal, blue for $80\lambda_{0} \times 80\lambda_{0}$ bullseye grating (not shown)); optimizing over cylindrical geometries, the lossless and dispersionless case converges to a bullseye geometry of radial periodicity $a = \frac{\lambda_{0}}{4}\frac{1 + \sqrt{1 + \chi}}{\sqrt{1 + \chi}}$ and ring thickness $h = \frac{\lambda_{0}}{4\sqrt{1 + \chi}}$ (a radial quarter-wave stack) supporting a pseudogap centered around $\omega_0$, see c); the lossy case converges to a radial chirped grating of initial spacing $a = \lambda_{0}/2$ and tapered thickness, causing the net absorbed power to come arbitrarily close to zero at $\omega_{0}$ as $L \to \infty$, in the vicinity of which the system exhibits square-root dispersion, see e). d) and f): The bandwidth-averaged LDOS of optimized bullseye structures within a design region of diameter $L$. The lossless (lossy) case demonstrates exponential (inverse square root) approach as a function of $L$ to a positive saturation value due to a nonzero bandwidth $\Delta\omega/\omega_{0} = 10^{-7}$.}
\label{fig:chi4finiteandinf}
\end{figure*}

\textit{Results---} While below we focus on 2d TM sources, similar observations hold for either 2d TE or 3d dipole sources, albeit with the electrostatic $\alpha$ term in those settings becoming relevant (and increasingly important at larger bandwidths).
As detailed in Ref.~\cite{chao_maximum_2022} and SM, bounds on Im[$\frac{s(\tilde{\omega})}{\tilde{\omega}\mathcal{N}}$] can be computed numerically for an arbitrary domain, here chosen as a square volume $V = L \times L$.
For an infinite design space ($L \to \infty$), all operators and fields can be expanded in a spectral basis conforming to the symmetry of the domain---vector cylindrical waves~\cite{tsang_scattering_2004,chew1999waves}---yielding the following semianalytic expression for the bound: 
\begin{multline}
    \langle\rho_{\textrm{sca}} \rangle_{L\to \infty}
    \geq 
    -\frac{1}{8\pi\mathcal{N}}\int_{0}^{\infty} \dd k \, k
    \bigg[\text{Im}\bigg(
    \frac{e^{i\theta}}{\chi^{*}} + \frac{e^{-i\theta}\tilde{\omega}^{2}}{k^{2} - \tilde{\omega}^{2}}
    \bigg)\bigg]^{-1}\\
    \times \text{Re}
    \Bigg[
    \bigg|\frac{\tilde{\omega} 
    }{k^{2} - \tilde{\omega}^{2}}\bigg|^{2}
    - 
    \frac{e^{-i\theta}\tilde{\omega}^{2}
    }{(k^{2} - \tilde{\omega}^{2})^{2}}
    \Bigg],
    \label{eq:rhoscaboundexplicit}
\end{multline}
where maximization over the parameter $\theta$ can be computed numerically (see SM).
To lowest order in the bandwidth, for $\text{Im}[\chi] \geq 0$, this integral can be further simplified to yield the following asymptotic expression:
\begin{align}
    \frac{\langle \rho\rangle_{L\to\infty}}{\langle\rho_{\textrm{vac}}\rangle} \geq \sqrt{\frac{2\text{Im}[\chi]}{|\chi|^{2}}\frac{\Delta\omega}{\omega_{0}}} + \mathcal{O}(\Delta\omega).
    \label{eq:bounds_dw_asymptotics}
\end{align}
Thus, the bounds suggest near-perfect suppression is possible as $\Delta\omega \to 0$ even for $\Im[\chi]>0$ (consistent with zero material and radiative losses). Below, we present a physical mechanism that confirms this finding.

Figure~\ref{fig:chi4finiteandinf} shows lower bounds on $\langle \rho\rangle$ obtained for finite $L\times L$ design regions along with achievable objective values discovered via inverse design. 
As the design footprint increases, inverse designs approach the limit given by the semi-analytical result of Eq.~\eqref{eq:rhoscaboundexplicit}, corresponding to $L\to\infty$.
For $L/\lambda_{0} = 10,$ where $\lambda_{0} = 2\pi c/\omega_{0}$ is the wavelength of the center frequency, the structures discovered via inverse design are within a factor of two of the infinite-space bounds for small $\text{Im}[\chi(\tilde{\omega})]$, but loosen by a couple of orders of magnitude for smaller ($L/\lambda_{0} \lesssim 4$) device footprints (partly due to the associated bound relaxations).
As seen, emission suppression for a TM source can be achieved through wave interference generated by either Bragg grating arrangements of dielectric material or 2d PhCs, resulting in either pseudogaps or complete band gaps centered around $\omega_{0}$, respectively.
For lossless materials, a 2d PhC of band gap size $\omega_{g}$ centered at $\omega_{0}$ has $\rho_\textrm{phc}(\omega)=0$ for $|\omega-\omega_{0}|~\leq~\omega_{g}/2$, yielding an average $\langle \rho \rangle = \int_{0}^{\infty} \rho_\textrm{phc}(\omega) W(\omega) \dd \omega \propto \Delta\omega$ as $\Delta\omega \to 0$, consistent with the scaling of the bounds seen in Fig. \ref{fig:chi4finiteandinf}a. 

In systems with material loss, $\text{Im}[\chi(\omega_{0})] > 0$, one might expect LDOS to saturate to a small but positive value $\rho(\omega_{0}) > 0$ due to absorption by the structure.
However, as confirmed in Fig. \ref{fig:chi4finiteandinf}, the bounds of Eq.~\eqref{eq:bounds_dw_asymptotics} suggest that complete emission cancellation is possible, with perfect suppression approached under vanishing bandwidth, $\Delta\omega \to 0$, albeit at a reduced $\sqrt{\Delta\omega}$ rate compared to the lossless case.
Topology optimization discovers radial tapers (chirped gratings) with an initial spacing $a=\lambda_0/2$ but increasing thickness, producing an LDOS spectrum $\rho_\textrm{taper}(\omega) \approx \sqrt{|\omega - \omega_{0}|}$ (Fig.~\ref{fig:chi4finiteandinf}b inset) that meets at a nonzero plateau around $\omega_0$ due to absorption in the medium.
Such structures initially yield $\langle \rho \rangle = \int_{0}^{\infty} \rho_\textrm{taper}(\omega) W(\omega) \dd \omega \propto \sqrt{\Delta\omega}$ but absorption ultimately leads to nonzero saturation as $\Delta\omega \rightarrow 0$. 

Since losses in the limit of vanishing bandwidth are ultimately dominated by absorption (as opposed to radiation), reducing the amount of material is crucial to LDOS minimization. However, material reduction is in conflict with the demands of nonzero bandwidth operation: in 2d PhCs, a minimum index contrast is required for the formation of a band gap~\cite{joannopoulos_photonic_2008,rechtsman_method_2009}, which implies a minimum fill fraction for any given index contrast (from perturbation theory). 
% This is much less of a problem for (quasi) 1D structures: 1D PhCs support band gaps for any nonzero index contrast, which intuitively explains the performance advantage of bullseye gratings over 2D PhCs in Fig. \ref{fig:chi4finiteandinf}b. 
This is not an issue for effective 1d structures capable of supporting band gaps for any nonzero index contrast~\cite{joannopoulos_photonic_2008}, which intuitively explains the performance advantage of bullseye gratings over 2d PhCs in Fig.~\ref{fig:chi4finiteandinf}. 
To further understand the engineering and performance of such quasi-1d designs, we exploit transfer matrices to study a 1d cavity design schema consisting of a central vacuum layer of thickness $d$ (containing the source at the center) sandwiched by two identical half-infinite PhC claddings of susceptibility $\chi$, unit cell size $a$, and material layer thickness $h$.
The electric field within such a cavity, $|z| \leq d/2$, has the form
$
E(z; \omega) = E_0 e^{i\omega |z|} + E_0 r e^{i \omega d} e^{-i\omega |z|},
$
where $r$ is the reflection coefficient of the cladding, % is determined as the ratio of the eigenvector components of the largest in norm eigenvalue of the transfer matrix (SM).
and the power output of the dipole is given by
\begin{align}
    \frac{\rho(\omega)}{\rho_{\textrm{vac}}(\omega)}
    % &= \Re{\frac{1+r e^{i k_v d}}{1-r e^{i k_v d}}}, \\
    &= \frac{1-|r|^2}{|1-r e^{i \omega d}|^2}.
    \label{eq:PhCCavityLDOS}
\end{align}
Perfect emission suppression can thus occur if $1 - |r|^{2} \to 0$ and the denominator does not also approach zero.

One plausible strategy to achieve this condition is to design the PhC claddings such that $\omega_0$ is at the band gap center $\omega_m$, and to take $h \rightarrow 0$.
For lossless $\chi$, first order perturbation theory gives the size of the band gap $\omega_{g}$ as $\omega_{g}/\omega_{m} \approx \text{Re}[\chi]\frac{h}{a} \propto h$ for $h/a~\ll~1$~\cite{joannopoulos_photonic_2008}.
An oscillating dipolar source with frequency within the band gap will excite localized evanescent states with an exponential field decay constant $\gamma~\propto~\sqrt{\omega_{g}/\kappa}$ where $\kappa$ is the band curvature at the band edge.
This state does not radiate and without material loss, the LDOS is strictly zero. With loss, the absorbed power $P_{\textrm{abs}} = \int \Im[\chi(z)] |E(z)|^2 \dd z \propto \Im[\chi] \frac{h}{a} \int e^{- 2\gamma z} \dd z \propto h/\gamma$ (strictly speaking material loss also increases $\gamma$, so this proportionality is an upper bound).
Since $\omega_{g}\propto h$, it may appear that $P_{\textrm{abs}} \propto h/\gamma \propto \sqrt{h}$ as $h \to 0$; however, the curvature at the band edge $\kappa \propto 1/h$ also depends on $h$ so ultimately $P_{\textrm{abs}}$ approaches a constant as $h\to 0$.
In other words, exponential decay on its own is insufficient since the decay length diverges as the band gap closes, which limits the minimum absorption possible.
Correspondingly, as $h\to 0$ the reflectivity at the midgap approaches
\begin{equation}
    r\left(\omega_m\right) \approx \frac{i\chi}{\Im[\chi] + \sqrt{ (\chi+i\Im[\chi])\cdot\Re[\chi] }},
    \label{eq:rmidgap}
\end{equation}
with $|r|<1$ given $\Im[\chi]>0$, i.e., no perfect suppression. 

A strategy that does produce near-perfect suppression is to engineer the cladding so $\omega_0$ is not at the midgap but at the \textit{upper band gap edge}; this can be achieved by setting the unit cell size $a=\lambda_0/2$ and taking $h \rightarrow 0$. In this case
\begin{equation}
    % r(\omega_0) \approx -1 + \frac{2\pi}{\sqrt{3}} \frac{h}{\lambda_{0}} - 2\pi \frac{h}{\lambda_{0}} i
    r(\omega_0) \approx -1 + \frac{\omega_{0}h}{\sqrt{3}} - i\omega_{0}h, \qquad h \rightarrow 0,
\end{equation}
independent of the material susceptibility, and we have $1-|r|^2 \rightarrow 0$; to avoid division by zero in Eq.~\eqref{eq:PhCCavityLDOS}, one may choose $d\not=(n + \frac{1}{2})\lambda_0$ where $n$ is a nonnegative integer, in which case 
$\rho(\omega_{0})/\rho_{\textrm{vac}}(\omega_{0}) \approx \frac{\pi}{\sqrt{3}} \sec^{2}(\frac{\omega_{0}d}{2})\frac{h}{\lambda_{0}}$ as $h \to 0$.
Intuitively, in the limit of $h \rightarrow 0$, $\omega_{0}$ approaches the band edge from above and the electric field profile approaches a standing wave (a slow light mode~\cite{baba2008slow}) with field nodes inside material layers. Thus, by making the thicknesses of the PhC mirrors arbitrarily thin, 
one can ensure arbitrarily small field overlap with the lossy medium and thus, vanishing absorption.
This is reminiscent of techniques for linear optical control of light in coherent tunable absorbers~\cite{zhang_controlling_2012,gutman_coherent_2013,baranov_coherent_2017}, with the difference that the standing wave is not the result of interfering multiple incident waves but is a mode of the PhC itself. 
% While the use of slow light to enhance light--matter interactions is well-documented~\cite{baba2008slow}, to our knowledge, this work is the first demonstration of its potential to suppress them.

The bandwidth-integrated LDOS for the 1d structure can also be computed and is given by
$\frac{\langle\rho\rangle}{\langle\rho_{\textrm{vac}}\rangle} = \Re{\frac{1+r e^{i \tilde{\omega} d}}{1-r e^{i \tilde{\omega} d}}\frac{1}{\tilde{\omega}\mathcal{N}}}$. For $\Delta\omega \ll \omega_0$ and $d \not\approx (n + \frac{1}{2})\lambda_{0}$, we find that the optimal slow light cavity design has thickness $h = \lambda_{0}\sqrt[3]{\frac{\Delta\omega}{\omega_{0}\beta}}$, which yields
\begin{equation}
    \frac{\langle\rho\rangle}{\langle\rho_{\textrm{vac}}\rangle} 
    =
    \frac{\pi}{\cos^{2}(\frac{\omega_{0}d}{2})}\Re\left[\sqrt{\frac{1}{3} + \frac{i\beta\chi^{*}}{2\pi^{2}|\chi|^{2}}}\right]\sqrt[3]{\frac{\Delta\omega}{\omega_{0}\beta}},
    \label{eq:dwbetahcubed}
\end{equation}
where $\beta = \frac{2\pi^{2}\text{Im}[\chi]}{3}\bigg[-1 +\sqrt{9 + 8 \left(\frac{\text{Re}[\chi]}{\text{Im}[\chi]}\right)^{2}}\bigg]$, proving the existence of structures exhibiting $\sqrt[3]{\Delta\omega}$ scaling, with
\begin{align}
    \frac{\langle\rho\rangle}{\langle\rho_{\textrm{vac}}\rangle}
    &=
    \frac{\sqrt[3]{\frac{\pi}{2\sqrt{3}|\text{Re}[\chi]|} \frac{\Delta\omega}{\omega_{0}}} }{\cos^{2}(\frac{\omega_{0}d}{2})} \bigg(
    1
    +
    \frac{\sqrt{2}}{3}\frac{\text{Im}[\chi]}{|\text{Re}[\chi]|}
    + \mathcal{O}(\text{Im}[\chi]^{2})
    \bigg),
\end{align}
% \begin{align}
%     \frac{\langle\rho\rangle}{\langle\rho_{\textrm{vac}}\rangle}
%     &=
%     \frac{\sqrt[3]{\frac{\pi}{2\sqrt{3}|\text{Re}[\chi]|}}}{\cos^{2}(\frac{\omega_{0}d}{2})} \bigg(
%     1
%     +
%     \frac{\sqrt{2}}{3}\frac{\text{Im}[\chi]}{|\text{Re}[\chi]|}
%     + \mathcal{O}(\text{Im}[\chi]^{2})
%     \bigg)
%     \sqrt[3]{\frac{\Delta\omega}{\omega_{0}}}.
% \end{align}
% \begin{align}
%     \frac{\langle\rho\rangle}{\langle\rho_{\textrm{vac}}\rangle}
%     =
%     \frac{1}{\cos^{2}(\frac{\omega_{0}d}{2})} 
%     \sqrt[3]{\frac{\pi}{2\sqrt{3}|\text{Re}[\chi]|}\frac{\Delta\omega}{\omega_{0}}} +
%     \mathcal{O}(\text{Im}[\chi]),
% \end{align}
in agreement with inverse designs (before saturation due to finite device footprint) in Fig.~\ref{fig:chi4finiteandinf}.
It remains an open question whether
the bounds are loose or there exist structures with infinitely long tapering that may indeed achieve the faster $\sqrt{\Delta\omega}$ scaling of Eq.~\eqref{eq:rhoscaboundexplicit}.

The proposed strategy for absorption cancellation is only useful in situations where extinction is dominated by absorption.
In particular, achieving near-perfect suppression also relies on an infinite device size where radiative losses can be eliminated: as seen in Fig.~\ref{fig:chi4finiteandinf}, $\rho(\omega_0) \to 0$ only as $L\to \infty$.
In the lossless case, $\rho(\omega_0) \propto \exp(-L/\xi)$ for some length-scale $\xi$ consistent with exponential localization~\cite{leistikow2011inhibited,asatryan2001two,yeganegi2014local}.
In the lossy case, inverse design converges on adiabatic tapers consisting of rings of vanishing thickness in the vicinity of the emitter, leading to $\rho(\omega_{0}) \propto 1/\sqrt{L}$, with tapering appearing to reduce reflections from the finite interface.
% While the exponential scaling of 2d PhCs with system size is far more favorable than that of the slow light mechanism, loss-dominated saturation in the former requires switching to the latter past a certain performance threshold.
% Thus, there is necessarily a tradeoff between desired performance and required system size.

\begin{figure}
   \centering
    \includegraphics[width=\linewidth]{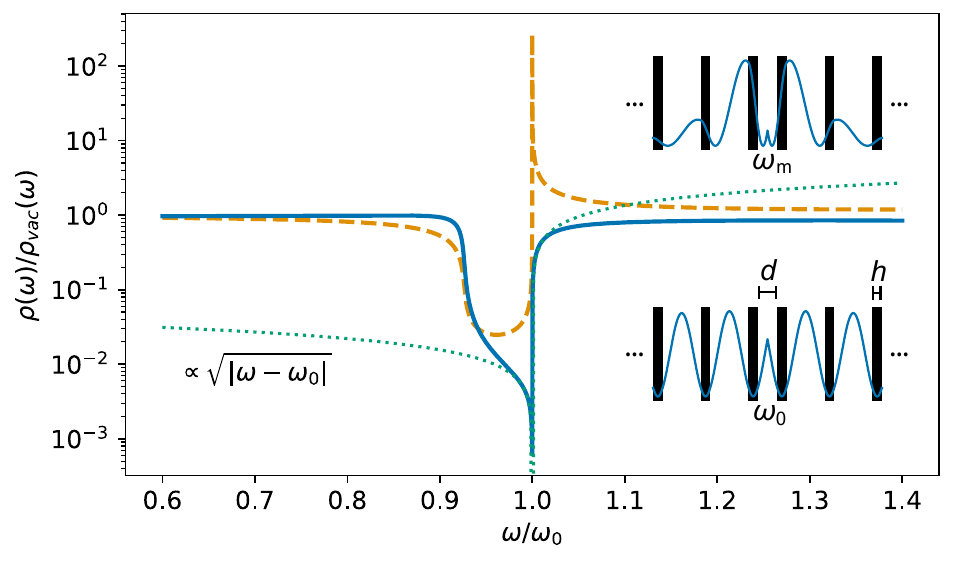}
    \caption{LDOS spectrum, Eq.~\eqref{eq:PhCCavityLDOS}, of a 1d PhC cavity with period $a = \lambda_{0}/2$, material layers of thickness $h = a/50$, and a central air gap of thickness $d = \{a/3, a - h\}$ (solid blue, dashed orange, respectively) with loss $\text{Im}[\chi(\omega)] = 0.1$.
    The dotted green curve shows a square root fit around $\omega_{0}.$
    Inset is a representative intensity profile $|E(z; \omega_{0})|^{2}$ at the band edge, exhibiting minima in the material layers (vertical black bars), as well as $|E(z; \omega_{m})|^{2}$ at the pseudo-midgap, exhibiting minima adjacent to the material layers for $d = a/3$.}
\label{fig:phc1d_plots}
\end{figure}

Finally, note that $d \not\approx (n + \frac{1}{2})\lambda_0$ can be understood as an off-resonance condition for the cavity formed by these slow light mirrors: if $d$ is chosen to be on resonance then instead of suppression the system produces enhancement.
Specifically, setting $d=a-h\approx\frac{\lambda_{0}}{2}$ forms a complete PhC and 
leads to $\rho(\omega_{0})/\rho_{\textrm{vac}}(\omega_{0}) \approx \frac{\sqrt{3}}{\pi}\big(\frac{h}{\lambda_{0}}\big)^{-1}$ as $h \to 0$, enhancing LDOS near $\omega_{0}$ (Fig.~\ref{fig:phc1d_plots} dashed orange curve).
For a fixed material layer thickness $h$, such structures initially yield $1/\sqrt{\Delta\omega}$ divergence in $\langle\rho\rangle$ but absorption ultimately leads to finite saturation as $\Delta\omega \rightarrow 0$.
Optimizing the material thickness at each $\Delta\omega$ produces an analogous diverging integrated LDOS enhancement $\frac{\langle\rho\rangle}{\langle\rho_{\textrm{vac}}\rangle}~\propto~1/\sqrt[3]{\Delta\omega}$ as $\Delta\omega \to 0$. 
This confirms the previously unexplained prediction in Ref.~\cite{chao_maximum_2022} that diverging extinction power is possible in the presence of loss without having to come arbitrarily close to field singularities at infinitely sharp tips~\cite{idemen2003confluent,budaev_electromagnetic_2007,schuck2005improving,kinkhabwala_large_2009,novotny_principles_2012}.
(Note that Ref.~\cite{chao_maximum_2022} finds a weaker $1/\sqrt[4]{\Delta\omega}$ bandwidth dependence as the emitter is adjacent to rather than surrounded by the structured medium.)
In practice, achievable extinction powers will ultimately rest on fabrication and material tolerances. 
Fortunately, the predicted cube root scaling of the optimal material thickness on the bandwidth suggests fabricable feature sizes for moderate bandwidths: considering $\Delta\omega/\omega_{0}=10^{-3}$ and silicon, with $\chi=11.8+3.6\times10^{-3}i$
at near-infrared $\lambda_0 = 1~\mu$m wavelengths, the optimal thickness $h\approx15$~nm, within reach of electron-beam lithography. 
Lastly, we remark once more that while we focused here on TM fields (applicable to thin film slab geometries), similar conclusions follow for 2d TE and 3d fields. Likewise, while we focused on dielectrics the formalism is applicable to metals as well.

%\begin{acknowledgments}
\textit{Acknowledgments---} This work was supported by the National Science Foundation under the Emerging Frontiers in Research and
Innovation (EFRI) program, EFMA-1640986, the Cornell Center for Materials Research (MRSEC) through award
DMR-1719875, the Defense Advanced Research Projects Agency (DARPA) under agreements HR00112090011,
HR00111820046, and HR0011047197. R.K.D. gratefully acknowledges financial support from the Princeton Presidential Postdoctoral Research Fellowship and from the National Academies of Science, Engineering, and Medicine Ford Foundation Postdoctoral Fellowship program. 
The simulations presented in this article were performed on computational
resources managed and supported by Princeton Research
Computing, a consortium of groups including the Princeton Institute for Computational Science and Engineering
(PICSciE) and the Office of Information Technology's
High Performance Computing Center and Visualization Laboratory at Princeton University.
The views, opinions, and findings expressed herein are those of the 
authors and should not be interpreted as representing the official 
views or policies of any institution. 
% \end{acknowledgments}

%REFERENCES
\bibliography{refs}
\end{document}

\title{Suppressing electromagnetic local density of states via \\ slow light in lossy quasi-1d gratings: supplemental material}

\author{Benjamin Strekha}
\affiliation{Department of Electrical and Computer Engineering, Princeton University, Princeton, New Jersey 08544, USA}

\author{Pengning Chao}
\affiliation{Department of Mathematics, Massachusetts Institute of Technology, Cambridge, Massachusetts 02139, USA}

\author{Rodrick Kuate Defo}
\affiliation{Department of Electrical and Computer Engineering, Princeton University, Princeton, New Jersey 08544, USA}

\author{Sean Molesky}
\affiliation{Department of Engineering Physics, Polytechnique Montréal, Montréal, Québec H3T 1J4, Canada}

\author{Alejandro W. Rodriguez}
\affiliation{Department of Electrical and Computer Engineering, Princeton University, Princeton, New Jersey 08544, USA}

\begin{abstract}
    This document provides supplemental material to ``Suppressing electromagnetic local density of states via slow light in lossy quasi-1d gratings'' to assist readers in reproducing our final results as well as extending the techniques to related problems.
\end{abstract}

\maketitle

\section{Relation of power to LDOS}

The local density of states $\text{LDOS}(\omega, \mathbf{r})$ is defined as~\cite{joulain2003definition} (our convention for $\mathbb{G}$ has an additional factor of $k^{2}$ compared to the standard convention)
\begin{align}
    \text{LDOS}(\omega, \mathbf{r})
    \equiv 
    \frac{1}{\pi\omega}\text{Tr}\text{Im}\mathbb{G}(\mathbf{r}, \mathbf{r}; \omega).
\end{align}
Comparing this to the expression for the 
time-averaged power $\rho(\omega, \mathbf{r})$ emitted by a harmonic dipole source
$\vb{J}(\vb{r};\omega) = \delta(\vb{r}-\vb{r}')
J_{x}(\omega)\hat{\vb{x}}$,
\begin{align}
    \rho(\omega, \mathbf{r}) &\equiv -\frac{1}{2} \int \Re{\vb{J}^*(\vb{r}) \cdot \vb{E}(\vb{r})} \,\dd\vb{r}, \\
    &=
    \text{Im}\bigg\{\frac{|J_{x}(\omega)|^{2}}{2\omega}\mathbb{G}_{xx}(\mathbf{r}, \mathbf{r}; \omega) \bigg\},
\end{align}
where we used $\mathbf{E}(\mathbf{r}) = \frac{i}{\omega}\mathbb{G}\mathbf{J},$ we see that $J_{x}(\omega)$ should be independent of frequency for the power radiated by this current to be proportional to the standard LDOS definition. Since we normalize our plots by the radiation in vacuum, we ignore carrying the ``correct'' constant for $J_{x}(\omega)$ (namely, $J_{x}(\omega) = \sqrt{2/\pi}$) to agree with the standard LDOS definition.%, and instead set it to 1 during intermediate calculations.

\section{The $W(\omega)$ function}

Here we discuss some properties of the weight function $W(\omega) \equiv \frac{L(\omega) - L(-\omega)}{\omega\mathcal{N}}$,
where $L(\omega) \equiv \frac{\Delta\omega/\pi}{(\omega-\omega_0)^2 + \Delta\omega^2}$
is a Lorentzian lineshape centered at $\omega_0$ with bandwidth $\Delta\omega$ and the normalization factor $\mathcal{N} \equiv \frac{\omega_{0}}{\omega_{0}^{2} +\Delta\omega^{2}}$ is chosen so that $\int_{0}^{\infty} W(\omega) \dd\omega = 1$. $W(\omega)$ is nonnegative and finite for $\omega\in[0,\infty)$ and integrates to 1, so $\int_{0}^{\infty} Q(\omega) W(\omega) \dd \omega$ can indeed be thought of as a weighted average of $Q(\omega).$ As $\omega \to 0,$
\begin{align}
    W(\omega)
    &=
    \frac{4\Delta\omega}{\pi(\omega_{0}^{2} + \Delta\omega^{2})}
    +
    \frac{8\Delta\omega(\omega_{0}^{2} - \Delta\omega^{2})}{\pi(\omega_{0}^{2} + \Delta\omega^{2})^{3}}\omega^{2}
    +
    \cdots
\end{align}
while if $\omega \to \infty$ then
\begin{align}
    W(\omega)
    &=
    \frac{4\Delta\omega(\omega_{0}^{2} + \Delta\omega^{2})}{\pi\omega^{4}}
    +
    \frac{8\Delta\omega(\omega_{0}^{4} - \Delta\omega^{4})}{\pi \omega^{6}}
    +
    \cdots.
\end{align}
Assuming a dipole source with unit norm, then $\rho_{vac,1d}(\omega) = \frac{1}{4},$ $\rho_{vac,2d}^{TM}(\omega) = \frac{|\omega|}{8}$, $\rho_{vac,2d}^{TE}(\omega) = \frac{|\omega|}{16}$, and $\rho_{vac,3d}(\omega) = \frac{\omega^{2}}{12\pi}$. Because $W(\omega) \sim \omega^{-4}$ as $\omega\to\infty,$ the integral $\int_{0}^{\infty} \rho_{\textrm{vac}}(\omega) W(\omega) \dd\omega$ converges in all cases. In particular, $\langle\rho_{vac,1d}^{TM}\rangle = \frac{1}{4}$, $\langle\rho_{vac,2d}^{TM}\rangle = \frac{1}{4\pi \mathcal{N}}\atan(\frac{\omega_{0}}{\Delta\omega})$, $\langle\rho_{vac,2d}^{TE}\rangle = \frac{1}{8\pi \mathcal{N}}\atan(\frac{\omega_{0}}{\Delta\omega})$, and $\langle\rho_{vac,3d}\rangle = \frac{\omega_{0}^{2} + \Delta\omega^{2}}{12\pi}$.

\section{Sum rule derivation}

In general,
\begin{align}
    \rho_{\textrm{sca}}(\omega, \mathbf{r}) &= \Im{\frac{1}{2\omega}\mathbb{G}_{\textrm{sca},\alpha\alpha}(\mathbf{r}, \mathbf{r}; \omega)},
    \\ &\equiv \Im{s(\omega)},
\end{align}
where $\mathbb{G}_{\textrm{sca}} = \mathbb{G} - \mathbb{G}_{0}$ is the total Green's function of the system minus the vacuum Green's function in the case where $\mathbf{J} \propto \mathbf{e}_{\alpha}$ has a constant unit amplitude.
Here, $\mathbb{G}_{\textrm{sca},\alpha\alpha} \equiv \mathbf{e}_{\alpha}\mathbb{G}_{\textrm{sca}}\mathbf{e}_{\alpha}$.
In our case, we are interested in integrals involving
\begin{align}
    \Im{s(\omega)/\omega} = \Im{\frac{1}{2\omega^{2}}\mathbb{G}_{\textrm{sca},\alpha\alpha}(\mathbf{r}, \mathbf{r}; \omega)}.
\end{align}
In order to evaluate a principal value integral by using the residue theorem, we need to integrate over an indented semicircle around $\omega = 0$ into the upper-half plane, and then take the radius of that detour to 0.
In such a case, we can expand $\mathbb{G}_{\textrm{sca}}(\mathbf{r}, \mathbf{r}; \omega)$ in a Laurent series in a small punctured disk about the origin,
\begin{align}
    \frac{1}{\omega^{2}}\mathbb{G}_{\textrm{sca},\alpha\alpha}(\mathbf{r}, \mathbf{r}; \omega)
    &=
    \frac{1}{\omega^{2}}\mathbb{G}_{\textrm{sca},\alpha\alpha}(\mathbf{r}, \mathbf{r}; 0) + \frac{1}{\omega}\frac{d \mathbb{G}_{\textrm{sca},\alpha\alpha}(\mathbf{r}, \mathbf{r}; 0)}{d\omega} + \cdots.
\end{align}
Switching to polar coordinates $\omega = \epsilon e^{it}$ for $\epsilon > 0, t \in [\pi, 0]$, doing the integral over the indented path and taking the limit of the indentation radius to 0, we find that the principal value of the integral is
\begin{align}
    P.V. 
    \bigg(
    \Im{s(\omega)/\omega}
    \bigg)
    &= 
    \lim_{\epsilon \to 0}
    \Im{
    2\mathbb{G}_{\textrm{sca},\alpha\alpha}(\mathbf{r}, \mathbf{r}; 0)/(2\epsilon)
    }
    +
    \Re{
    \frac{\pi}{2}
    \frac{d \mathbb{G}_{\textrm{sca},\alpha\alpha}(\mathbf{r}, \mathbf{r}; 0)}{d\omega}
    },
    \\
    % &\stackrel{?}{=} \Re{
    % \frac{d \mathbb{G}_{\textrm{sca}}(\mathbf{r}, \mathbf{r}; 0)}{d\omega}
    % }
    % \\
    &= 0.
\end{align}
Introducing a Lorentzian function $L(\omega) \equiv \frac{\Delta\omega/\pi}{(\omega-\omega_0)^2 + \Delta\omega^2}$ in the integrand leads to
\begin{align}
    P.V. 
    \bigg(
    \Im{s(\omega)/\omega}L(\omega)
    \bigg)
    &= 
    \Im{s(\tilde{\omega})/\tilde{\omega}}
    +
    \lim_{\epsilon \to 0}
    \Im{
    2\mathbb{G}_{\textrm{sca},\alpha\alpha}(\mathbf{r}, \mathbf{r}; 0)L(0)/(2\epsilon)
    }
    +
    \Re{
    \frac{\pi}{2}
    \frac{d (\mathbb{G}_{\textrm{sca},\alpha\alpha}L)}{d\omega}\bigg|_{\omega=0}
    },
    \\
    &=
    \Im{s(\tilde{\omega})/\tilde{\omega}}
    +
    \Re{
    \frac{\pi}{2}
    \mathbb{G}_{\textrm{sca},\alpha\alpha}(\mathbf{r}, \mathbf{r}; 0)\frac{dL(0)}{d\omega}
    },
    \\
    &=
    \Im{s(\tilde{\omega})/\tilde{\omega}} 
    +
    \frac{\omega_{0}\Delta\omega}{(\omega_{0}^2 + \Delta\omega^2)^2}\Re{
    \mathbb{G}_{\textrm{sca},\alpha\alpha}(\mathbf{r}, \mathbf{r}; 0)
    },
\end{align}
where $\tilde{\omega} \equiv \omega_{0} + i\Delta\omega.$
It was shown in the SI of Ref.~\cite{shim_fundamental_2019} that the electrostatic ($\omega = 0$) term is nonnegative.%,so for minimization bounds it can be set to 0 and attention can be focused on minimizing $\Im{s(\tilde{\omega})/\tilde{\omega}}.$

The above leads to the ``weighted sum-rule''
\begin{equation}
    \langle \rho_{\textrm{sca}} \rangle%_{\omega_0, \Delta\omega} 
    = \Im\left[\frac{s(\tilde{\omega})}{\tilde{\omega}\mathcal{N}}\right]
    + \frac{2\omega_{0}\Delta\omega}{|\tilde{\omega}|^4\mathcal{N}}\alpha
    \label{eq:cplx_freq_rho_sca}
\end{equation}
stated in the main text.
In the limit of zero bandwidth, only the first term in Eq.~\eqref{eq:cplx_freq_rho_sca} is in general nonzero,
which represents single-frequency LDOS: $\langle\rho\rangle = \rho_{\textrm{vac}}(\omega_{0}) + \rho_{\textrm{sca}}(\omega_{0})$.
As the bandwidth goes to infinity, $\Im\left[\frac{s(\tilde{\omega})}{\tilde{\omega}\mathcal{N}}\right]$
decays rapidly and the second term yields an all-frequency sum rule $\int_{0}^{\infty}\rho_{\textrm{sca}}(\omega)\dd\omega = \pi\alpha/2$~\cite{shim_fundamental_2019}.
To see this, note that the weight function is normalized as $\int_{0}^{\infty} W(\omega) \dd\omega = 1$ so to preserve this the weight function goes to 0 in the limit $\Delta\omega \to \infty;$ in the infinite bandwidth limit, one should rescale the weight function by a factor of $\pi\Delta\omega/4$ so that it approaches 1 everywhere, which then rescales the $\frac{2\omega_{0}\Delta\omega}{|\tilde{\omega}|^4\mathcal{N}}$ prefactor multiplying $\alpha$ to $\pi/2$ in the infinite bandwidth limit.

\section{The electrostatic contribution to bandwidth-averaged LDOS}

For TM and TE, the P.V. converges and so the average can be written as
\begin{align}
    \langle \rho_{\textrm{tot}} \rangle 
    &=
    \lim_{\epsilon\to 0^{+}}\int_{\epsilon}^{\infty} -\frac{1}{2} \int \Re{\vb{J}^*(\vb{r}; \omega) \cdot \vb{E}_{\textrm{tot}}(\vb{r}; \omega)} \,\dd\vb{r} \left[\frac{L(\omega) - L(-\omega)}{\omega\mathcal{N}}\right] \dd\omega, \\
    &= P.V. \bigg( -\frac{1}{2} \int \Re{\vb{J}^*(\vb{r}; \omega) \cdot \vb{E}_{\textrm{tot}}(\vb{r}; \omega)} \,\dd\vb{r}  \frac{L(\omega)}{\omega\mathcal{N}} \bigg),
\end{align}
which can be written as evaluations of terms at $\tilde{\omega}$ (the complex pole in the upper-half plane of the Lorentzian $L(\omega)$) and a zero frequency term by using the residue theorem.
The contribution from the indented path around $\omega = 0$ may or may not be 0.
Separating $\vb{E}_{\textrm{tot}}(\vb{r}; \omega) = \vb{E}_{\textrm{vac}}(\vb{r}; \omega) + \vb{E}_{\textrm{sca}}(\vb{r}; \omega)$, the contribution of the $\mathbf{E}_{\textrm{sca}}$ term was worked out in the previous section.
The vacuum term is similar ($\vb{J}(\vb{r};\omega) = \delta(\vb{r}-\vb{r}')\mathbf{e}_{a}$):
\begin{multline}
P.V. \bigg( -\frac{1}{2} \int \Re{\vb{J}^*(\vb{r}; \omega) \cdot \vb{E}_{\textrm{vac}}(\vb{r}; \omega)} \,\dd\vb{r} \frac{L(\omega)}{\omega\mathcal{N}} \bigg)
    =\\
    -\frac{1}{2} \int \Re{\vb{J}^*(\vb{r}; \tilde{\omega}) \cdot \vb{E}_{\textrm{vac}}(\vb{r}; \tilde{\omega}) \frac{1}{\tilde{\omega}\mathcal{N}} }\,\dd\vb{r}
    +
    \frac{\omega_{0}\Delta\omega}{(\omega_{0}^2 + \Delta\omega^2)^2}\Re{\mathbf{e}_{a}\cdot
    \mathbb{G}_{0}(\mathbf{r}, \mathbf{r}; 0)\mathbf{e}_{a}
    }.
\end{multline}

For the TM case there are no fields generated at $\omega = 0$ (trivial electrostatics) so the indented path term vanishes for $\mathbb{G}_{0}$ and for $\mathbb{G}_{\textrm{sca}},$ which implies that
\begin{align}
    \langle \rho_{\textrm{tot}} \rangle^{TM} 
    %&= -\frac{1}{2} \Re{\int \vb{J}^*(\vb{r}; \tilde{\omega}) \cdot \vb{E}_{\textrm{vac}}(\vb{r}; \tilde{\omega}) \,\dd\vb{r} \frac{1}{\tilde{\omega}\mathcal{N}}  } - \frac{1}{2} \Re{\int \vb{J}^*(\vb{r}; \tilde{\omega}) \cdot \vb{E}_{\textrm{sca}}(\vb{r}; \tilde{\omega}) \,\dd\vb{r} \frac{1}{\tilde{\omega}\mathcal{N}} } \\
    &= P.V. \bigg( -\frac{1}{2} \int \Re{\vb{J}^*(\vb{r}; \omega) \cdot \vb{E}_{\textrm{tot}}(\vb{r}; \omega)} \,\dd\vb{r}  \frac{L(\omega)}{\omega\mathcal{N}} \bigg), \\
    &= -\frac{1}{2} \int \Re{ \vb{J}^*(\vb{r}; \tilde{\omega}) \cdot \vb{E}_{\textrm{tot}}(\vb{r}; \tilde{\omega})\frac{1}{\tilde{\omega}\mathcal{N}} } \,\dd\vb{r}.
\end{align}
This can be explicitly confirmed for the vacuum portion by using $\mathbf{E}_{\textrm{vac}}^{TM}(\mathbf{r}; \tilde{\omega}) = -\frac{\omega}{4}H_{0}^{(1)}(\tilde{\omega}\rho)\mathbf{e}_{z}$ along with Mathematica to find
\begin{align}
    \lim_{\rho\to 0} \Re{-\frac{1}{2} \bigg(-\frac{\tilde{\omega}}{4}H_{0}^{(1)}(\tilde{\omega}\rho) \bigg)\frac{1}{\tilde{\omega}\mathcal{N}}} 
    &= \frac{\pi - 2\text{Arg}[\text{Re}(\tilde{\omega}) + i\text{Im}(\tilde{\omega})]}{8\pi} \frac{1}{\mathcal{N}}, \\
    &= \frac{\pi - 2 \arctan(\frac{\text{Im}(\tilde{\omega})}{\text{Re}(\tilde{\omega})})}{8\pi} \frac{1}{\mathcal{N}}, \\
    &= \frac{\arctan(\frac{\text{Re}(\tilde{\omega})}{\text{Im}(\tilde{\omega})})}{4\pi} \frac{1}{\mathcal{N}},
\end{align}
where the last line follows from the identity $\arctan(x) + \arctan(1/x) = \pi/2.$
This agrees precisely with the result $\langle \rho_{\textrm{vac}}^{\text{TM}} \rangle 
= \frac{1}{4\pi \mathcal{N}}\atan(\frac{\omega_{0}}{\Delta\omega})$ provided in the main text and calculated by evaluating $\lim_{\epsilon\to 0^{+}}\int_{\epsilon}^{\infty} \frac{\omega}{8} \left[\frac{L(\omega) - L(-\omega)}{\omega\mathcal{N}}\right] \dd\omega.$

For TE, the electrostatics is nontrivial and one also needs to keep track of the terms in the above expressions resulting from the indented path around $\omega = 0.$ 
Namely, we know that
\begin{align}
    \langle \rho_{\textrm{vac}}^{TE} \rangle
    %&= -\frac{1}{2} \Re{\int \vb{J}^*(\vb{r}; \tilde{\omega}) \cdot \vb{E}_{\textrm{vac}}(\vb{r}; \tilde{\omega}) \,\dd\vb{r} \frac{1}{\tilde{\omega}\mathcal{N}}  } - \frac{1}{2} \Re{\int \vb{J}^*(\vb{r}; \tilde{\omega}) \cdot \vb{E}_{\textrm{sca}}(\vb{r}; \tilde{\omega}) \,\dd\vb{r} \frac{1}{\tilde{\omega}\mathcal{N}} } \\
    &=
    P.V. \bigg( -\frac{1}{2} \int \Re{\vb{J}^*(\vb{r}; \omega) \cdot \vb{E}_{\textrm{vac}}(\vb{r}; \omega)} \,\dd\vb{r}  \frac{L(\omega)}{\omega\mathcal{N}} \bigg), \\
    &= \frac{\arctan(\frac{\omega_{0}}{\Delta\omega})}{8\pi\mathcal{N}},
\end{align}
but contour methods require more care.
From Section 5.4 in Ref.~\cite{felsen1994radiation}, for $\rho > 0$ the field is given by
\begin{align}
    \mathbf{E}_{\textrm{vac}} &=
    -\frac{1}{4\rho}\sin\phi H_{1}^{(1)}(\tilde{k}\rho)\mathbf{e}_{\rho}
    -
    \frac{\tilde{k}}{8}\cos\phi
    (H_{0}^{(1)}(\tilde{k}\rho) - H_{2}^{(1)}(\tilde{k}\rho))\mathbf{e}_{\phi}.
    % \\
    % &=
    % -\frac{1}{4\rho}\sin\phi H_{1}^{(1)}(\tilde{k}\rho)\mathbf{e}_{\rho}
    % -
    % \frac{\tilde{k}}{4}\cos\phi
    % H_{1}^{(1)'}(\tilde{k}\rho)\mathbf{e}_{\phi} \\
    % &=
    % -\frac{1}{4\rho}\sin\phi H_{1}^{(1)}(\tilde{k}\rho)\mathbf{e}_{\rho}
    % -
    % \frac{1}{4}\cos\phi
    % [\frac{d}{d\rho}H_{1}^{(1)}(\tilde{k}\rho)]\mathbf{e}_{\phi}.
    \label{eq:TElinesourcefieldFelsen}
\end{align}
Using this closed form expression, we find that using the contour method is not useful for evaluating the vacuum contribution in the TE case since there are divergences in both terms of
\begin{align}
    \langle \rho_{\textrm{vac}}^{TE} \rangle 
    %&= -\frac{1}{2} \Re{\int \vb{J}^*(\vb{r}; \tilde{\omega}) \cdot \vb{E}_{\textrm{vac}}(\vb{r}; \tilde{\omega}) \,\dd\vb{r} \frac{1}{\tilde{\omega}\mathcal{N}}  } - \frac{1}{2} \Re{\int \vb{J}^*(\vb{r}; \tilde{\omega}) \cdot \vb{E}_{\textrm{sca}}(\vb{r}; \tilde{\omega}) \,\dd\vb{r} \frac{1}{\tilde{\omega}\mathcal{N}} } \\
    &=
    P.V. \bigg( -\frac{1}{2} \int \Re{\vb{J}^*(\vb{r}; \omega) \cdot \vb{E}_{\textrm{vac}}(\vb{r}; \omega)} \,\dd\vb{r}  \frac{L(\omega)}{\omega\mathcal{N}} \bigg), \\
    &= -\frac{1}{2} \int \Re{ \vb{J}^*(\vb{r}; \tilde{\omega}) \cdot \vb{E}_{\textrm{vac}}(\vb{r}; \tilde{\omega})\frac{1}{\tilde{\omega}\mathcal{N}} } \,\dd\vb{r}
    +
    \frac{\omega_{0}\Delta\omega}{(\omega_{0}^2 + \Delta\omega^2)^2}\frac{1}{\mathcal{N}}
    \Re{\mathbb{G}_{vac,yy}(\mathbf{r}, \mathbf{r}, 0)
    }.
\end{align}
Explicitly, we find that
\begin{align}
    \frac{\omega_{0}\Delta\omega}{(\omega_{0}^2 + \Delta\omega^2)^2}
    \frac{1}{\mathcal{N}}
    \Re{
    \mathbb{G}_{vac,yy}(\mathbf{r}, \mathbf{r}, 0)
    }
    &=
    \lim_{\rho \to 0}\frac{\omega_{0}\Delta\omega}{(\omega_{0}^2 + \Delta\omega^2)^2} \frac{1}{2\pi\mathcal{N}\rho^{2}}
\end{align}
and also (expanding $\tilde{\omega} \equiv \omega_{0} + i\Delta\omega$)
\begin{multline}
    -\frac{1}{2} \int \Re{ \vb{J}^*(\vb{r}; \tilde{\omega}) \cdot \vb{E}_{\textrm{vac}}(\vb{r}; \tilde{\omega})\frac{1}{\tilde{\omega}\mathcal{N}} } \,\dd\vb{r}
    \\
    =
    \lim_{\rho\to 0}
    \Re{
    -\frac{i}{4\tilde{\omega}^{2}\pi\mathcal{N}\rho^{2}} 
    + i\frac{(-1 + 2\text{EulerGamma} - i\pi + 2\text{Log}[\tilde{\omega}/2] + 2\log{\rho})}{16\pi\mathcal{N}}
     + \cdots},
     \\
     =
    \lim_{\rho\to 0}
    \Re{
    -\frac{i\tilde{\omega}^{*2}}{4|\tilde{\omega}|^{4}\pi\mathcal{N}\rho^{2}} 
    + \frac{(\pi - 2\text{Arg}[\tilde{\omega}])}{16\pi\mathcal{N}}}, \\
     =
    \lim_{\rho\to 0}
    -\frac{\omega_{0}\Delta\omega}{(\omega_{0}^{2} + \Delta\omega^{2})^{2}}\frac{1}{2\pi\mathcal{N}\rho^{2}} 
    + \frac{\arctan(\frac{\omega_{0}}{\Delta\omega})}{8\pi\mathcal{N}}.
\end{multline}
Thus, we see a diverging term $\lim_{\rho \to 0}\frac{\omega_{0}\Delta\omega}{(\omega_{0}^2 + \Delta\omega^2)^2} \frac{1}{2\pi\mathcal{N}\rho^{2}}$ appearing in the complex frequency $\tilde{\omega}$ term and also in the electrostatic $\omega=0$ term when trying to use contour methods on the vacuum contribution. To be safe, one should use contour methods only on the terms involving $\mathbb{G}_{\textrm{sca}}$ in analytics and numerics. Since a dipole in 3d also has nontrivial electrostatics, similar remarks hold in that setting.

\section{Formulation of the optimization problem}

Since we are seeking lower bounds on $\langle \rho \rangle$, the nonnegative electrostatic term $\alpha$ can be ignored and thus set to zero, allowing us to focus on the remaining term $\Im[\frac{s(\tilde{\omega})}{\tilde{\omega}\mathcal{N}}].$
(For TM line sources, the electrostatic term is always exactly 0 for finite $\mathbf{J}(\mathbf{r}; \omega=0)$.)
The structural design problem for minimizing bandwidth-integrated LDOS can therefore be formulated as follows:
\begin{subequations}
\begin{align}
    \min_{\chi(\vb{r}; \tilde{\omega})} \,\, \Im\left[\frac{s(\tilde{\omega})}{\tilde{\omega}\mathcal{N}}\right] &= -\frac{1}{2\mathcal{N}} \Im\bigg[ \int \vb{E}_{\textrm{vac}}(\vb{r}) \cdot \vb{P}(\vb{r}) \,\dd\vb{r} \bigg]
    \nonumber \\
% \textrm{s.t.} \quad & \forall \vb{r} \in V \nonumber \\
\textrm{s.t.} \quad & \nonumber \\
    \curl \curl \vb{E}(\vb{r}; \tilde{\omega}) &- \tilde{\omega}^2 [1+\chi(\vb{r}; \tilde{\omega})] \vb{E}(\vb{r}; \tilde{\omega}) = i\tilde{\omega}\vb{J}(\vb{r}; \tilde{\omega}) \label{eq:maxwell_constraint} \\
    \chi(\vb{r}; \tilde{\omega}) &= 
    \begin{cases}
    0 \text{ or } \chi(\tilde{\omega}) & \vb{r} \in V \\
    0 &  \vb{r} \notin V 
    \end{cases} \\
    \vb{P}(\vb{r}; \tilde{\omega})&=\chi(\vb{r}; \tilde{\omega}) \vb{E}(\vb{r}; \tilde{\omega})
\label{eq:structural_opt} 
\end{align}
\end{subequations}
where $V$ is a prescribed design region that the structure resides
within and $\chi(\tilde{\omega})$ is the Fourier transform of the bulk material
susceptibility evaluated at complex frequency $\tilde{\omega}$.
Due to the high dimensionality of the
structural degrees of freedom $\chi(\vb{r})$ and the nonconvex dependence of the field $\vb{E}(\vb{r})$ on $\chi(\vb{r})$,
it is generally not possible to solve for the global optimum~\cite{christiansen_inverse_2021}. % of Eq.~\eqref{eq:TO_objective}~\cite{christiansen_inverse_2021}. 
Fortunately, it is possible to exploit an alternative parametrization of the
problem in which the polarization density
$\vb{P}(\vb{r})=\chi(\vb{r}) \vb{E}(\vb{r})$ rather than
the susceptibility $\chi(\vb{r})$ serves as the optimization degree of freedom to
obtain bounds on $\langle\rho\rangle$ applicable to \emph{any} possible
structure made of susceptibility $\chi$ and contained in the design region $V$~\cite{molesky_global_2020,chao_physical_2022}.

The key to formulating a shape-independent
bound on $\langle\rho\rangle$ is to forgo the need for structural information by
relaxing the requirement that $\vb{P}$ satisfies Maxwell's equations
everywhere (only possible if the structure is known), and to instead impose a smaller subset of wave
constraints~\cite{chao_physical_2022}. Such constraints can indeed be derived from
Maxwell's equations through a complex frequency generalization of the
time-harmonic Poynting's theorem~\cite{chao_physical_2022,chao_maximum_2022}. 
Following the procedure laid out in Refs.~\cite{molesky_global_2020,chao_physical_2022,chao_maximum_2022}, one arrives at an optimization problem:
\begin{align}
\min_{\vb{P}(\vb{r}; \tilde{\omega})\in V} \, \Im\left[\frac{s(\tilde{\omega})}{\tilde{\omega}\mathcal{N}}\right] &= -\frac{1}{2\mathcal{N}} \Im\bigg[ \int \vb{E}_{\textrm{vac}}(\vb{r}) \cdot \vb{P}(\vb{r}) \,\dd\vb{r} \bigg] \nonumber \\
\textrm{s.t.} \quad  &\forall V_k \subseteq V, \nonumber \\
\int_{V_k} \vb{E}_{\textrm{vac}}^{*}(\vb{r}) \cdot \vb{P}(\vb{r}) \,\dd\vb{r} &= \int_{V_k} \vb{P}^*(\vb{r}) \cdot \int_{V_k} \mathbb{U}(\vb{r},\vb{r}') \cdot \vb{P}(\vb{r}') \,\dd \vb{r}'\, \dd \vb{r},
\label{eq:primal_problem}
\end{align}
where we have defined the composite operator $\U(\vb{r},\vb{r}') \equiv \chi^{*-1} \delta(\vb{r}-\vb{r}') - \G_{0}^*(\vb{r},\vb{r}')$, and where $V_k \subseteq V$ is any spatial region within the design domain
$V$. Equation~\eqref{eq:primal_problem} constitutes a quadratically constrained linear
program with the optimization variable $\vb{P}(\mathbf{r};\tilde{\omega})$.
Although the nonconvexity of some of the constraints precludes any guarantee of finding globally optimal solutions,
a bound on the problem can be efficiently computed via the Lagrange dual
function~\cite{angeris_heuristic_2021,chao_physical_2022,boyd_convex_2004,strekha2022tracenoneq}.
The constraints in Eq.~\eqref{eq:primal_problem} can be interpreted as a statement of the
conservation of energy (per unit time) over each spatial region $V_{k}$, requiring only
specification of the allotted design footprint $V$ and material
susceptibility $\chi$. Note also that geometric information is
contained only implicitly in $\vb{P}$, with material properties
specified by the known complex scalar $\chi$, so that solution of the optimization problem constitutes a shape-independent bound.

\section{Lagrangian duality given global constraints}

In this section we detail how the dual problem for the two global power conservation constraints is equivalent to the dual problem for a single constraint with an additional parameter in the form of a complex phase rotation. 

We are interested in LDOS minimization, so we maximize the negative of LDOS.
In particular, we are interested in placing dual bounds on the minimum power extracted from a dipole source (the vacuum part is independent of the polarization current) given global conservation of power:
\begin{subequations}
\begin{align}
\text{maximize} \quad & -\Im\left[\frac{s(\tilde{\omega})}{\tilde{\omega}\mathcal{N}}\right] = 
\frac{1}{2\mathcal{N}} \Im[\bra{\vb{E}^*_v} \ket{\vb{P}}] \\
\text{such that} \quad &\Re(\bra{\vb{E}_v}\ket{\vb{P}}) - \expval{\Sym\U}{\vb{P}} = 0 \\
&\Im(\bra{\vb{E}_v}\ket{\vb{P}}) - \expval{\Asym\U}{\vb{P}} = 0
\end{align}
\end{subequations}
where $\U \equiv \chi^{-1\dagger} - \mathbb{G}_{0}^{\dagger}.$ 
The Lagrangian for this constrained optimization problem is
\begin{align*}
    \mathcal{L}(\vb{P}, \alpha_{Re}, \alpha_{Im}) &= \frac{1}{2\mathcal{N}} \Im(\bra{\vb{E}^*_v} \ket{\vb{P}}) + \alpha_{Re} \big[ \Re(\bra{\vb{E}_v}\ket{\vb{P}}) - \expval{\Sym\U}{\vb{P}} \big] + \alpha_{Im} \big[\Im(\bra{\vb{E}_v}\ket{\vb{P}}) - \expval{\Asym\U}{\vb{P}} \big], \\
    &= \frac{1}{2\mathcal{N}} \Im(\bra{\vb{E}_v^*} \ket{\vb{P}}) + \big(\frac{\alpha_{Re}}{2} + \frac{\alpha_{Im}}{2i}\big) \bra{\vb{E}_v}\ket{\vb{P}} + \big(\frac{\alpha_{Re}}{2} - \frac{\alpha_{Im}}{2i}\big) \bra{\vb{P}}\ket{\vb{E}_v} \\
    & - \expval{\big[\big(\frac{\alpha_{Re}}{2} + \frac{\alpha_{Im}}{2i}\big)\U + \big(\frac{\alpha_{Re}}{2} - \frac{\alpha_{Im}}{2i}\big)\U^\dagger  \big]  }{\vb{P}} \numthis ,
\end{align*}
with the corresponding dual function
\begin{equation}
    \mathcal{D}(\alpha_{Re}, \alpha_{Im}) = \max_{\vb{P}} \mathcal{L}(\vb{P}, \alpha_{Re}, \alpha_{Im}) .
\end{equation}
Defining $\alpha \equiv \sqrt{\alpha_{Re}^2+\alpha_{Im}^2}$ and a complex phase rotation $p \equiv e^{i\theta} = (\alpha_{Im} + i\alpha_{Re})/\alpha,$ the Lagrangian can be rewritten as
\begin{align}
    \mathcal{L}(\vb{P}, \alpha; \theta) &= \frac{1}{2\mathcal{N}} \Im(\bra{\vb{E}_v^*} \ket{\vb{P}}) + \alpha \bigg( \frac{e^{i\theta}\bra{\vb{E}_v}\ket{\vb{P}} - e^{-i\theta} \bra{\vb{P}}\ket{\vb{E}_v}}{2i} - \expval{ \frac{e^{i\theta}\U - e^{-i\theta}\U^\dagger}{2i} }{\vb{P}} \bigg), \\
    &= \frac{1}{2\mathcal{N}} \Im(\bra{\vb{E}_v^*} \ket{\vb{P}}) + \alpha \big[ \Im(p \bra{\vb{E}_v}\ket{\vb{P}}) - \expval{\Asym(p \U)}{\vb{P}} \big],
\end{align}
which is exactly the Lagrangian of the single constraint optimization
\begin{subequations}
\begin{align}
\text{maximize} \quad &  \frac{1}{2\mathcal{N}}\Im(\bra{\vb{E}_v^*} \ket{\vb{P}}) \label{eq:theta_primal_objective}\\
\text{such that} \quad &\Im(p \bra{\vb{E}_v}\ket{\vb{P}}) - \expval{\Asym(p \U)}{\vb{P}} = 0 \label{eq:theta_primal_constraint}
\end{align}
\label{eq:theta_primal}
\end{subequations}
with corresponding dual function
\begin{equation}
    \mathcal{D}(\alpha; \theta) = \max_{\vb{P}} \mathcal{L}(\vb{P}, \alpha; \theta) .
    \label{def:theta_dual}
\end{equation}
In particular, $(e^{i\theta}, \alpha)$ is just an alternate parametrization of the multiplier space of $(\alpha_{Re}, \alpha_{Im})$, and the tightest dual bound is
\begin{equation}
    \min_{\alpha_{Re}, \alpha_{Im}} \mathcal{D}(\alpha_{Re}, \alpha_{Im}) = \min_\theta \min_\alpha \mathcal{D}(\alpha; \theta) .
\end{equation}
We can now derive an expression for $\min_\alpha \mathcal{D}(\alpha; \theta)$ for fixed phase rotation $\theta$. First, note that $\max_{\vb{P}} \mathcal{L}(\vb{P}, \alpha; \theta)$ only has a finite maximum when $\Asym(p\U) \succ 0$. The requirement that $\Asym(p\U) \succ 0$ restricts the range of complex phase rotation $p$ can take. We evaluate Eq.~\eqref{def:theta_dual} by calculating the stationary point of $\mathcal{L}$ with respect to $\ket{\vb{P}}$:
\begin{equation}
    \pdv{\mathcal{L}}{\bra{\vb{P}}} = 0 \quad \Rightarrow \quad \ket{\vb{P}} = \frac{i}{4\alpha\mathcal{N}} \Asym(p\U)^{-1} \ket{\vb{E}_v^*} + \frac{i}{2} p^* \Asym(p\U)^{-1} \ket{\vb{E}_v}. 
    \label{eq:theta_Popt}
\end{equation}
This stationary point maximizes $\mathcal{L}$ when $\Asym(p\U) \succ 0$. Given this positive definite condition, the primal problem Eq.~\eqref{eq:theta_primal} is also a convex problem with a nonempty feasible set, so strong duality holds by Slater's condition~\cite{boyd_convex_2004}.
Thus to solve for the optimal $\alpha_{\textrm{opt}}$ that minimizes $\mathcal{D}(\alpha;\theta)$, we substitute Eq.~\eqref{eq:theta_Popt} into Eq.~\eqref{eq:theta_primal_constraint} and obtain
\begin{align}
    \alpha_{\textrm{opt}} &= \sqrt{\frac{1}{4\lvert p \rvert^2\mathcal{N}^{2}} \frac{\expval{\Asym(p\U)^{-1}}{\vb{E}_v^*} }{\expval{\Asym(p\U)^{-1}}{\vb{E}_v}}}, \\
    &= \frac{1}{2\mathcal{N}},
    \label{eq:theta_alphaopt}
\end{align}
where we used $\expval{\Asym(p\U)^{-1}}{\vb{E}_v^*} = \expval{\Asym(p\U)^{-1}}{\vb{E}_v}.$
Now Eq.~\eqref{eq:theta_Popt} and Eq.~\eqref{eq:theta_alphaopt} can be simultaneously substituted back into Eq.~\eqref{def:theta_dual} to get the bound
\begin{align*}
    -\text{Im}\left[\frac{s(\tilde{\omega})}{\tilde{\omega}\mathcal{N}}\right] \leq \min_{\alpha} \mathcal{D}(\alpha; \theta) 
    =& \, 
    \frac{1}{4\mathcal{N}} \Re\Big\{ \expval{\Asym(p\U)^{-1}}{\vb{E}_v} + p^* \bra{\vb{E}_v^*}\Asym(p\U)^{-1}\ket{\vb{E}_v} \Big\} \numthis
    \label{eq:theta_bound}
\end{align*}
or
\begin{align*}
    \text{Im}\left[\frac{s(\tilde{\omega})}{\tilde{\omega}\mathcal{N}}\right] \geq -\min_{\alpha} \mathcal{D}(\alpha; \theta) =& \, -\frac{1}{4\mathcal{N}} 
    \Re\Big\{\expval{\Asym(p\U)^{-1}}{\vb{E}_v} + p^* \bra{\vb{E}_v^*}\Asym(p\U)^{-1}\ket{\vb{E}_v} \Big\}. \numthis
    \label{eq:theta_bound2}
\end{align*}

% \begin{align*}
%     -\text{Im}\left[\frac{s(\tilde{\omega})}{\tilde{\omega}\mathcal{N}}\right] \leq \min_{\alpha} \mathcal{D}(\alpha; \theta) =& \, \frac{1}{4\mathcal{N}}  \sqrt{ \expval{\Asym(p\U)^{-1}}{\vb{E}_v^*} \expval{\Asym(p\U)^{-1}}{\vb{E}_v} } \\
%     &+ \frac{1}{4\mathcal{N}} \Re\Big\{ p^* \bra{\vb{E}_v^*}\Asym(p\U)^{-1}\ket{\vb{E}_v} \Big\} \numthis
%     \label{eq:theta_bound}
% \end{align*}
% or
% \begin{align*}
%     \text{Im}\left[\frac{s(\tilde{\omega})}{\tilde{\omega}\mathcal{N}}\right] \geq -\min_{\alpha} \mathcal{D}(\alpha; \theta) =& \, -\frac{1}{4\mathcal{N}}  \sqrt{ \expval{\Asym(p\U)^{-1}}{\vb{E}_v^*} \expval{\Asym(p\U)^{-1}}{\vb{E}_v} } \\
%     &- \frac{1}{4\mathcal{N}} \Re\Big\{ p^* \bra{\vb{E}_v^*}\Asym(p\U)^{-1}\ket{\vb{E}_v} \Big\}. \numthis
%     \label{eq:theta_bound2}
% \end{align*}
The tightest bound on $\text{Im}\left[\frac{s(\tilde{\omega})}{\tilde{\omega}\mathcal{N}}\right]$ in this framework follows by doing the remaining optimization over $\theta$ which satisfy $\Asym(e^{i\theta}\U) \succ 0.$

We remark that one could instead consider window functions comprising ``higher-order Lorentzians'' (for example, $f(\omega) \equiv \frac{\sqrt{2}(\Delta\omega)^{3}}{\pi}\frac{1}{(\omega-\omega_{0})^{4} + (\Delta\omega)^{4}}$) in place of $\frac{L(\omega)}{\omega\mathcal{N}}$ to regularize the ultra-violet divergence in $\rho_{\textrm{vac}}(\omega)$.
For the case of a different weight function, with multiple poles $\{\omega_{k}\}$ in the upper-half plane, 
we note that each pole contributes a term $\frac{1}{2\mathcal{N}} \Im{c_{k}\bra{\vb{E}^*_v} \ket{\vb{P}}}$ for some complex coefficient $c_{k}$
so that the independent problems (no cross-constraints) at each pole instead have a dual of the form
\begin{align*}
    \text{Im}\left[c_{k}\frac{s(\tilde{\omega}_{k})}{\tilde{\omega}_{k}\mathcal{N}}\right] \geq -\min_{\alpha} \mathcal{D}(\alpha; \theta) =& \,
    -\frac{1}{4\mathcal{N}}\Re\Big\{\lvert c_{k} \rvert  \expval{\Asym(p\U)^{-1}}{\vb{E}_v} + p^* c_{k}\bra{\vb{E}_v^*}\Asym(p\U)^{-1}\ket{\vb{E}_v} \Big\}. \numthis
    \label{eq:theta_bound_ck}
\end{align*}
However, preliminary numerical bounds seem to indicate that adding multiple poles without cross-constraints in general leads to trivial bounds.
This explains the choice of the weight function in the main text, which decays faster than a Lorentzian but still only has one complex pole in the upper-half plane but changes the order of the pole at $\omega = 0.$
The electrostatic term vanishes for TM line sources, so no cross-constraints are needed. 
For TE line sources and point dipoles in 3d, ignoring cross-constraints with the electrostatic contribution still results in nontrivial bounds for small bandwidths (but approach the trivial bound $\geq 0$ and then exit the domain of feasibility for large bandwidths).
However, adding more poles in the upper-half plane without cross-constraints in general appears to lead to trivial bounds, even for small bandwidths (and even for TM line sources). Essentially, $\omega = 0$ is the case of electrostatics while the $\omega \to \infty$ limit is a perturbative regime~\cite{shim_fundamental_2019}.
At all other frequencies on the real line or in the upper-half plane, one typically cannot make general statements.

\section{Global constraint bounds - vector cylindrical waves}

\subsection{Green's function identities}
Consider the vector cylindrical wave functions~\cite{tsang_scattering_2004,chew1999waves}
\begin{align}
    Rg\mathbf{L}_{n}(k_{\rho}, k_{z}, \mathbf{r}) &= \nabla J_{n}(k_{\rho}\rho)e^{ik_{z}z + in\phi}, \\
    Rg\mathbf{M}_{n}(k_{\rho}, k_{z}, \mathbf{r}) &= \nabla \times [\mathbf{e}_{z}J_{n}(k_{\rho}\rho)e^{ik_{z}z + in\phi}], \\
    Rg\mathbf{N}_{n}(k_{\rho}, k_{z}, \mathbf{r}) &= \frac{1}{\sqrt{k_{\rho}^{2} + k_{z}^{2}}} \nabla\times Rg\mathbf{M}_{n}(k_{\rho}, k_{z}, \mathbf{r}), \\
    &= \frac{1}{\sqrt{k_{\rho}^{2} + k_{z}^{2}}} \nabla\times\nabla\times [\mathbf{e}_{z}J_{n}(k_{\rho}\rho)e^{ik_{z}z + in\phi}].
\end{align}
If there is no Rg, then this is the same as above but where $J_{n}$, the Bessel function of order $n$, is replaced with $H_{n}^{(1)},$ the Hankel function of the first kind. To separate the $\rho$ dependence from $z$ and $\phi$ dependence, write
\begin{align}
    Rg\mathbf{L}_{n}(k_{\rho}, k_{z}, \mathbf{r}) &=
    Rg\mathbf{l}_{n}(k_{\rho}, k_{z}, \boldsymbol{\rho})e^{ik_{z}z + in\phi}, \\
    Rg\mathbf{M}_{n}(k_{\rho}, k_{z}, \mathbf{r}) &=
    Rg\mathbf{m}_{n}(k_{\rho}, k_{z}, \boldsymbol{\rho})e^{ik_{z}z + in\phi}, \\
    Rg\mathbf{N}_{n}(k_{\rho}, k_{z}, \mathbf{r}) &=
    Rg\mathbf{n}_{n}(k_{\rho}, k_{z}, \boldsymbol{\rho})e^{ik_{z}z + in\phi},
\end{align}
with
\begin{align}
    Rg\mathbf{l}_{n}(k_{\rho}, k_{z}, \boldsymbol{\rho}) &= 
    \mathbf{e}_{\rho}k_{\rho}J'_{n}(k_{\rho}\rho) + \mathbf{e}_{\phi}\frac{in}{\rho}J_{n}(k_{\rho}\rho) + \mathbf{e}_{z}ik_{z}J_{n}(k_{\rho}\rho), \\
    Rg\mathbf{m}_{n}(k_{\rho}, k_{z}, \boldsymbol{\rho}) &= 
    \mathbf{e}_{\rho}\frac{in}{\rho}J_{n}(k_{\rho}\rho) - \mathbf{e}_{\phi}k_{\rho}J_{n}'(k_{\rho}\rho), \\
    Rg\mathbf{n}_{n}(k_{\rho}, k_{z}, \boldsymbol{\rho}) &=
    \mathbf{e}_{\rho}\frac{ik_{\rho}k_{z}}{\sqrt{k_{\rho}^{2} + k_{z}^{2}}}J_{n}'(k_{\rho}\rho) - \mathbf{e}_{\phi}\frac{nk_{z}}{\rho\sqrt{k_{\rho}^{2} + k_{z}^{2}}}J_{n}(k_{\rho}\rho) + \mathbf{e}_{z}\frac{k_{\rho}^{2}}{\sqrt{k_{\rho}^{2} + k_{z}^{2}}}J_{n}(k_{\rho}\rho).
\end{align}
The vector cylindrical wave functions obey the following completeness relation
\begin{align}
    \mathbb{I}\delta(\mathbf{r} - \mathbf{r}')
    &= \sum_{n=-\infty}^{\infty}
\frac{1}{(2\pi)^{2}}
\int_{-\infty}^{\infty}
\dd k_{z}\int_{0}^{\infty} \dd k_{\rho}k_{\rho} \nonumber \\
&\bigg[
\frac{Rg\mathbf{M}_{n}(k_{\rho}, k_{z}, \mathbf{r}) Rg\mathbf{M}_{-n}(-k_{\rho}, -k_{z}, \mathbf{r}')}{k_{\rho}^{2}} 
\nonumber \\
&+
\frac{Rg\mathbf{N}_{n}(k_{\rho}, k_{z}, \mathbf{r}) Rg\mathbf{N}_{-n}(-k_{\rho}, -k_{z}, \mathbf{r}')}{k_{\rho}^{2}}
\nonumber \\
&+
\frac{Rg\mathbf{L}_{n}(k_{\rho}, k_{z}, \mathbf{r}) Rg\mathbf{L}_{-n}(-k_{\rho}, -k_{z}, \mathbf{r}')}{k_{\rho}^{2} + k_{z}^{2}}
\bigg].
\end{align}
Using this identity with the Dyadic Green's function equation
\begin{align}
    \nabla\times\nabla\times\mathbb{G}_{0}(\mathbf{r}, \mathbf{r}') - \tilde{k}^{2}\mathbb{G}_{0}(\mathbf{r}, \mathbf{r}') = \tilde{k}^{2}\mathbb{I}\delta(\mathbf{r} - \mathbf{r}')
\end{align}
one can verify that the Green's function for 3d spaces is given by 
\begin{align}
\mathbb{G}_{0}(\mathbf{r}, \mathbf{r}')
&= \tilde{k}^{2} \sum_{n=-\infty}^{\infty}
\frac{1}{(2\pi)^{2}}
\int_{-\infty}^{\infty}
\dd k_{z}\int_{0}^{\infty} \dd k_{\rho}k_{\rho} \nonumber \\
&\bigg[
\frac{Rg\mathbf{M}_{n}(k_{\rho}, k_{z}, \mathbf{r}) Rg\mathbf{M}_{-n}(-k_{\rho}, -k_{z}, \mathbf{r}')}{k_{\rho}^{2}(k_{\rho}^{2} + k_{z}^{2} - \tilde{k}^{2})} 
\nonumber \\
&+
\frac{Rg\mathbf{N}_{n}(k_{\rho}, k_{z}, \mathbf{r}) Rg\mathbf{N}_{-n}(-k_{\rho}, -k_{z}, \mathbf{r}')}{k_{\rho}^{2}(k_{\rho}^{2} + k_{z}^{2} - \tilde{k}^{2})}
\nonumber \\
&+
\frac{Rg\mathbf{L}_{n}(k_{\rho}, k_{z}, \mathbf{r}) Rg\mathbf{L}_{-n}(-k_{\rho}, -k_{z}, \mathbf{r}')}{(-\tilde{k}^{2})(k_{\rho}^{2} + k_{z}^{2})}
\bigg].
\end{align}
The advantage of working in such a basis is that this is a diagonal basis, leading to easy evaluations of inner products and inverse operators. For instance,
\begin{align}
    \mathbb{G}_{0}^{-1}(\mathbf{r}, \mathbf{r}')
    &= \frac{1}{\tilde{k}^{2}} \sum_{n=-\infty}^{\infty}
    \frac{1}{(2\pi)^{2}}
    \int_{-\infty}^{\infty}
    \dd k_{z}\int_{0}^{\infty} \dd k_{\rho}k_{\rho} \nonumber \\
    &\bigg[
    Rg\mathbf{M}_{n}(k_{\rho}, k_{z}, \mathbf{r}) Rg\mathbf{M}_{-n}(-k_{\rho}, -k_{z}, \mathbf{r}')
    \frac{(k_{\rho}^{2} + k_{z}^{2} - \tilde{k}^{2})}{k_{\rho}^{2}} 
    \nonumber \\
    &+
    Rg\mathbf{N}_{n}(k_{\rho}, k_{z}, \mathbf{r}) Rg\mathbf{N}_{-n}(-k_{\rho}, -k_{z}, \mathbf{r}')
    \frac{(k_{\rho}^{2} + k_{z}^{2} - \tilde{k}^{2})}{k_{\rho}^{2}} 
    \nonumber \\
    &+
    Rg\mathbf{L}_{n}(k_{\rho}, k_{z}, \mathbf{r}) Rg\mathbf{L}_{-n}(-k_{\rho}, -k_{z}, \mathbf{r}')
    \frac{(-\tilde{k}^{2})}{k_{\rho}^{2} + k_{z}^{2}} 
    \bigg].
\end{align}
That is,
\begin{align}
    \int d^{3}\mathbf{u}~
    \mathbb{G}_{0}^{-1}(\mathbf{r}, \mathbf{u})\mathbb{G}_{0}(\mathbf{u}, \mathbf{r}')
    =
    \mathbb{I}\delta(\mathbf{r} - \mathbf{r}')
\end{align}
as can be verified by making use of the orthogonality relations (from Chapter 7 of Ref.~\cite{chew1999waves})
\begin{align}
    \int_{0}^{2\pi}d\phi\int_{-\infty}^{\infty}dz\int_{0}^{\infty}d\rho \rho Rg\mathbf{M}_{n}(k_{\rho}, k_{z}, \mathbf{r})\cdot Rg\mathbf{M}_{-n'}(-k_{\rho}', -k_{z}', \mathbf{r}) \nonumber \\
    =(2\pi)^{2}k_{\rho}\delta_{nn'}\delta(k_{z} - k_{z}')\delta(k_{\rho} - k_{\rho}'),
\end{align}

\begin{align}
    \int_{0}^{2\pi}d\phi\int_{-\infty}^{\infty}dz\int_{0}^{\infty}d\rho \rho Rg\mathbf{N}_{n}(k_{\rho}, k_{z}, \mathbf{r})\cdot Rg\mathbf{N}_{-n'}(-k_{\rho}', -k_{z}', \mathbf{r}) \nonumber \\
    =(2\pi)^{2}k_{\rho}\delta_{nn'}\delta(k_{z} - k_{z}')\delta(k_{\rho} - k_{\rho}'),
\end{align}

\begin{align}
    \int_{0}^{2\pi}d\phi\int_{-\infty}^{\infty}dz\int_{0}^{\infty}d\rho \rho Rg\mathbf{L}_{n}(k_{\rho}, k_{z}, \mathbf{r})\cdot Rg\mathbf{L}_{-n'}(-k_{\rho}', -k_{z}', \mathbf{r}) \nonumber \\
    =(2\pi)^{2}(k_{\rho}^{2} + k_{z}^{2})\delta_{nn'}\delta(k_{z} - k_{z}')\frac{\delta(k_{\rho} - k_{\rho}')}{k_{\rho}},
\end{align}
along with the fact that the $Rg\mathbf{M},Rg\mathbf{N},Rg\mathbf{L}$ functions are mutually orthogonal.

\subsection{Electric field due to a line source}

Suppose we have a line source  $\vb{J}(\rho',\phi', z') = \delta(x')\delta(y')\mathbf{p}.$ Then (in our convention $Z = \sqrt{\frac{\mu_{0}}{\epsilon_{0}}} = 1$),
\begin{align}
\mathbf{E}_{\textrm{vac}}(\mathbf{r})
&=
\frac{iZ}{\tilde{k}}\mathbb{G}_{0}\mathbf{J}, \\
&= i\tilde{k}\sum_{n=-\infty}^{\infty}
\frac{1}{2\pi}
\int_{0}^{\infty} \dd k_{\rho}k_{\rho} \nonumber \\
&\bigg[
\frac{Rg\mathbf{M}_{n}(k_{\rho}, 0, \mathbf{r})
\big(\lim_{\mathbf{r}'\to 0}Rg\mathbf{M}_{-n}(-k_{\rho}, 0, \mathbf{r}') \cdot\mathbf{p}\big)
}{k_{\rho}^{2}(k_{\rho}^{2} - \tilde{k}^{2})} 
\nonumber \\
&+
\frac{Rg\mathbf{N}_{n}(k_{\rho}, 0, \mathbf{r}) 
\big(\lim_{\mathbf{r}'\to 0}Rg\mathbf{N}_{-n}(-k_{\rho}, 0, \mathbf{r}') \cdot\mathbf{p}\big)
}{k_{\rho}^{2}(k_{\rho}^{2} - \tilde{k}^{2})}
\nonumber \\
&+
\frac{Rg\mathbf{L}_{n}(k_{\rho}, 0, \mathbf{r}) 
\big(\lim_{\mathbf{r}'\to 0}Rg\mathbf{L}_{-n}(-k_{\rho}, 0, \mathbf{r}') \cdot\mathbf{p}\big)
}{(-\tilde{k}^{2})(k_{\rho}^{2})}
\bigg], \\
&\equiv 
\sum_{n=-\infty}^{\infty}
\frac{1}{2\pi}
\int_{0}^{\infty} \dd k_{\rho}\nonumber \\
&\bigg[
Rg\mathbf{M}_{n}(k_{\rho}, 0, \mathbf{r})
e_{M,n}(\tilde{k}, k_{\rho})
\nonumber \\
&+
Rg\mathbf{N}_{n}(k_{\rho}, 0, \mathbf{r}) 
e_{N,n}(\tilde{k}, k_{\rho})
\nonumber \\
&+
Rg\mathbf{L}_{n}(k_{\rho}, 0, \mathbf{r}) 
e_{L,n}(\tilde{k}, k_{\rho})
\bigg].
\end{align}

\subsubsection{TM vacuum field}
Consider $\mathbf{p} = \mathbf{e}_{z}.$ This should reduce to the worked out TM case. 
\begin{align}
    e_{M,n} 
    &=
    \frac{i\tilde{k}
    \big(\lim_{\mathbf{r}'\to 0}Rg\mathbf{M}_{-n}(-k_{\rho}, 0, \mathbf{r}') \cdot\mathbf{e}_{z}\big)
    }{k_{\rho}(k_{\rho}^{2} - \tilde{k}^{2})} 
    \\
    &=
    0
\end{align}
and
\begin{align}
    e_{N,n} 
    &=
    \frac{i\tilde{k}
    \big(\lim_{\mathbf{r}'\to 0}Rg\mathbf{N}_{-n}(-k_{\rho}, 0, \mathbf{r}') \cdot\mathbf{e}_{z}\big)
    }{k_{\rho}(k_{\rho}^{2} - \tilde{k}^{2})} 
    \\
    &=
    \frac{i\tilde{k}
    }{(k_{\rho}^{2} - \tilde{k}^{2})} 
    \delta_{n,0}
\end{align}
and
\begin{align}
    e_{L,n} 
    &=
    \frac{i\tilde{k}
    \big(\lim_{\mathbf{r}'\to 0}Rg\mathbf{L}_{-n}(-k_{\rho}, 0, \mathbf{r}') \cdot\mathbf{e}_{z}\big)
    }{(-\tilde{k}^{2})(k_{\rho})} 
    \\
    &=
    0.
\end{align}
Therefore,
\begin{align}
\mathbf{E}_{\textrm{vac}}^{TM}(\mathbf{r})
&= 
\frac{1}{2\pi}
\int_{0}^{\infty} \dd k_{\rho}\nonumber 
\bigg[
Rg\mathbf{N}_{0}(k_{\rho}, 0, \mathbf{r})
e_{N,0}(\tilde{k}, k_{\rho}, 0)
\bigg] \\
&=
\frac{i\tilde{k}}{2\pi}
\int_{0}^{\infty} dk_{\rho}
\frac{Rg\mathbf{N}_{0}(k_{\rho}, 0, \mathbf{r})
}{(k_{\rho}^{2} - \tilde{k}^{2})} \\
&=
\frac{i\tilde{k}}{2\pi}
\int_{0}^{\infty} dk_{\rho}
\frac{
J_{0}(k_{\rho}\rho)k_{\rho}
}{(k_{\rho}^{2} - \tilde{k}^{2})}
\mathbf{e}_{z} \\
&= 
\frac{i\tilde{k}}{2\pi}
K(0,-i\tilde{k}\rho)
\mathbf{e}_{z} \\
&=
-\frac{\tilde{k}}{4}H_{0}^{(1)}(\tilde{k}\rho)\mathbf{e}_{z}.
\end{align}
The radiated power per unit length is given by (for real wavevector $k_{0}$)
$
    \rho_{\textrm{vac}}^{TM}(k_{0}) = \frac{k_{0}}{8}.
$

\subsubsection{TE vacuum field - Line source polarized along $\mathbf{e}_{y}$}

Consider now the case where, for example, $\mathbf{p} = \mathbf{e}_{y}.$ Then
\begin{align}
    e_{M,|n|} 
    &=
    \frac{i\tilde{k}
    \big(\lim_{\mathbf{r}'\to 0}Rg\mathbf{M}_{-|n|}(-k_{\rho}, 0, \mathbf{r}') \cdot\mathbf{e}_{y}\big)
    }{k_{\rho}(k_{\rho}^{2} - \tilde{k}^{2})} 
    \\
    &=
    e^{-i|n|\pi/2}\lim_{\rho'\to 0}\frac{i\tilde{k}}
    {k_{\rho}(k_{\rho}^{2} - \tilde{k}^{2})}
    (-i|n|)\frac{\rho^{'|n|-1}(k_{\rho})^{|n|}}{2^{|n|}\Gamma(|n|+1)} \\
    &=
    -\frac{i\tilde{k}}
    {2(k_{\rho}^{2} - \tilde{k}^{2})}
    \delta_{|n|,1}
\end{align}
and
\begin{align}
    e_{M,-|n|} 
    &=
    \frac{i\tilde{k}
    \big(\lim_{\mathbf{r}'\to 0}Rg\mathbf{M}_{|n|}(-k_{\rho}, 0, \mathbf{r}') \cdot\mathbf{e}_{y}\big)
    }{k_{\rho}(k_{\rho}^{2} - \tilde{k}^{2})} 
    \\
    &=
    e^{i|n|\pi/2}\lim_{\rho'\to 0}\frac{i\tilde{k}}
    {k_{\rho}(k_{\rho}^{2} - \tilde{k}^{2})}
    (i|n|)(-1)^{|n|}\frac{\rho^{'|n|-1}(k_{\rho})^{|n|}}{2^{|n|}\Gamma(|n|+1)} \\
    &=
    \frac{i\tilde{k}}
    {2(k_{\rho}^{2} - \tilde{k}^{2})}
    \delta_{|n|,1}
\end{align}
Thus, the $e_{M,n}$ coefficients are nonzero only for $n=\pm 1.$ Similarly,
\begin{align}
    e_{N,n} 
    &=
    \frac{i\tilde{k}
    \big(\lim_{\mathbf{r}'\to 0}Rg\mathbf{N}_{-n}(-k_{\rho}, 0, \mathbf{r}') \cdot\mathbf{e}_{y}\big)
    }{k_{\rho}(k_{\rho}^{2} - \tilde{k}^{2})} 
    \\
    &=
    0
\end{align}
for all $n.$ Also, we find that
\begin{align}
    e_{L,|n|} 
    &=
    \frac{i\tilde{k}
    \big(\lim_{\mathbf{r}'\to 0}Rg\mathbf{L}_{-n}(-k_{\rho}, 0, \mathbf{r}') \cdot\mathbf{e}_{y}\big)
    }{(-\tilde{k}^{2}k_{\rho})} 
    \\
    &=
    \lim_{\rho'\to 0}\frac{i\tilde{k}}
    {(-\tilde{k}^{2}k_{\rho})}
    (-i|n|)\frac{\rho^{'|n|-1}(k_{\rho})^{|n|}}{2^{|n|}\Gamma(|n|+1)} \\
    &=
    -\frac{1}
    {2\tilde{k}}
    \delta_{|n|,1}
\end{align}
and, likewise,
\begin{align}
    e_{L,-|n|} 
    &=
    \frac{i\tilde{k}
    \big(\lim_{\mathbf{r}'\to 0}Rg\mathbf{L}_{|n|}(-k_{\rho}, 0, \mathbf{r}') \cdot\mathbf{e}_{y}\big)
    }{(-\tilde{k}^{2}k_{\rho})} 
    \\
    &=
    \lim_{\rho'\to 0}\frac{i\tilde{k}}
    {(-\tilde{k}^{2}k_{\rho})}
    (i|n|)(-1)^{|n|}\frac{\rho^{'|n|-1}(k_{\rho})^{|n|}}{2^{|n|}\Gamma(|n|+1)} \\
    &=
    -\frac{1}
    {2\tilde{k}}
    \delta_{|n|,1}
\end{align}
The radiated power per unit length is given by (for real wavevector $k_{0}$)
$
    \rho_{\textrm{vac}}^{TE}(k_{0}) = \frac{k_{0}}{16}.
$

\subsection{$\Asym(p\U)$ and $\Asym(p\U)^{-1}$}

In the calculation of the bounds, we need to evaluate $\Asym(p\U)^{-1} \ket{\vb{E}_{\textrm{vac}}}.$ As a first step, we need to find $\Asym(p\U)$ in this cylindrical wave vector basis, and then from that we can find $\Asym(p\U)^{-1}$ in that same basis as well.
Using $\U \equiv \chi^{-1\dagger} - \mathbb{G}_{0}^{\dagger}$ we find that
\begin{align}
    \Asym(p\U)(\mathbf{r}, \mathbf{r}')
    &= \Asym(p\chi^{-1\dagger})(\mathbf{r}, \mathbf{r}') + \Asym(p^{*}\mathbb{G}_{0})(\mathbf{r}, \mathbf{r}') \\
    &=\sum_{n=-\infty}^{\infty}
    \frac{1}{(2\pi)^{2}}
    \int_{-\infty}^{\infty}
    \dd k_{z}\int_{0}^{\infty} \dd k_{\rho}k_{\rho} \nonumber \\
    &\bigg[
    \frac{Rg\mathbf{M}_{n}(k_{\rho}, k_{z}, \mathbf{r}) Rg\mathbf{M}_{-n}(-k_{\rho}, -k_{z}, \mathbf{r}')}{k_{\rho}^{2}} 
    \text{Im}\bigg(
    \frac{p}{\chi^{*}} + \frac{p^{*}\tilde{k}^{2}}{k_{\rho}^{2} + k_{z}^{2} - \tilde{k}^{2}}
    \bigg)
    \nonumber \\
    &+
    \frac{Rg\mathbf{N}_{n}(k_{\rho}, k_{z}, \mathbf{r}) Rg\mathbf{N}_{-n}(-k_{\rho}, -k_{z}, \mathbf{r}')}{k_{\rho}^{2}}
    \text{Im}\bigg(
    \frac{p}{\chi^{*}} + \frac{p^{*}\tilde{k}^{2}}{k_{\rho}^{2} + k_{z}^{2} - \tilde{k}^{2}}
    \bigg)
    \nonumber \\
    &+
    \frac{Rg\mathbf{L}_{n}(k_{\rho}, k_{z}, \mathbf{r}) Rg\mathbf{L}_{-n}(-k_{\rho}, -k_{z}, \mathbf{r}')}{k_{\rho}^{2} + k_{z}^{2}}
    \text{Im}\bigg(
    \frac{p}{\chi^{*}} + \frac{p^{*}\tilde{k}^{2}}{(-\tilde{k}^{2})}
    \bigg)
    \bigg].
\end{align}
It follows that
\begin{align}
    \Asym(p\U)^{-1}(\mathbf{r}, \mathbf{r}')
    &=\sum_{n=-\infty}^{\infty}
    \frac{1}{(2\pi)^{2}}
    \int_{-\infty}^{\infty}
    \dd k_{z}\int_{0}^{\infty} \dd k_{\rho}k_{\rho} \nonumber \\
    &\bigg[
    \frac{Rg\mathbf{M}_{n}(k_{\rho}, k_{z}, \mathbf{r}) Rg\mathbf{M}_{-n}(-k_{\rho}, -k_{z}, \mathbf{r}')}{k_{\rho}^{2}} 
    \bigg(\text{Im}\bigg(
    \frac{p}{\chi^{*}} + \frac{p^{*}\tilde{k}^{2}}{k_{\rho}^{2} + k_{z}^{2} - \tilde{k}^{2}}
    \bigg)\bigg)^{-1}
    \nonumber \\
    &+
    \frac{Rg\mathbf{N}_{n}(k_{\rho}, k_{z}, \mathbf{r}) Rg\mathbf{N}_{-n}(-k_{\rho}, -k_{z}, \mathbf{r}')}{k_{\rho}^{2}}
    \bigg(\text{Im}\bigg(
    \frac{p}{\chi^{*}} + \frac{p^{*}\tilde{k}^{2}}{k_{\rho}^{2} + k_{z}^{2} - \tilde{k}^{2}}
    \bigg)\bigg)^{-1}
    \nonumber \\
    &+
    \frac{Rg\mathbf{L}_{n}(k_{\rho}, k_{z}, \mathbf{r}) Rg\mathbf{L}_{-n}(-k_{\rho}, -k_{z}, \mathbf{r}')}{k_{\rho}^{2} + k_{z}^{2}}
    \bigg(\text{Im}\bigg(
    \frac{p}{\chi^{*}} + \frac{p^{*}\tilde{k}^{2}}{(-\tilde{k}^{2})}
    \bigg)\bigg)^{-1}
    \bigg] \\
    &\equiv\sum_{n=-\infty}^{\infty}
    \frac{1}{(2\pi)^{2}}
    \int_{-\infty}^{\infty}
    \dd k_{z}\int_{0}^{\infty} \dd k_{\rho}k_{\rho} \nonumber \\
    &\bigg[
    \frac{Rg\mathbf{M}_{n}(k_{\rho}, k_{z}, \mathbf{r}) Rg\mathbf{M}_{-n}(-k_{\rho}, -k_{z}, \mathbf{r}')}{k_{\rho}^{2}} 
    (\text{Im}(pU)_{M,n}(p, \chi, \tilde{k}, k_{\rho}, k_{z}))^{-1}
    \nonumber \\
    &+
    \frac{Rg\mathbf{N}_{n}(k_{\rho}, k_{z}, \mathbf{r}) Rg\mathbf{N}_{-n}(-k_{\rho}, -k_{z}, \mathbf{r}')}{k_{\rho}^{2}}
    (\text{Im}(pU)_{N,n}(p, \chi, \tilde{k}, k_{\rho}, k_{z}))^{-1}
    \nonumber \\
    &+
    \frac{Rg\mathbf{L}_{n}(k_{\rho}, k_{z}, \mathbf{r}) Rg\mathbf{L}_{-n}(-k_{\rho}, -k_{z}, \mathbf{r}')}{k_{\rho}^{2} + k_{z}^{2}}
    (\text{Im}(pU)_{L,n}(p, \chi, \tilde{k}, k_{\rho}, k_{z}))^{-1}
    \bigg].
\end{align}

\subsection{Calculation of the dual objective}

We find that
\begin{align}
    \Asym(p\U)^{-1} \ket{\vb{E}_{\textrm{vac}}} 
    &=
    \sum_{n=-\infty}^{\infty}\frac{1}{2\pi}\int_{0}^{\infty}\dd k_{\rho}Rg\mathbf{M}_{n}(k_{\rho}, 0, \mathbf{r})(\text{Im}(pU)_{M,n})^{-1}e_{M,n}(\tilde{k}, k_{\rho}) \nonumber \\
    &+\sum_{n=-\infty}^{\infty}\frac{1}{2\pi}\int_{0}^{\infty}\dd k_{\rho}Rg\mathbf{N}_{n}(k_{\rho}, 0, \mathbf{r})(\text{Im}(pU)_{N,n})^{-1}e_{N,n}(\tilde{k}, k_{\rho}) \nonumber \\
    &+\sum_{n=-\infty}^{\infty}\frac{1}{2\pi}\int_{0}^{\infty}\dd k_{\rho}Rg\mathbf{L}_{n}(k_{\rho}, 0, \mathbf{r})(\text{Im}(pU)_{L,n})^{-1}e_{L,n}(\tilde{k}, k_{\rho}).
\end{align}
Therefore,
\begin{align}
    \bra{\vb{E}_{\textrm{vac}}} \Asym(p\U)^{-1} \ket{\vb{E}_{\textrm{vac}}} 
    &=
    \delta(0)\sum_{P\in\{M,N,L\}}\sum_{n=-\infty}^{\infty}\int_{0}^{\infty}\dd k_{\rho}k_{\rho} (\text{Im}(pU)_{P,n})^{-1}e_{P,n}(\tilde{k}, k_{\rho})e_{P,n}^{*}(\tilde{k}, k_{\rho})
\end{align}
and
\begin{align}
    \bra{\vb{E}_{\textrm{vac}}^{*}} \Asym(p\U)^{-1} \ket{\vb{E}_{\textrm{vac}}} 
    &=
    \delta(0)\sum_{P\in\{M,N,L\}}\sum_{n=-\infty}^{\infty}\int_{0}^{\infty}\dd k_{\rho}k_{\rho} (\text{Im}(pU)_{P,n})^{-1}e_{P,n}(\tilde{k}, k_{\rho})e_{P,-n}(\tilde{k}, k_{\rho})(-1)^{-n}.
\end{align}
All the above inner products were over 3d space, so they diverge due to the translational invariance along $z.$
Normalizing by length $L_{z}$ replaces $\lim_{L_{z}\to\infty} \delta(0)/L_{z}$ with $\frac{1}{2\pi}.$
Combining all this together, we find that (normalized by length)
\begin{align}
    -\text{Im}\left[\frac{s(\tilde{\omega})}{\tilde{\omega}\mathcal{N}}\right] &\leq \frac{1}{4\mathcal{N}} \expval{\Asym(p\U)^{-1}}{\vb{E}_{\textrm{vac}}} + \frac{1}{4\mathcal{N}} \Re\Big\{ p^{*} \bra{\vb{E}_{\textrm{vac}}^{*}}\Asym(p\U)^{-1}\ket{\vb{E}_{\textrm{vac}}} \Big\}, \\
    &\leq
    \text{Re}
    \Bigg[
    \frac{1}{8\pi\mathcal{N}}\sum_{P\in\{M,N,L\}}\sum_{n=-\infty}^{\infty}\int_{0}^{\infty}\dd k_{\rho}k_{\rho} (\text{Im}(pU)_{P,n}(p, \chi, \tilde{k}, k_{\rho}, 0))^{-1}\bigg(e_{P,n}(\tilde{k}, k_{\rho})e_{P,n}^{*}(\tilde{k}, k_{\rho}) \nonumber \\
    &\qquad\qquad\qquad\qquad
    + (-1)^{-n}p^{*}e_{P,n}(\tilde{k}, k_{\rho}) e_{P,-n}(\tilde{k}, k_{\rho})\bigg)
    \bigg].
    \label{eq:vcw_bound_expression}
\end{align}

\subsubsection{TM case}

Evaluating Eq.~\eqref{eq:vcw_bound_expression} for the TM case results in
\begin{align}
    -\langle\rho_{\textrm{sca}}^{TM}\rangle
    &\leq 
    \text{Re}
    \Bigg[
    \frac{1}{8\pi\mathcal{N}}\int_{0}^{\infty}\dd k_{\rho}k_{\rho} (\text{Im}(pU)_{N,0}(p, \chi, \tilde{k}, k_{\rho}, 0))^{-1}(e_{N,0}(\tilde{k}, k_{\rho})e_{N,0}^{*}(\tilde{k}, k_{\rho}) 
    + p^{*}e_{N,0}(\tilde{k}, k_{\rho}) e_{N,0}(\tilde{k}, k_{\rho})
    \bigg], \\
    &\leq
    \frac{1}{8\pi\mathcal{N}}\int_{0}^{\infty}\dd k_{\rho}k_{\rho} 
    \bigg(\text{Im}\bigg(
    \frac{p}{\chi^{*}} + \frac{p^{*}\tilde{k}^{2}}{k_{\rho}^{2} - \tilde{k}^{2}}
    \bigg)\bigg)^{-1}
    \text{Re}
    \Bigg[
    \bigg|\frac{\tilde{k}
    }{k_{\rho}^{2} - \tilde{k}^{2}}\bigg|^{2}
    - 
    \frac{p^{*}\tilde{k}^{2}
    }{(k_{\rho}^{2} - \tilde{k}^{2})^{2}}
    \Bigg].
    \label{eq:TM_vcw_bound_expression}
\end{align}

\subsubsection{TE case}

Note that $(pU)^{-1}_{L,n}$ and the coefficients $e_{L,n}$ are independent of $k_{\rho}$ so that, in general, the $Rg\mathbf{L}$ terms lead to a divergence in Eq.~\eqref{eq:vcw_bound_expression}. However, when plugging the expression for $e_{L,n},$ we note that the integrand involves
\begin{align}
    e_{L,n}(\tilde{k}, k_{\rho})e_{L,n}^{*}(\tilde{k}, k_{\rho}) 
    + (-1)^{-n}p^{*}e_{L,n}(\tilde{k}, k_{\rho}) e_{L,-n}(\tilde{k}, k_{\rho})
    &= \frac{1}{4} (\delta_{n,1} + \delta_{n,-1}) \bigg(\frac{1}{|\tilde{k}|^{2}} - \frac{p^{*}}{\tilde{k}^{2}}\bigg),
\end{align}
which vanishes for $p_{L} \equiv (\frac{\tilde{k}^{*}}{|\tilde{k}|})^{2}.$ The minimization over $\theta$ for the TM case is done numerically using, say, a bisection method. However, doing the minimization numerically for the TE case will fail as there is only one value $p_{L}$ which leads to a finite integral and so one should explicitly discard the diverging contribution and then do any remaining integrals numerically with $p$ set to $p_{L}.$
Note that for general values of $\chi$ and $\Delta\omega$, it might be that $\Asym(p_{L}\U) \not\succ 0.$  Evaluating Eq.~\eqref{eq:vcw_bound_expression} for the TE case results in
\begin{align}
    -\langle\rho_{\textrm{sca}}^{TE}\rangle
    % &\leq 
    % \text{Re}
    % \Bigg[
    % \frac{1}{8\pi}\sum_{n=-1,1}\int_{0}^{\infty}\dd k_{\rho}k_{\rho} (pU)^{-1}_{M,n}(p, \chi, \tilde{k}, k_{\rho}, n)(|\tilde{k}|e_{M,n}(\tilde{k}, k_{\rho})e_{M,n}^{*}(\tilde{k}, k_{\rho}) \nonumber \\
    % &\qquad\qquad\qquad + (-1)^{-n}p^{*}\tilde{k}e_{M,n}(\tilde{k}, k_{\rho}) e_{M,-n}(\tilde{k}, k_{\rho})
    % \bigg], \\
    &\leq
    \frac{1}{16\pi\mathcal{N}}\int_{0}^{\infty}\dd k_{\rho}k_{\rho} 
    \bigg(\text{Im}\bigg(
    \frac{\tilde{k}^{*2}}{|\tilde{k}|^{2}\chi^{*}} + \frac{\tilde{k}^{4}}{|\tilde{k}|^{2}(q^{2} - \tilde{k}^{2})}
    \bigg)\bigg)^{-1}
    \text{Re}
    \Bigg[
    \bigg|\frac{\tilde{k}
    }{k_{\rho}^{2} - \tilde{k}^{2}}\bigg|^{2}
    - 
    \frac{\tilde{k}^{4}
    }{|\tilde{k}|^{2}(k_{\rho}^{2} - \tilde{k}^{2})^{2}}
    \Bigg].
    \label{eq:TE_vcw_bound_expression}
\end{align}

\subsubsection{Dipole at the origin}

Consider a dipole at the origin $\mathbf{J}(\mathbf{r}) = \delta(\mathbf{r})\mathbf{e}_{z}.$ Repeating the same work as above but now working in spherical wave vectors, we eventually find an equation analogous to Eq.~\eqref{eq:vcw_bound_expression} from which we find
\begin{align}
    -\text{Im}\left[\frac{s(\tilde{\omega})}{\tilde{\omega}\mathcal{N}}\right]
    &\leq
    \text{Re}
    \Bigg[
    \frac{1}{12\pi^{2}\mathcal{N}}\int_{0}^{\infty}\dd q~q^{2}
    \bigg(\text{Im}\bigg(
    \frac{p}{\chi^{*}} + \frac{p^{*}\tilde{k}^{2}}{q^{2} - \tilde{k}^{2}}
    \bigg)\bigg)^{-1}
    \bigg(
    \bigg|\frac{\tilde{k}}{q^{2} - \tilde{k}^{2}}\bigg|^{2} - p^{*} \frac{\tilde{k}^{2}}{(q^{2} - \tilde{k}^{2})^{2}}
    \bigg)
    \bigg] \nonumber \\
    &~~~~~~~~+ \text{Re}
    \Bigg[
    \frac{1}{24\pi^{2}\mathcal{N}}\int_{0}^{\infty}\dd q~q^{2}
    \bigg(\text{Im}\bigg(
    \frac{p}{\chi^{*}} + \frac{p^{*}\tilde{k}^{2}}{(- \tilde{k}^{2})}
    \bigg)\bigg)^{-1}
    \bigg(
    \frac{1}{|\tilde{k}|^{2}} - \frac{p^{*}}{\tilde{k}^{2}}
    \bigg)
    \bigg].
\end{align}
As in the TE line source case, the $Rg\mathbf{L}$ terms lead to divergences in the bounds expression for general $\theta$ in $p \equiv e^{i\theta}.$ However, the divergence disappears if the phase is chosen so that $p_{L} \equiv (\tilde{k}^{*}/|\tilde{k}|)^{2},$ which is the same value as in the 2D TE case. 
At this value of $p_{L},$ assuming that $\Asym(p_{L}\U) \succ 0$ for the chosen material parameters, then
\begin{align}
    \text{Im}\left[\frac{s(\tilde{\omega})}{\tilde{\omega}\mathcal{N}}\right]
    &\geq
    -\text{Re}
    \Bigg[
    \frac{1}{12\pi^{2}\mathcal{N}}\int_{0}^{\infty}\dd q~q^{2}
    \bigg(\text{Im}\bigg(
    \frac{\tilde{k}^{*2}}{|\tilde{k}|^{2}\chi^{*}} + \frac{\tilde{k}^{4}}{|\tilde{k}|^{2}(q^{2} - \tilde{k}^{2})}
    \bigg)\bigg)^{-1}
    \bigg(
    \bigg|\frac{\tilde{k}}{q^{2} - \tilde{k}^{2}}\bigg|^{2} - \frac{\tilde{k}^{4}}{|\tilde{k}|^{2}(q^{2} - \tilde{k}^{2})^{2}}
    \bigg)
    \bigg].
\end{align}

The requirement that $\text{Asym}(e^{i\theta}\mathbb{U}) \succ 0$ puts restrictions on $\theta.$
For example, the requirement that $\text{Im}\big(\frac{e^{i\theta}}{\chi^{*}} + \frac{e^{-i\theta}\tilde{\omega}^{2}}{(-\tilde{\omega}^{2})} \big) > 0$ forces $\tan(\theta) > \frac{\text{Im}[\chi(\tilde{\omega})]}{\text{Re}[\chi(\tilde{\omega})] + |\chi(\tilde{\omega})|^{2}}$.
For the dual bound to be valid and finite for the TE line source and dipole case, $\theta_{\textrm{opt}}$ in $e^{i\theta_{\textrm{opt}}} = (\tilde{\omega}^{*}/|\tilde{\omega}|)^{2}$ must be in the domain of feasibility, so $\frac{2\omega_{0}\Delta\omega}{\omega_{0}^{2} - \Delta\omega^{2}} < \frac{\text{Im}[\chi(\tilde{\omega})]}{\text{Re}[\chi(\tilde{\omega})] + |\chi(\tilde{\omega})|^{2}}$ which can be rewritten as $\text{sgn}(\text{Re}[\tilde{\omega}^{2}])\,\text{Im}\left[\tilde{\omega}^{2} (\chi(\tilde{\omega})^{*}  + |\chi(\tilde{\omega})|^{2})\right] < 0$. 
For example, the maximum bandwidth is given by $\Delta\omega_{\textrm{max}} \equiv -\omega_{0}\zeta + \sqrt{(\omega_{0}\zeta)^{2} + \omega_{0}^{2}}$
where $\zeta \equiv \frac{\text{Re}[\chi(\tilde{\omega})] + |\chi(\tilde{\omega})|^{2}}{\text{Im}[\chi(\tilde{\omega})]}.$
Here, $\chi(\tilde{\omega})$ is evaluated at a complex frequency with $\text{Im}[\tilde{\omega}] > 0$,
in which case for any realistic $\chi$ causality and loss requires that $\text{Im}[\chi(\tilde{\omega})] > 0$ be strictly positive~\cite{hashemi_diameter-bandwidth_2012}. In particular, we see that considering the limit $\text{Im}[\chi(\tilde{\omega})] \to 0^{+}$
rules out any nonzero bandwidth value for which our formulas place bounds for the TE line source and dipole cases.
The $\text{Im}[\chi(\tilde{\omega})] \to 0^{+}$ limit, however, exists for nonzero bandwidths and leads to finite dual bounds for the TM case.
The reason for the bounds becoming lose for large bandwidths is that the weight function $W(\omega)$ gives more weight to quasistatic terms, so the electrostatic term $\alpha$, which is in general nonzero for TE and dipole in 3d, is important and setting it to 0 is too loose of a relaxation to yield nontrivial bounds. For TM, $\alpha = 0$ always so this issue never arises in that case.

\subsubsection{Compact expression of all results}

In summary, 
\begin{align}
    \langle\rho_{\textrm{sca}} \rangle
    \geq 
    -\frac{\eta}{\mathcal{N}}\int_{0}^{\infty} \dd k_\rho \, k_\rho^\nu
    \bigg[\text{Im}\bigg(
    \frac{e^{i\theta}}{\chi^{*}} + \frac{e^{-i\theta}\tilde{\omega}^{2}}{k_\rho^{2} - \tilde{\omega}^{2}}
    \bigg)\bigg]^{-1}
    \text{Re}
    \Bigg[
    \bigg|\frac{\tilde{\omega}
    }{k_\rho^{2} - \tilde{\omega}^{2}}\bigg|^{2}
    - 
    \frac{e^{-i\theta}\tilde{\omega}^{2}
    }{(k_\rho^{2} - \tilde{\omega}^{2})^{2}}
    \Bigg]
    \label{eq:rhoscaboundexplicit}
\end{align}
where for a TM line source the maximization over $\theta$ can only be computed numerically (e.g., using bisection), but for the TE line source and dipole cases it is carried out explicitly and in both cases fixes $e^{i\theta_{\textrm{opt}}} = \big(\tilde{\omega}^{*}/|\tilde{\omega}|\big)^{2}$ above.
As written above, the normalization and exponent factors $\eta$ and $\nu$, respectively, depend on the dimensionality and polarization of the source. 
In particular, for a TM line source, $\eta =  \frac{1}{8\pi}$ and $\nu = 1$,
with the TE case acquiring an additional factor of $1/2$ in the normalization, while for a dipole in 3d, $\eta = \frac{1}{12\pi^2}$ and $\nu = 2$. As derived further below, for a 1d dipole (a sheet in 3d) $\eta = \frac{1}{4\pi}$ and $\nu = 0$. For $\text{Im}[\chi] > 0$, taking the limit $\Delta\omega/\omega_{0} \ll 1$, we find
\begin{align}
\frac{\langle \rho\rangle_{L\to\infty}}{\langle\rho_{\textrm{vac}}\rangle} \geq \sqrt{\frac{2\text{Im}[\chi]}{|\chi|^{2}}\frac{\Delta\omega}{\omega_{0}}} + \mathcal{O}(\Delta\omega).
\end{align}

% \section{Band gap and Lorentzian averaging}

% Here, we explore the density of states (DOS) near a band edge, as well as look into what the averaging of it with a Lorentzian window function centered at the middle of the band gap.
% Assume a weight-function of the form
% \begin{align}
%     L(\omega) = \frac{1}{\pi}\frac{\Delta\omega}{(\omega-\omega_{0})^{2} + (\Delta\omega)^{2}}.
% \end{align}
% Let $g(\omega)$ denote the density of states at frequency $\omega.$ Assuming approximately constant DOS $g(\omega) \approx \rho_{c}$ for $\omega$ outside the band gap, one finds
% \begin{align}
%     \int_{\omega_{0} + \omega_{g}/2}^{\infty}\dd\omega
%     \rho_{c} \frac{1}{\pi}\frac{\Delta\omega}{(\omega-\omega_{0})^{2} + (\Delta\omega)^{2}}
%     =
%     \rho_{c}\arctan\bigg(\frac{2\Delta\omega}{\omega_{g}}\bigg)
% \end{align}

% \begin{align}
%     \int_{0}^{\omega_{0} - \omega_{g}/2}\dd\omega 
%     \rho_{c}\frac{1}{\pi}\frac{\Delta\omega}{(\omega-\omega_{0})^{2} + (\Delta\omega)^{2}}
%     = 
%     -\rho_{c}\arctan\bigg(\frac{\Delta\omega(\omega_{g}-2\omega_{0})}{2\Delta\omega^2 + \omega_{g}\omega_{0}}\bigg).
% \end{align}
% Here, we assume $\omega_{0} - \omega_{g}/2 > 0.$ Adding a taking a Taylor series in powers of $\Delta\omega > 0 $ gives
% \begin{align}
%     \rho_{c}\frac{(4\omega_{0} - \omega_{g})\Delta\omega}{\pi\omega_{0}\omega_{g}} 
%     +
%     \rho_{c}\frac{(-16\omega_{0}^{3} + \omega_{g}^{3})\Delta\omega^3}{3\pi\omega_{0}^{3}\omega_{g}^{3}} + \cdots
% \end{align}
% This gives linear scaling with the bandwidth $\Delta\omega$ for small bandwidths.

\section{Asymptotics of bandwidth averaged LDOS using $W(\omega)$}

Here, we explore scaling relations in the limit of small bandwidths when simple models for the density of states (DOS) near $\omega_{0}$ are used in averaging with the window function $W(\omega) \equiv \frac{L(\omega) - L(-\omega)}{\omega\mathcal{N}}$ where $\mathcal{N} \equiv \frac{\omega_{0}}{\omega_{0}^{2} +\Delta\omega^{2}}$.
Let $\rho(\omega)$ denote the density of states at frequency $\omega.$
\begin{itemize}
    \item 
Assuming approximately constant LDOS $\rho(\omega) \approx \rho_{c}$ for $\omega$ outside the band gap (or pseudogap), one finds
\begin{align}
    \langle\rho\rangle
    &= 
    \int_{0}^{\omega_{0} - \omega_{g}/2}\dd\omega 
    \rho_{c} W(\omega)
    +
    \int_{\omega_{0} + \omega_{g}/2}^{\infty}\dd\omega
    \rho_{c} W(\omega), \\
    &=
    \rho_{c} - \rho_{c}\int_{\omega_{0} - \omega_{g}/2}^{\omega_{0} + \omega_{g}/2} W(\omega)\dd\omega.
    \label{eq:bandgapWomega}
\end{align}
We find
\begin{align}
    \int_{\omega_{0} - \omega_{g}/2}^{\omega_{0} + \omega_{g}/2} W(\omega)\dd\omega 
    =
    \frac{\omega_{0}
    (
    - \arccot(\frac{2\Delta\omega}{4\omega_{0} - \omega_{g}})
    + 2\arccot(\frac{2\Delta\omega}{\omega_{g}})
    + \arccot(\frac{2\Delta\omega}{4\omega_{0} + \omega_{g}})
    )
    +
    \Delta\omega \text{arctanh}(\frac{8\omega_{0}\omega_{g}}{4\Delta\omega^{2} + 16\omega_{0}^{2} + \omega_{g}^{2}})
    }{\pi\omega_{0}}
\end{align}
where we assume $\omega_{0} - \omega_{g}/2 > 0.$
Taking a Taylor series in powers of $\Delta\omega > 0$ yields
\begin{align}
    \langle\rho\rangle
    &=
    \rho_{c}\frac{
    \bigg(
    \frac{-64\omega_{0}^{3} + 8\omega_{0}\omega_{g}^{2}}{-16\omega_{0}^{2}\omega_{g} + \omega_{g}^{2}} 
    -
    \text{arctanh}(\frac{8\omega_{0}\omega_{g}}{16\omega_{0}^{2} + \omega_{g}^{2}})
    \bigg)\Delta\omega}{\pi\omega_{0}} + \cdots, \\
    &=
    \rho_{c}
    \bigg(
    \frac{4}{\pi}
    - \frac{3}{4\pi}\bigg(\frac{\omega_{g}}{\omega_{0}}\bigg)^{2} + \mathcal{O}(\omega_{g}^{4})
    \bigg)\frac{\Delta\omega}{\omega_{g}} + \mathcal{O}(\Delta\omega^{3}).
\end{align}
This gives linear scaling with the bandwidth $\Delta\omega$ for small bandwidths.

\item
Assuming LDOS of the form $\rho(\omega) = \rho_{R}\sqrt{\omega - \omega_{0}}$ for $\omega \in [\omega_{0}, \omega_{R}]$ yields
\begin{align}
    \int_{\omega_{0}}^{\omega_{R}} \rho(\omega) W(\omega)\dd\omega
    =
    \rho_{R}\sqrt{\Delta\omega/2} + \cdots
\end{align}
for $\Delta\omega > 0$ small.
Assuming LDOS of the form $\rho(\omega) = \rho_{L}\sqrt{\omega_{0} - \omega}$ for $\omega \in [\omega_{L}, \omega_{0}]$ does not as straightforwardly lead to a closed-form series expansion in Mathematica, but it can easily be verified numerically that
\begin{align}
    \int_{\omega_{L}}^{\omega_{0}} \rho(\omega) W(\omega)\dd\omega
    =
    \rho_{L}\sqrt{\Delta\omega/2} + \cdots
\end{align}
for $\omega_{L}$ near $\omega_{0}$ and for small bandwidths.
\end{itemize} 

% \section{Transfer Matrix Method}

% We consider the transfer matrix method applied in 1D, where the fields and LDOS can be analytically calculated.
% We are interested in a Bragg stack where the thickness of the layers differs, starting from thin slice and then tapered to $d(n) = d_{0} \alpha n / (1 + \alpha n)$.

\section{Periodic structures}

\begin{figure}
   \centering
       \includegraphics[width=1.0\linewidth]{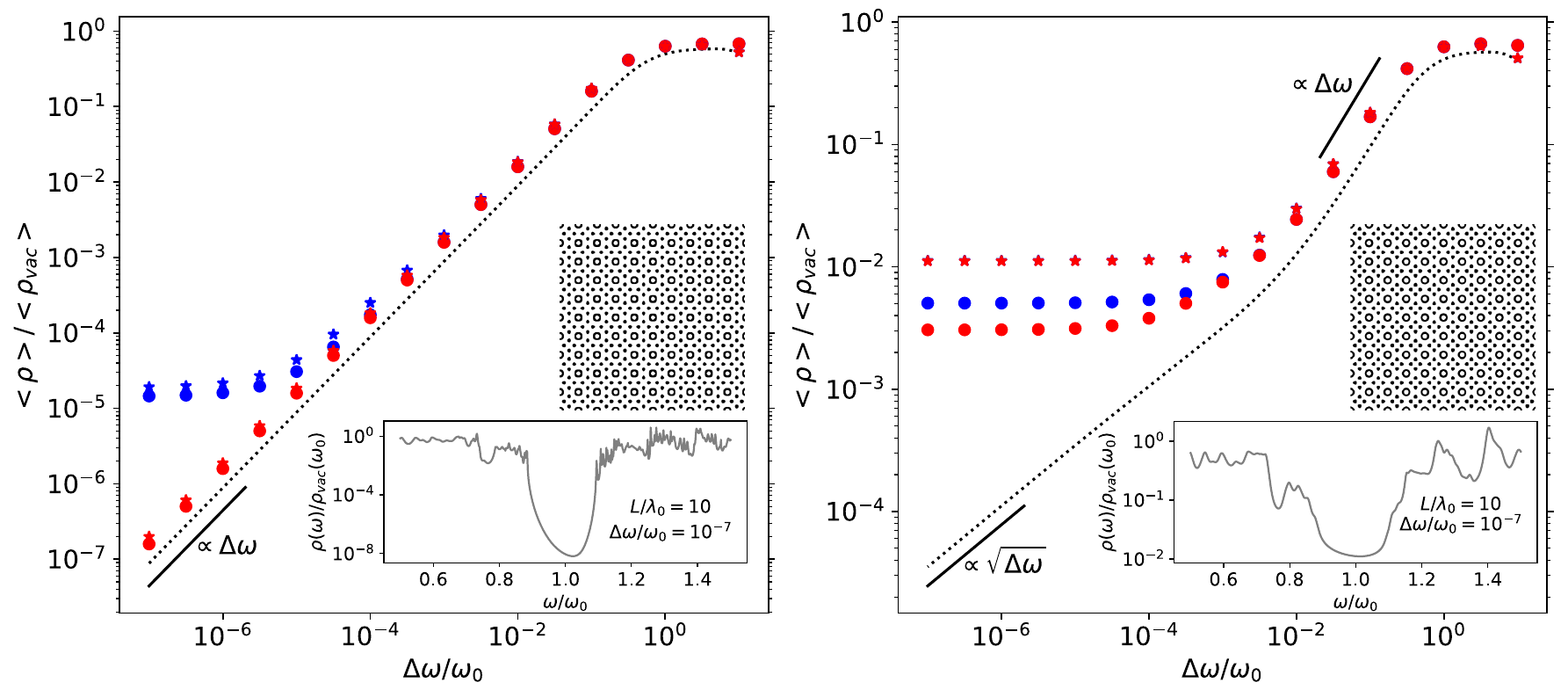}
      \caption{
        \textbf{Lower bounds pertaining to maximum suppression of bandwidth-integrated LDOS.}
        All curves and markers are for TM line sources at the center of $L\times L$ design domains. 
        Stars are for truncated 2d square-lattice photonic crystals of lattice constant $\lambda_{0} = 2\pi c/\omega_0$ while circles are for rotationally symmetric structures (shown for comparison), with $\chi = 4$ (left) and $\chi = 4 + 0.1i$ (right) for $L/\lambda_{0} = \{5, 10\}$ (blue, red).
        All results are normalized by the corresponding bandwidth-integrated LDOS in vacuum $\langle \rho_{\textrm{vac}} \rangle = \frac{\omega_{0}^{2} + \Delta\omega^{2}}{4\pi \omega_{0}}\atan(\frac{\omega_{0}}{\Delta\omega})$. Note that doubling the size in the lossy case for the finite periodic crystal leads to minor changes in the bandwidth-integrated LDOS (red and blue stars nearly overlap).}
\label{fig:chi4and4+1e-1jperiodicFULL}
\end{figure}

We also considered finite structures with periodic cells, see Fig.~\ref{fig:chi4and4+1e-1jperiodicFULL}.
When $\text{Im}[\chi(\tilde{\omega})] > 0$, periodic 2d structures seem to saturate to a minimum average LDOS for $L/\lambda_{0} \gtrsim 5,$ with minor changes when doubling the design footprint.
However, rotationally symmetric structures (Bragg fibers) continue to lead to improvements, albeit at a rate proportional to $1/\sqrt{L}$ (see main text) for small bandwidths before saturating due to a nonzero bandwidth.

\section{TM bounds and bullseye structures}

\begin{figure}
    \centering
        \includegraphics[width=\linewidth]{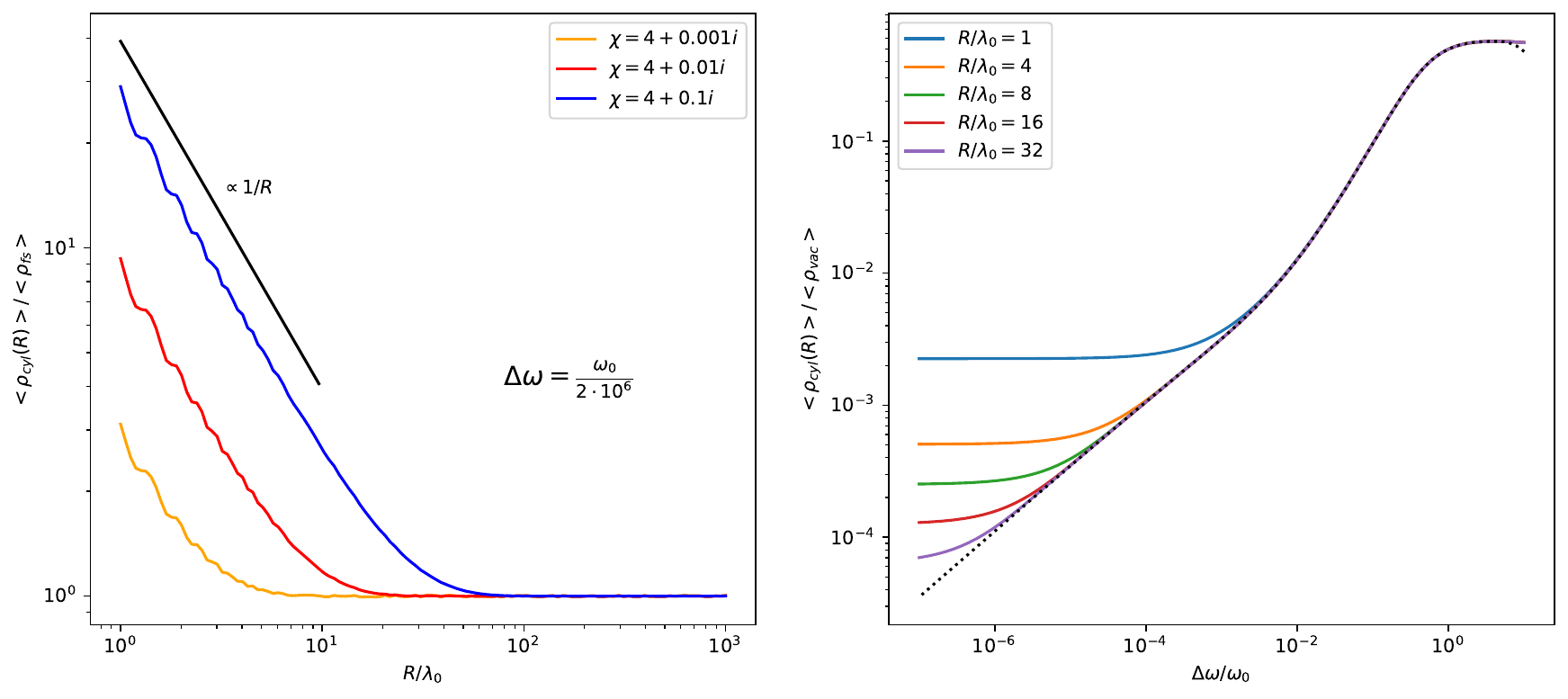}
        \caption{
        \textbf{Lower bounds on bandwidth-integrated LDOS for cylindrical design domains of radius $R$ containing at TM source at the center assuming cylindrical symmetry of the structures.} 
        (Left) The bounds are normalized by full-space bounds which do not assume cylindrical symmetry of the structure. Calculations were performed with $\Delta\omega = \frac{\omega_{0}}{2\cdot 10^{6}}.$
        (Right) Curves for $\chi = 4 + 0.1i$ showing that bounds for cylindrically symmetric structures approach the infinite design domain bounds which do not assume rotational symmetry of the structure (dotted line) as $R/\lambda_{0}$ increases.
        }
\label{fig:cylfscomparison}
\end{figure}

\begin{figure}
    \centering
        \includegraphics[width=\linewidth]{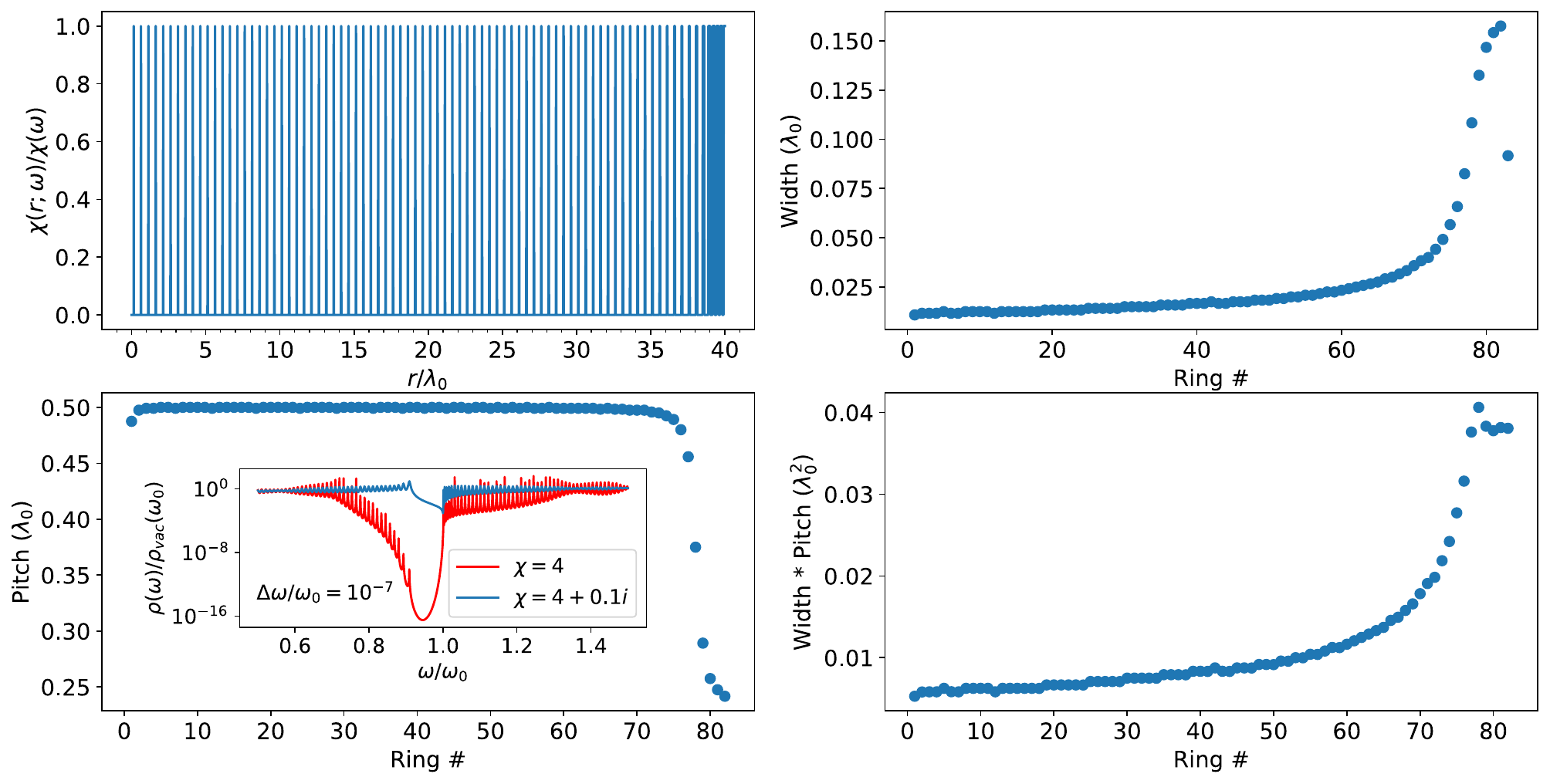}
        \caption{
        Plot of the tapered and chirped grating profile for $\chi = 4 + 0.1i$, $\Delta\omega/\omega_{0} = 10^{-7}$, $L/\lambda_{0} = 80$ discovered via topology optimization. Shown as an inset is LDOS$(\omega)$ of the same structure assuming the susceptibility is $\chi(\tilde{\omega}) = 4$ and $\chi(\tilde{\omega}) = 4 + 0.1i,$ which inspires the study of 1d photonic crystal structures with vanishing thickness of lossy material layers as a way to suppress (or enhance) LDOS at a band edge (see discussion further below).
        }
\label{fig:Braggfiberlossandnoloss}
\end{figure}

We ran bounds calculations for cylindrical design domains of radius $R$ for cylindrically symmetric structures. 
Figure~\ref{fig:cylfscomparison} shows that for large enough $R,$ bounds calculations $\langle\rho_{cyl}(R)\rangle$ assuming cylindrical structures approach the full-space bounds $\langle\rho_{fs}\rangle$ that allow for arbitrary structures in the design domain.
This is support for the view that cylindrically symmetric structures like bullseye structures with pseudogaps perform near optimally for large enough domain designs.
Also, shown in Fig.~\ref{fig:Braggfiberlossandnoloss} are plots showing the properties of the structures discovered in topology optimization for $\chi = 4 + 0.1i$ and $\Delta\omega/\omega_{0} = 10^{-7}.$

\section{Inverse design methodology}

To calculate the inverse design structures, we made use of the NLopt package~\cite{johnson_nlopt_2019} and followed standard topology optimization algorithms~\cite{molesky_inverse_2018,christiansen_inverse_2021} where the electric susceptibility value of each pixel in the specified design region is considered as an independent design parameter, based on the method of moving asymptotes~\cite{svanberg_class_2002}. Each pixel, indexed by $k$, in the design region is allowed to explore a continuous range of susceptibility values varying between the vacuum value $\chi_{k} = 0$ and the prescribed material $\chi_{k} = \chi_{mat}$. This can be interpreted as a filling fraction of the pixel with the material with susceptbility $\chi_{mat}$.

To enforce cylindrical symmetry, we wrote a uniform grid axi-symmetric 2D TM solver, simplifying the problem into an effective 1d problem with only radial degrees of freedom. This allows for rapid convergence ($\leq 5$ minutes) to cylindrically symmetric inverse designs.
More generally, we implemented a 2d Maxwell solver which could be used for arbitrary structuring (allowing the pixels within any arbitrary 2d design region to vary). To enforce periodic boundary conditions, we restricted the variable pixels to a single unit-cell and enforced periodicity onto other pixels within the design region that are periodic translations of that single unit-cell. Doing a thousand function evaluations (iterations of optimization over the susceptibility profile for fixed material parameters, design region, bandwidth, etc) in 2d takes between 1 to 2 hours.
The primary challenge in such optimization problems lies in the computational cost of evaluating the objective function, which is evaluated several hundred if not thousand times during the course of any one optimization.
The gradient of $f(\{\chi_{k}\};\tilde{\omega}) = -\frac{1}{2}\Re{\int \mathbf{J}^{*}(\mathbf{r};\tilde{\omega})\cdot\mathbf{E}(\mathbf{r};\tilde{\omega})\frac{1}{\tilde{\omega}\mathcal{N}}\dd\mathbf{r}}$ for a TM polarization with respect to the susceptibility degrees of freedom is given by
\begin{align}
    \frac{\partial f}{\partial \chi_{k}} = \frac{1}{2}
    \Re{\frac{i\chi_{mat}E_{k}^{2}}{\mu_{0}c^{2}\mathcal{N}}},
\end{align}
where $\chi_{k}$ is the value of the susceptibility at the $k$-th pixel and $E_{k}$ is the out-of-plane component of the electric field at the $k$-th pixel.
Hence, only a single additional Maxwell solve is required to obtain complete gradient information for the optimization objective at each iteration of the local optimization. 
Details on the computational complexities in 3d and solution methods can be found in Appendix C of Ref.~\cite{molesky_global_2020} and the therein cited references.

\section{Half-space design domain}

\begin{figure}
   \centering
       \includegraphics[width=0.6\linewidth]{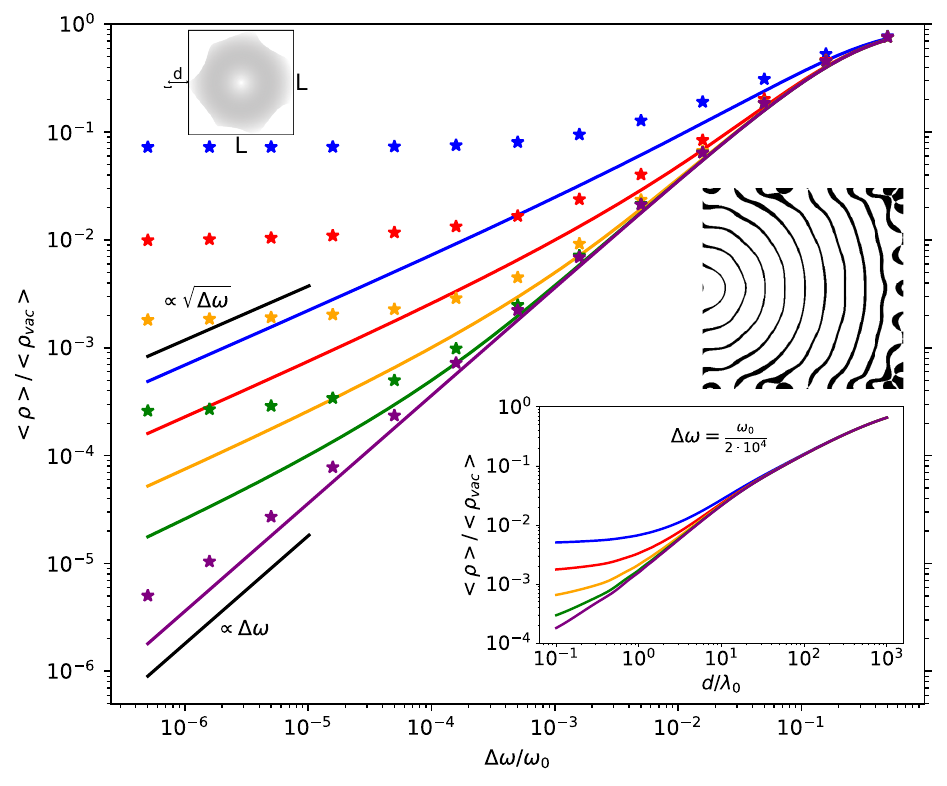}
       \caption{
       \textbf{Lower bounds on bandwidth-integrated LDOS for a half-space design region.}
       Inset is a schematic showing a source a distance $d$ away from an $L \times L$ design domain.
       Plots (log-log scale): Solid lines correspond to a TM line source a distance $d = \lambda_{0}/10$ away from a half-space design region, by which we mean the case where $L/\lambda_{0} \to \infty.$
       $\text{Im}[\chi(\tilde{\omega})] = \{10^{0}, 10^{-1}, 10^{-2}, 10^{-3}, 0\}$ for blue, red, orange, green, purple curves, respectively, and $\text{Re}[\chi(\tilde{\omega})] = 4$ in all cases.
       Stars correspond to structures discovered via topology optimization for a $5\lambda_{0} \times 5\lambda_{0}$ design region.
       Lower bounds with positive $\text{Im}[\chi]$ asymptotically scale slower than linear with bandwidth.
       Also inset is an inverse design for $\chi = 4 + 1i$ and $\Delta\omega = \frac{\omega_{0}}{2\cdot 10^{6}}$, as well as a log-log plot of lower bounds with fixed bandwidth $\Delta\omega = \frac{\omega_{0}}{2\cdot 10^{4}}$ (and $L/\lambda_{0} \to \infty$) but varying separation $d$.
       }
\label{fig:totLDOSminHSchi4vsQsrc}
\end{figure}

To evaluate bounds, explicit evaluation of $\Asym(p\mathbb{U})^{-1}$ is not required since one only needs $\Asym(p\mathbb{U})^{-1}\ket{\vb{E}_{\textrm{vac}}}$.
Given a source $\mathbf{J}$ and working in the case where $\mathbb{G}_{0}$ is the vacuum Green's function, then typically $\mathbb{U}$ and $\text{Asym}(p\mathbb{U})$ are known.
In the full-space case, one can then work out $\text{Asym}(p\mathbb{U})^{-1}$ explicitly by working in well-known bases (spherical or cylindrical wave vectors, or Fourier basis), and then evaluate $\Asym(p\mathbb{U})^{-1}\ket{\vb{E}_{\textrm{vac}}}.$
Another common design domain of interest is a dipole near the side of an object.
For the half-space case, one can use a Laplace/Fourier transform to evaluate $\Asym(p\mathbb{U})^{-1}\ket{\vb{E}_{\textrm{vac}}}$ without ever explicitly calculating the inverse operator~\cite{chao_maximum_2022}.
Following this procedure as detailed in Ref.~\cite{chao_maximum_2022}, Fig.~\ref{fig:totLDOSminHSchi4vsQsrc} shows bounds and designs for the setting where the source is a distance $d$ away from a half-space design domain. 
We find that the bounds exhibit the same scaling behavior as in the case of the full-space design domain.

\section{Bounds in 1d}

\begin{figure}
        \centering
        \includegraphics[width=\linewidth]{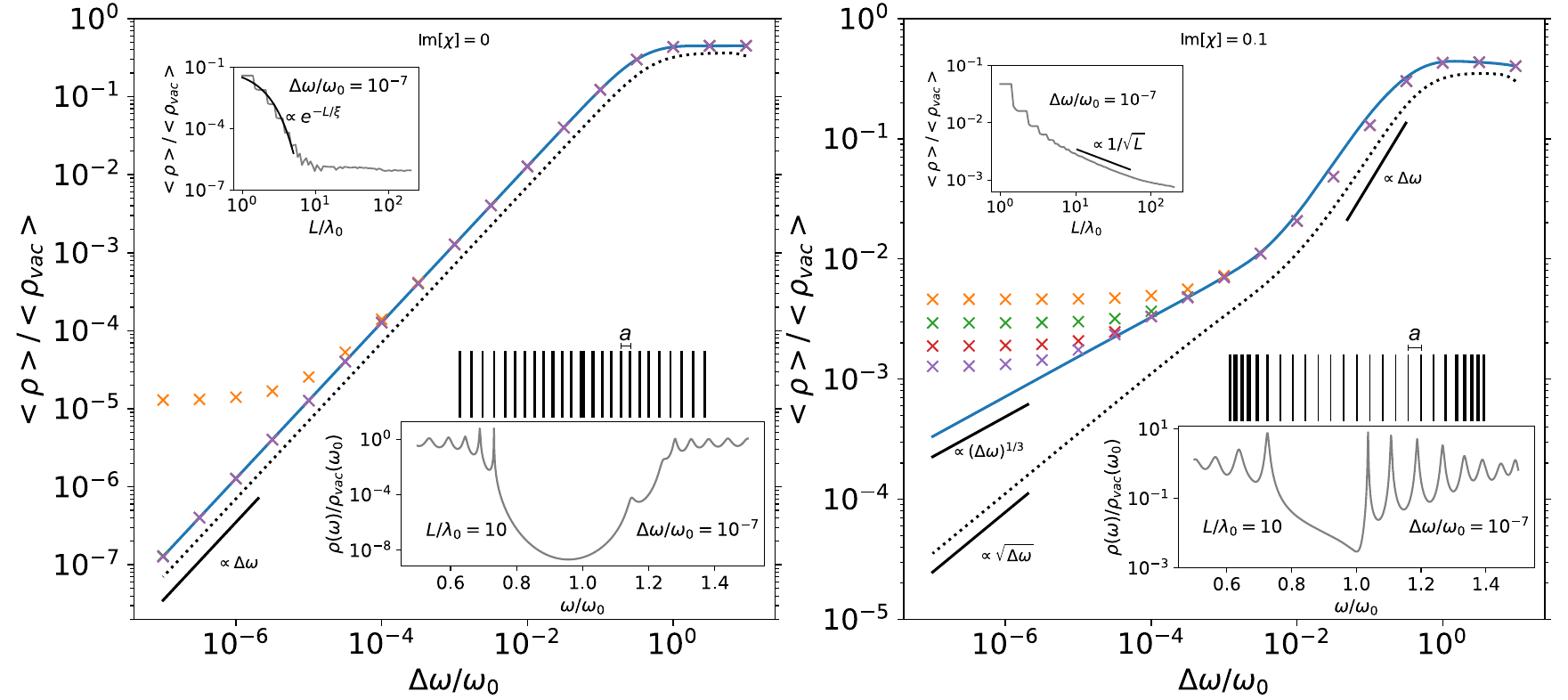}
        \caption{
        \textbf{Lower bounds on bandwidth-integrated LDOS for a full-space design region.}
        All markers correspond to structures discovered via topology optimization for 1d TM source (sheet source in 3d) at the center of design domain of length $L$ with $\chi = 4$ (left) and $\chi = 4 + 0.1i$ (right) for $L/\lambda_{0} = \{5, 10, 20, 40\}$ (orange, green, red, purple, respectively) where $\lambda_0 = 2\pi c/\omega_0$.
        Dashed lines are bounds which incorporate only global power conservation for infinitely extended structures, $L/\lambda_{0} \to \infty$, and are computed via Eq.~\eqref{eq:boundTM1d}.
        The lossless and dispersionless limit (left) converges to a geometry with periodicity $a = \frac{\lambda_0}{4}\frac{1 + \sqrt{1 + \text{Re}[\chi]}}{\sqrt{1 + \text{Re}[\chi]}}$ and thickness $h = \frac{\lambda_0}{4\sqrt{1+\text{Re}[\chi]}}$ of each material layer (a quarter-wave stack on both sides of the source) supporting a pseudogap centered around $\omega_0$ (bottom inset). The lossy case (right) converges to chirped gratings sandwiching the source, each initially of spacing $a = \frac{\lambda_{0}}{2}$ and a tapered thickness that grows as the distance to the source increases, leading to loss at $\omega_0$ arbitrarily close to 0 in the limit $L\to\infty$.
        The solid lines are from calculations of $\langle \rho \rangle$ for an infinite 1d PhC with period $a = \frac{\lambda_0}{4}\frac{1 + \sqrt{1 + \text{Re}[\chi]}}{\sqrt{1 + \text{Re}[\chi]}}$ (left) and $a = \frac{\lambda_{0}}{2}$ (right) and a source at the center of an air defect cavity, 
        Eq.~\eqref{eq:PhCCavityavgLDOSRe}, numerically optimized over the thickness of the material layers for vanishing central air defect cavity. Interesting, this results in $(\Delta\omega)^{1/3}$ scaling for this class of structures with loss, 
        in agreement with the inverse designs for finite $L$ before saturation.
        }
\label{fig:totLDOSmin2DTM1DvsQsrc}
\end{figure}

In this section, we derive the semianalytical bound expression for a dipole in 1d.
The Green's function satisfying $(\frac{\partial^{2}}{\partial x ^{2}} + \tilde{\omega}^{2})G_{0}(x, x'; \tilde{\omega}) = -\tilde{\omega}^{2}\delta(x-x')$ is given by $G_{0}(x, x'; \tilde{\omega}) = i\tilde{\omega} e^{i\tilde{\omega}|x-x'|}/2.$ Thus, for a source $J(x) = \delta(x)$ the field in vacuum is $E_{\textrm{vac}}(x) = -e^{i\tilde{\omega}|x-x'|}/2,$ which has a Fourier expansion of
\begin{align}
    E_{\textrm{vac}}(x) = \int_{-\infty}^{\infty} \frac{\dd q}{2\pi} \frac{-i\tilde{\omega}}{(\tilde{\omega} + q)(\tilde{\omega} - q)}e^{iqx}.
\end{align}
Also,
\begin{align}
    \Asym(p\U)(q)
    =
    \text{Im}
    \bigg(
    \frac{p}{\chi^{*}} - p^{*}\frac{\tilde{\omega}^{2}}{(\tilde{\omega} + q)(\tilde{\omega} - q)}
    \bigg).
\end{align}
The core calculation in Eq.~\eqref{eq:theta_bound} is the evaluation of 
\begin{equation}
    \Asym(p\U)^{-1} \ket{\vb{E}_{\textrm{vac}}}
    =
    \int_{-\infty}^{\infty} \frac{\dd k}{2\pi} \int_{-\infty}^{\infty}
    \dd x'
    \Asym(p\U)^{-1}(x, x')
    E_{\textrm{vac}}(k)
    e^{ikx'}.
    \label{eq:AsympUinv_Evac1d}
\end{equation}
Defining
\begin{equation}
    f(x, k) = \int_{-\infty}^{\infty} \dd x'
    \Asym(p\U)^{-1}(x, x')
    E_{\textrm{vac}}(k)
    e^{ikx'}
\end{equation}
(which is the currently unknown quantity we wish to work out)
we have
\begin{equation}
    \int_{-\infty}^{\infty}
    \Asym(p\U)(x, x') f(x', k) \dd x'
    = 
    E_{\textrm{vac}}(k)
    e^{ikx}.
    \label{eq:Fourier_target_1d}
\end{equation}
The action of $\Asym(p\U)(x, x') = \Asym(p\U)(x - x')$ is a convolution over $x'$, which suggests that a Fourier transform $\mathscr{F}\{f(x)\}(q) = \int_{-\infty}^{\infty} \dd x f(x) e^{-iqx}$ may be used to solve for $f(x, k)$ in Eq.~\eqref{eq:Fourier_target_1d}.
This leads to
\begin{align}
    \Asym(p\U)(q)F(q) 
    &=
    E_{\textrm{vac}}(q)
    (2\pi)\delta(k - q), \\
    F(q) 
    &=
    \Asym(p\U)^{-1}(q)
    E_{\textrm{vac}}(q)
    (2\pi)\delta(k - q).
\end{align}
Therefore,
\begin{align}
    f(x, k) &=
    \int_{-\infty}^{\infty} \frac{\dd q}{2\pi} F(q)e^{iqx}
     \\
     &=
    \Asym(p\U)^{-1}(k)
    E_{\textrm{vac}}(k)
    e^{ikx}
\end{align}
so that
\begin{equation}
    \Asym(p\U)^{-1} \ket{\vb{E}_{\textrm{vac}}}
    =
    \int_{-\infty}^{\infty} \frac{\dd k}{2\pi}
    \Asym(p\U)^{-1}(k)
    E_{\textrm{vac}}(k)
    e^{ikx}.
\end{equation}
With this, it then follows that
\begin{equation}
    \bra{\vb{E}_{\textrm{vac}}}\Asym(p\U)^{-1} \ket{\vb{E}_{\textrm{vac}}}
    =
    \int_{-\infty}^{\infty} \frac{\dd k}{2\pi}
    E_{\textrm{vac}}(k)^{*}
    \Asym(p\U)^{-1}(k)
    E_{\textrm{vac}}(k)
\end{equation}
and
\begin{equation}
    \bra{\vb{E}_{\textrm{vac}}^{*}}\Asym(p\U)^{-1} \ket{\vb{E}_{\textrm{vac}}}
    =
    \int_{-\infty}^{\infty} \frac{\dd k}{2\pi}
    E_{\textrm{vac}}(-k)
    \Asym(p\U)^{-1}(k)
    E_{\textrm{vac}}(k).
\end{equation}
Thus,
\begin{align}
    -\text{Im}\left[\frac{s(\tilde{\omega})}{\tilde{\omega}\mathcal{N}}\right] &\leq \frac{1}{4\mathcal{N}}
    \Re
    \Big\{
    \expval{\Asym(p\U)^{-1}}{\vb{E}_{\textrm{vac}}} +  p^{*} \bra{\vb{E}_{\textrm{vac}}^{*}}\Asym(p\U)^{-1}\ket{\vb{E}_{\textrm{vac}}}
    \Big\}
\end{align}
becomes
\begin{align}
    -\text{Im}\left[\frac{s(\tilde{\omega})}{\tilde{\omega}\mathcal{N}}\right] &\leq 
    \frac{1}{4\mathcal{N}}\Re\int_{-\infty}^{\infty} \frac{\dd k}{2\pi}
    (E_{\textrm{vac}}(k)^{*} + p^{*} E_{\textrm{vac}}(-k))
    \Asym(p\U)^{-1}(k)
    E_{\textrm{vac}}(k),
    \label{eq:boundFourier1D}
\end{align}
which written out explicitly is
\begin{align}
    \text{Im}\left[\frac{s(\tilde{\omega})}{\tilde{\omega}\mathcal{N}}\right] &\geq 
    -\frac{1}{4\mathcal{N}}\Re\int_{-\infty}^{\infty} \frac{\dd k}{2\pi}
    \bigg(
    \text{Im}
    \bigg(
    \frac{p}{\chi^{*}} - \frac{p^{*}\tilde{\omega}^{2}}{(\tilde{\omega}^{2} - k^{2})}
    \bigg)\bigg)^{-1}
    \bigg(
    \frac{|\tilde{\omega}|^{2}}{|\tilde{\omega}^{2} - k^{2}|^{2}}
    -
    p^{*}\frac{\tilde{\omega}^{2}}{(\tilde{\omega}^{2} - k^{2})^{2}}
    \bigg).
    \label{eq:boundTM1d}
\end{align}
This is similar to Eq.~\eqref{eq:rhoscaboundexplicit} after folding the integral in half, but now with $\eta = \frac{1}{4\pi}$ and $\nu = 0$. 
Figure~\ref{fig:totLDOSmin2DTM1DvsQsrc} shows calculations of the lower bounds on $\langle \rho\rangle$ for finite-sized $L$ design regions along with achievable objective values discovered via inverse design.
We remark that Eq.~\eqref{eq:boundFourier1D} can be generalized to $d$ dimensions. Doing so for TM will reduce the 2d Fourier transform to a Hankel transform (of order zero) and lead to Eq.~\eqref{eq:rhoscaboundexplicit}.
For TE in 2d, the 2D integral $\iint \dd k_{x} \dd k_{y}$ diverges in general except for specially chosen $p$.
The best way to see the divergence is to switch into a $k_{\perp}, k_{\parallel}$ basis, with the divergence coming from the $k_{\parallel}$ contribution.
The benefit of working with spherical or cylindrical wave vectors is that the divergence from the longitudinal waves is more explicit.

Note that for any $p$ such that $\text{Asym}(p\mathbb{U})$ is positive-definite Eq.~\eqref{eq:boundTM1d} is a valid bound, and an optimization over $p$ is done to find the tightest bound within this framework. For $\text{Im}[\chi]>0,$ $p=1$ is always a feasible point. Using $p=1$ in Eq.~\eqref{eq:boundTM1d} we find that for $\Delta\omega \ll \omega_{0}$,
\begin{align}
    \frac{\langle \rho \rangle}{\langle \rho_{\textrm{vac}}\rangle} &\geq 1 + \text{Im}\left[\frac{s(\tilde{\omega})}{\tilde{\omega}\mathcal{N}}\right]\frac{1}{\langle\rho_{\textrm{vac}}\rangle} \\
    &= 
    \sqrt{\frac{2\text{Im}[\chi]}{|\chi|^{2}}\frac{\Delta\omega}{\omega_{0}}}
    + 
    \bigg(
    -\frac{3}{2}\sqrt{\frac{|\chi|^{2}}{2\text{Im}[\chi]}}
    +
    \frac{(|\chi|^{4} -4\text{Im}[\chi]^2)}{2\sqrt{2}|\chi|^{3}\sqrt{\text{Im}[\chi]}}
    \bigg)(\Delta\omega/\omega_{0})^{3/2} + \cdots \\
    &=
    \sqrt{\frac{2}{\zeta}\frac{\Delta\omega}{\omega_{0}}}
    - 
    \bigg(1+\frac{4}{\zeta^{2}}\bigg)\sqrt{\frac{\zeta}{2}}(\Delta\omega/\omega_{0})^{3/2} + \cdots
\end{align}
where $\zeta \equiv \frac{|\chi|^{2}}{\text{Im}[\chi]}$ is a material loss figure of merit~\cite{miller_fundamental_2016,molesky_mathbbt-operator_2019,strekha2022tracenoneq}.
% \begin{figure}
%    \centering
%        \includegraphics[width=0.6\linewidth]{totLDOSmin2DTM1DvsQsrc.pdf}
%       \caption{\textbf{Lower bounds on bandwidth-integrated LDOS for a full-space design region.}
%        Blue lines correspond to a TM line source (a dipole in 2d), while red lines correspond to a sheet source in 3d (or a dipole in 1d). 
%        Solid and dashed lines correspond to $\chi = 4$ and $\chi = 4 + 0.1i,$ respectively.
%        }
% \label{fig:totLDOSmin2DTM1DvsQsrc}
% \end{figure}

\section{1d photonic crystal cavities with material loss approach 0 single-frequency LDOS in the limit of vanishing thickness}

This section contains all technical details of using transfer matrices to determine reflection coefficients and LDOS suppression associated with 1d photonic crystal devices. 

Specifically, we are interested in multilayer stacks that have continuous translational symmetry in the $x$ and $y$ directions, with fields excited by current sheets perpendicular to the $z$ direction and uniform in the $x$ and $y$ directions.
A harmonic time dependence of $e^{-i\omega t}$ will be assumed, with dimensionless units $\epsilon_0=\mu_0=c=1$. In general $\omega$ can be complex as in the main text; for certain discussions we will set $\omega$ to be real, in which case it will be written as $\omega_0$. 

\subsection{Transfer matrix formalism for multilayer stacks}
Within any uniform layer of material, the electric field is a superposition of forward and backward waves, $E = b_-e^{-ikz} + b_+e^{ikz}$, with the coefficients $b_-$ and $b_+$ completely determining the field profile.
At the interface between adjacent layers of materials, physical boundary conditions (the continuity of the electric and magnetic fields) specify linear relations that connect the coefficients of the adjacent layers: the coefficient vector $(b_-,b_+)^T$ of any layer can thus be written as a transfer matrix multiplied by the coefficient vector of any other layer.
These transfer matrices thus provide a complete description of EM physics in 1D, and will be the mathematical tool we use to analyze the physical characteristics of 1d devices with potentially infinite length. 

\subsection{The half-infinite photonic crystal}
Consider the half-infinite photonic crystal (PhC) formed by alternating layers of material with thickness $h$ and vacuum with thickness $\Delta$.
The interfaces of this structure are at $z=-d_1,-d_2,\cdots$, with $0=d_0<d_1<d_2<\cdots$: material is situated in the intervals $[-d_{2m},-d_{2m-1}]$, with $d_{2m} - d_{2m-1} = h$ and vacuum is situated in the intervals $[-d_{2m+1},-d_{2m}]$ as well as all of $z>0$, with $d_{2m+1}-d_{2m}=\Delta$.
The field within the layer $[-d_{j},-d_{j-1}]$ is represented by the coefficient vector $[b_{j,-}, b_{j,+}]^T$: $E = b_{j,-} e^{-i k_j d_j} e^{-i k_j z} + b_{j,+} e^{i k_j d_j} e^{i k_j z}$, where $k_j=k_v=\omega_0$ in vacuum and $k_j=k_m=\sqrt{\epsilon} \omega_0$ in material; the explicit phase factors are specified for notational convenience. 

Because the different layers of the PhC are identical, the transfer matrix $D$ relating coefficients in one region of vacuum $b_{2m-1,\pm}$ with those in the adjacent region of vacuum $b_{2m+1,\pm}$ are identical for all $m$; we derive $D$ via the interface conditions at $z=-d_{2m-1}$ and $z=-d_{2m}$:
\begin{equation}
    C_v \mqty(b_{2m-1,-} \\ b_{2m-1,+}) = C_m P_m \mqty(b_{2m,-} \\ b_{2m,+})
\end{equation}
\begin{equation}
    C_m \mqty(b_{2m,-} \\ b_{2m,+}) = C_v P_v \mqty(b_{2m+1,-} \\ b_{2m+1,+})
\end{equation}
where we have defined the boundary matrices
\begin{equation}
    C_v = \mqty(1 & 1 \\ -k_v & k_v) \qquad C_m = \mqty(1 & 1 \\ -k_m & k_m)
\end{equation}
and phase-shift matrices
\begin{equation}
    P_v = \mqty(\dmat{e^{-i k_v \Delta}, e^{i k_v \Delta}}) \qquad P_m = \mqty(\dmat{e^{-i k_m h}, e^{i k_m h}})
\end{equation}
which yields for a single PhC unit cell
\begin{equation}
    \mqty(b_{2m-1,-} \\ b_{2m-1,+}) = D \mqty(b_{2m+1,-} \\ b_{2m+1,+})
\end{equation}
with transfer matrix
\begin{equation}
    D = C_v^{-1} C_m P_m C_m^{-1} C_v P_v. 
\end{equation}
The transfer matrices for multiple layers of PhC simply correspond to taking powers of $D$, which can be analyzed through its eigendecomposition, as discussed in the following subsection. 

\subsubsection{Reflection Coefficient of a half-infinite photonic crystal}
Suppose that $D$ is nondefective, i.e., $D$ has an eigendecomposition (this assumption may be verified explicitly for specific examples of $D$):
\begin{equation}
    D = \mqty(D_{11} & D_{12} \\ D_{21} & D_{22}) = Q \Lambda Q^{-1} = \mqty(v_{1,1} & v_{2,1} \\ v_{1,2} & v_{2,2}) \mqty(\dmat{w_1, w_2}) \mqty(v_{1,1} & v_{2,1} \\ v_{1,2} & v_{2,2})^{-1}.
\end{equation}
The transfer matrix for $n$ contiguous layers of PhC is just
\begin{equation}
    D^n = \mqty(v_{1,1} & v_{2,1} \\ v_{1,2} & v_{2,2}) \mqty(\dmat{w_1^n, w_2^n}) \mqty(v_{1,1} & v_{2,1} \\ v_{1,2} & v_{2,2})^{-1}. 
\end{equation}
The reflection $r$ and transmission $t$ coefficients of such a structure, as seen from the ``upper'' (smaller $|z|$) side can then be obtained by solving for $r(n)$ and $t(n)$ via the following linear relation
\begin{equation}
    \mqty(1 \\ r(n)) = D^n \mqty(t(n) \\ 0),
\end{equation}
i.e., given a unit amplitude incident wave on side and no incident wave on the other, $r(n)$ and $t(n)$ are the coefficients of the reflected and transmitted waves. This gives
\begin{equation}
    r(n) = \frac{(D^n)_{21}}{(D^n)_{11}} = \frac{(w_1^n-w_2^n)v_{1,2}v_{2,2}}{w_1^n v_{1,1}v_{2,2} - w_2^n v_{1,2}v_{2,1}}.
    \label{eq:reflDn}
\end{equation}
Conservation of energy implies that $|r(n)| \leq 1$; for lossless $\epsilon$ we have $|r(n)|^2 + |t(n)|^2 = 1$.
For a lossless PhC multilayer outside of a band gap $|w_1|=|w_2|=1$ and the value of $r(n)$ oscillates with $n$ so there is no half-infinite $n\rightarrow\infty$ limit.
Within a band gap, and/or if there is loss, $|w_1| \neq |w_2|$ and $r(n \rightarrow \infty)$ converges; denoting this half-infinite crystal reflection coefficient as just $r$ and without loss of generality specifying $w_1$ as the larger magnitude eigenvalue we have
\begin{equation}
    r = \frac{v_{1,2}}{v_{1,1}}.
\end{equation}

\subsection{LDOS of a 1d photonic crystal cavity}
In this section we investigate a specific class of 1d designs, namely a cavity formed by sandwiching a central vacuum layer with two identical half-infinite PhC claddings with lossy susceptibility $\chi$. We show that when the real frequency $\omega_0$ falls precisely on the upper band edge of a band gap of a hypothetical PhC with the same unit cell as the cladding but made of lossless material $\Re{\chi}$, the center of the lossy cavity has small LDOS at $\omega_0$. This small LDOS further reduces linearly with material layer thickness $h$ as $h \rightarrow 0$, providing a recipe for approaching arbitrarily close to the predicted 0 single-frequency LDOS lower bound for infinite domains as shown in the main text and Fig.~\ref{fig:totLDOSmin2DTM1DvsQsrc} above.

Notation: the coordinate system is specified such that $z=0$ is the symmetry center of the design. Let $d$ denote the length of the central air cavity, which may be different from the air gap size of the claddings $\Delta$. The reflection coefficient of the PhC claddings $r$ is understood to have an implicit dependence on frequency.

\subsubsection{LDOS expression at the center of the cavity}
We derive an expression for the LDOS at the center of the cavity by explicitly calculating the field: by reflection symmetry, the electric field in the cavity as the form
\begin{equation}
    E(z; k_{v}) = E_0 e^{ik_{v}|z|} + E_0 r e^{i k_{v} d} e^{-ik_{v}|z|}.
    \label{eq:E_center}
\end{equation}
The amplitude $E_0$ is found by substituting Eq.~\eqref{eq:E_center} into the dipole driven wave equation
\begin{equation}
    \pdv[2]{E}{z} + \omega^2 E = -i\omega\delta(z)
\end{equation}
which yields
\begin{equation}
    E_0 = -\frac{1}{2(1 - r e^{ik_v d})}.
\end{equation}
The power output of the dipole source is $\rho(\omega) = -\frac{1}{2}\Re{E(0)}$, which scaled by the vacuum power of the source $\rho_{\textrm{vac}}(\omega) = \frac{1}{4}$ gives a LDOS enhancement of
\begin{align}
    \frac{\rho(\omega)}{\rho_{\textrm{vac}}(\omega)}
    &= \Re{\frac{1+r e^{i k_v d}}{1-r e^{i k_v d}}},
    \label{eq:PhCCavityLDOSRe}\\
    &= \frac{1-|r|^2}{|1 - r e^{i k_v d}|^2}.
    \label{eq:PhCCavityLDOS}
\end{align}

% TODO: Add in nice geometric interpretation of (\ref{eq:PhCCavityLDOS}) on complex plane.

\subsubsection{Reflectivity of lossy half-infinite photonic crystal mirror}
\begin{figure}
\centering
\begin{subfigure}[t]{0.48\textwidth}
    \centering
        \includegraphics[width=\linewidth]{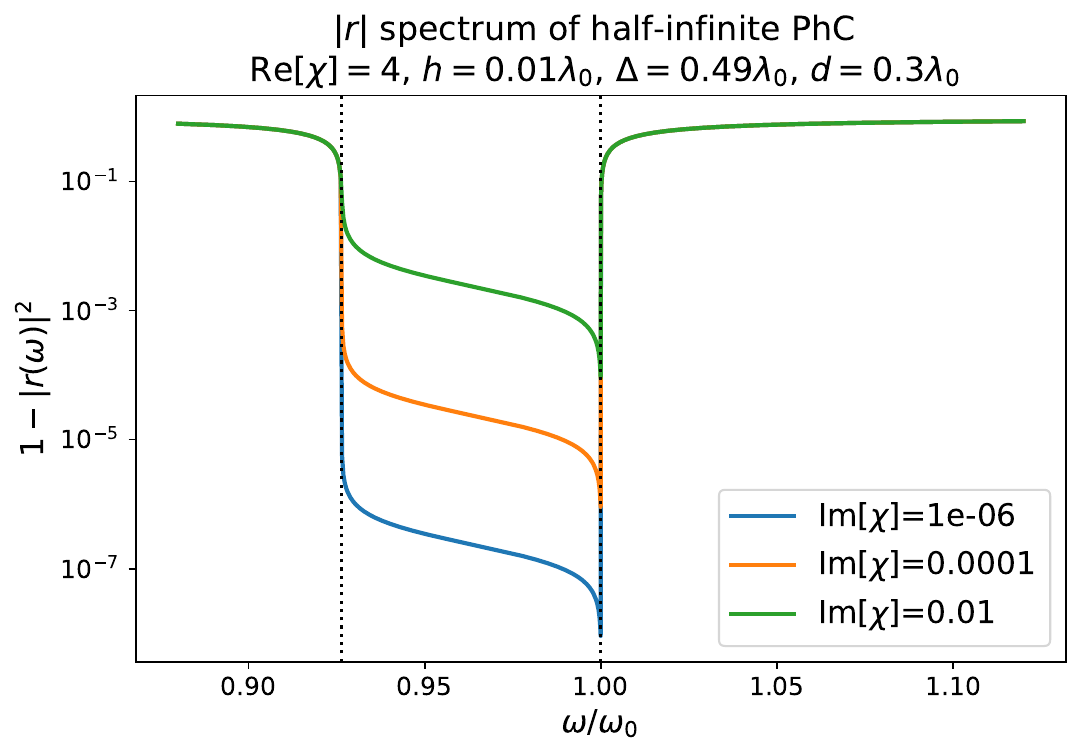}
        \caption{Change of $|r|$ with frequency for a half-infinite PhC slab with lossy $\chi$ where $\Re[\chi] = 4$, and $h=0.01\lambda_{0}$, $\Delta=0.49\lambda_{0}$ so that the periodicity is exactly $0.5\lambda_{0}$ and the first band gap is close to vacuum frequency $\omega_{0}$. Dotted vertical lines show the band gap of the lossless complete PhC. \label{fig:normRplot}}
\end{subfigure}
\hfill
\begin{subfigure}[t]{0.5\textwidth}
    \centering
        \includegraphics[width=\linewidth]{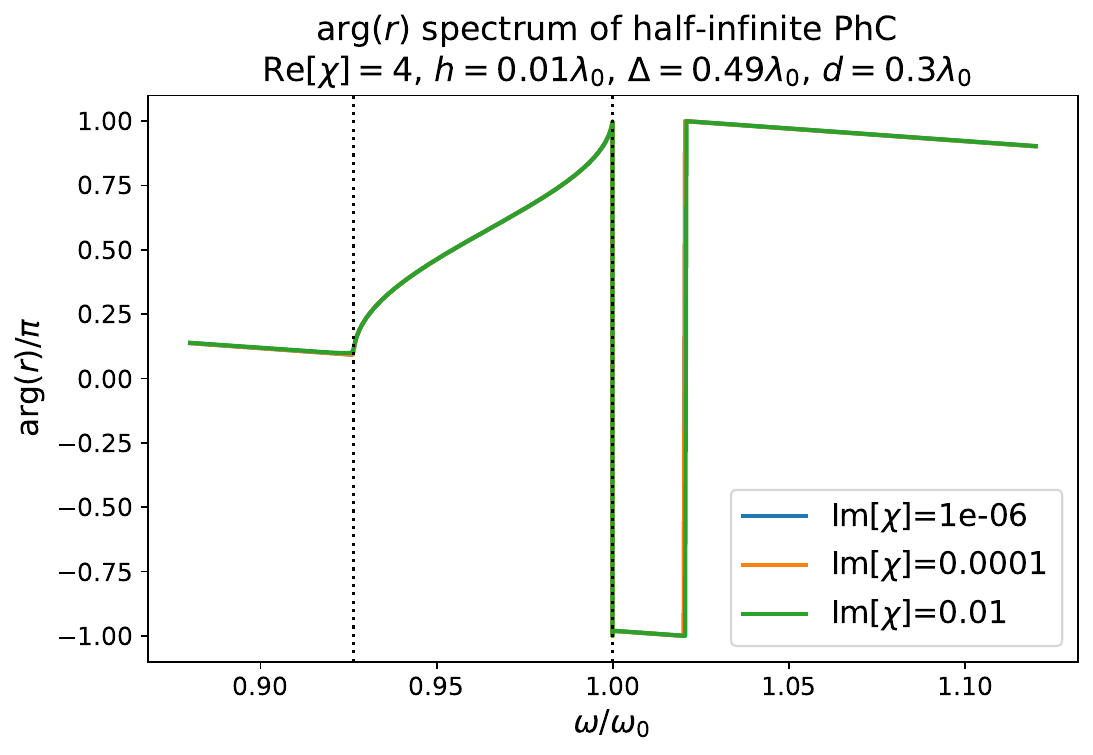}
        \caption{Argument of $r$ with frequency for a half-infinite PhC slab with lossy $\chi$ with $\Re[\chi]=4$, and $h=0.01\lambda_{0}$, $\Delta=0.49\lambda_{0}$ so that the periodicity is exactly $0.5\lambda_{0}$ and the first band gap is close to vacuum frequency $\omega_{0}$. Dotted vertical lines show the band gap of the lossless complete PhC.\label{fig:argRplot}}
\end{subfigure}
\caption{Reflectivity of a lossy half-infinite PhC slab. }
\end{figure}
\begin{figure}
   \centering
       \includegraphics[width=0.6\linewidth]{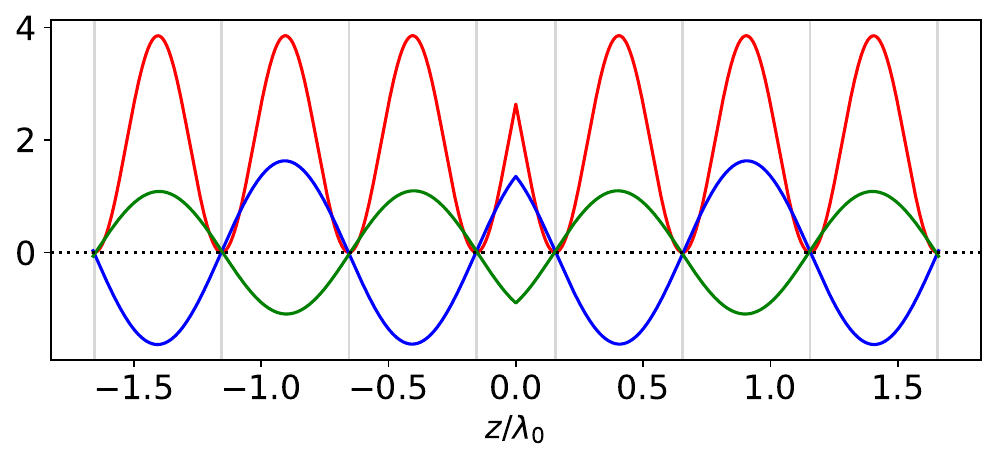}
      \caption{\textbf{Electric field profile at frequency $\omega_{0}$.}
       $|E(z;\omega_{0})/E_{0}|^{2}$ (red), $\text{Re}[E(z;\omega_{0})/E_{0}]$ (blue), and $\text{Im}[E(z;\omega_{0})/E_{0}]$ (green). We set $\chi = 4 + 0.1i$, $h = 0.01\lambda_{0}$, $\Delta = 0.49\lambda_{0}$, and $d = 0.3\lambda_{0}$.
        Vertical gray lines denote the material layers composed with susceptibility $\chi.$
       }
\label{fig:phc1dEsqfield}
\end{figure}

It is clear that for small enhancement, $|r|$ should be close to 1. How does a photonic crystal do? See Fig.~\ref{fig:normRplot}: $|r|$ comes close to 1 in the band gap of the hypothetical lossless PhC, with a prominent further approach exactly at the upper band edge. 
We shall see that this near unity $|r|$ is key to achieving either extreme LDOS enhancement or suppression in the cavity formed by sandwiching a central vacuum gap with two such PhC mirrors at the edge of the band: exactly which occurs is determined by the length of the central vacuum gap $d$. For convenience, let us set the period of the PhC such that the band edge occurs near $\omega_{0} = \frac{2\pi}{\lambda_{0}}$; this is the case when the period is $\lambda_{0}/2$ and the material layer thickness $h \rightarrow 0$. As seen in Fig.~\ref{fig:argRplot}, $\arg(r)$ has a rapid switch between $\pi$ and $-\pi$ at the band-edge, which together with $|r|$ implies that $r$ is close to $-1$. With the help of Mathematica, the leading orders of $r(\omega=2\pi/\lambda_0)$ for a PhC mirror with period $a=0.5\lambda_0$ as $h\rightarrow 0$ are
\begin{equation}
    r\left(\omega=\frac{2\pi}{\lambda_{0}}\right) \approx (-1 + \frac{2\pi}{\sqrt{3}} (h/\lambda_0)) - 2\pi (h/\lambda_0) i \qquad h \rightarrow 0
\end{equation}
and are independent of the material susceptibility.
Per Eq.~\eqref{eq:PhCCavityLDOS}, $\rho(\omega_{0})$ will be small due to the small numerator $1-|r|^{2}$, unless the denominator $|1- r e^{i k_v d}|^2$ is also small, which occurs when $e^{i k_v d} \approx -1$, or equivalently when $d = (n + \frac{1}{2})\lambda_{0}$ where $n$ is a nonnegative integer. As we shall see in later sections, when $d$ is far from such ``resonant'' values the LDOS at the band edge is suppressed, and approaches $0$ linearly as a function of $h$ as $h/\lambda_{0} \rightarrow 0$. Otherwise, the reverse happens: LDOS is enhanced at the band edge, and diverges linearly as a function of $h^{-1}$ as $h/\lambda_{0} \rightarrow 0$. For minimization, the absorption by the material occurs in regions where the $|E(z)|$ is at a minimum, see Fig.~\ref{fig:phc1dEsqfield}.
In contrast, at the middle of the band gap, $r$ asymptotes to a nonzero constant value as $h \rightarrow 0$:
\begin{equation}
    r\left(\omega=\frac{2\pi}{\lambda_0}(1 - \Re{\chi}h) \right) \approx \frac{i\chi}{\Im{\chi} + \sqrt{ (\chi+i\Im{\chi})\cdot\Re{\chi} } } \qquad h \rightarrow 0
\end{equation}
From the form it is clear that $|r|<1$ in the limit $h \rightarrow 0$.

\begin{figure}
  \centering
   \includegraphics[width=0.6\linewidth]{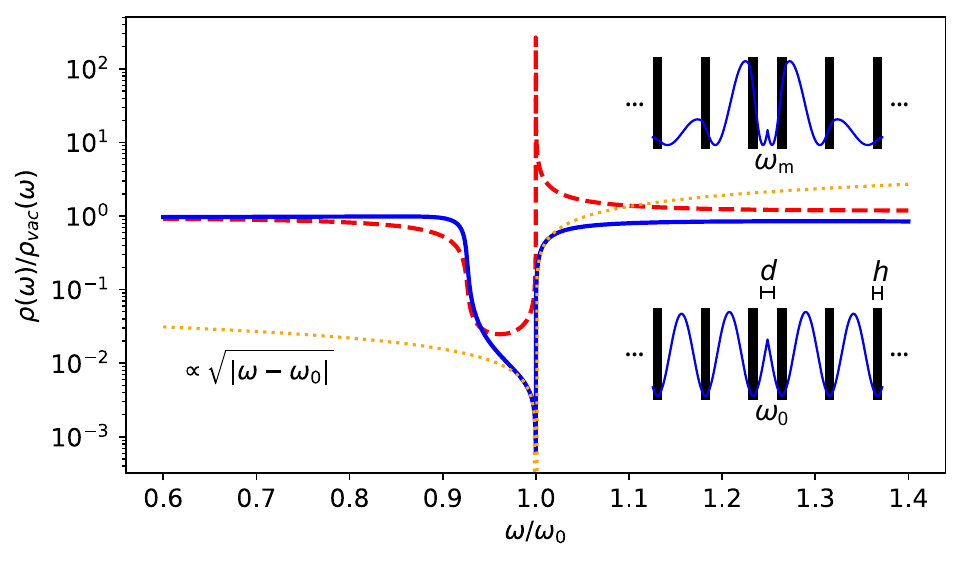}
   \caption{LDOS spectrum, Eq.~\eqref{eq:PhCCavityLDOS}, of a 1d PhC cavity with $\chi=4 + 0.1i$, period $a = 0.5\lambda_{0}$, material layers of thickness $h = a/50$, and a central air gap of thickness $d=a/3$ (solid blue curve) or $d=a-h$ (dashed red curve). 
   The dotted curve shows a square root fit around $\omega_{0}.$
   Inset is a representative intensity profile $|E(z; \omega_{0})|^{2}$ at the band edge, exhibiting minima in the material layers (vertical black bars), as well as $|E(z; \omega_{m})|^{2}$ at the pseudo-midgap.}
\label{fig:phc1d_plots}
\end{figure}

In sum, in the limit of $h \rightarrow 0$, $\omega_{0}$ approaches the band edge from above and the electric field profile approaches a standing wave (a slow-light mode) with field nodes inside material layers.
By making the thicknesses of the PhC mirrors arbitrarily thin, 
one can ensure arbitrarily small field overlap with the lossy medium and, thus, vanishing absorption. 
See Fig.~\ref{fig:phc1d_plots} insets for comparison of the field intensity profiles for the midgap vs. upper band edge strategy.

\subsubsection{LDOS spectrum of photonic crystal cavity with central cavity deviating from cladding air layer}

\begin{figure}
    \centering
\begin{subfigure}[t]{0.48\textwidth}
    \centering
        \includegraphics[width=\linewidth]{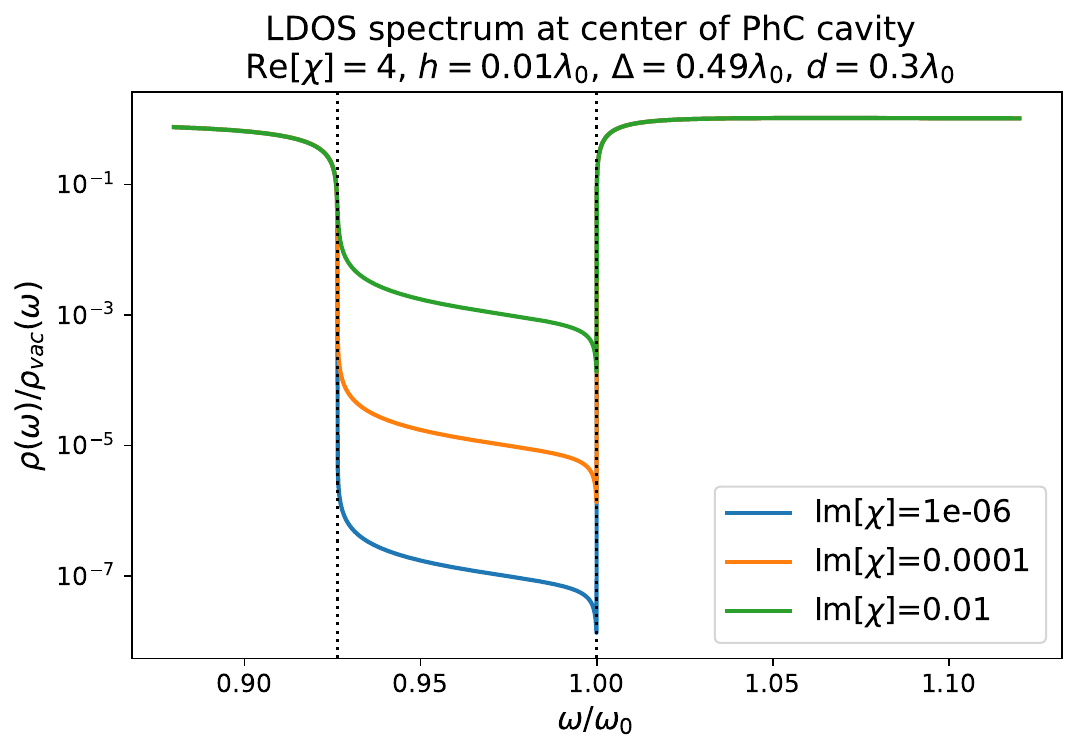}
        \caption{LDOS spectrum of a 1d PhC cavity with $d=0.3\lambda_{0}$ and varying loss. Dotted vertical lines show the band gap of the hypothetical lossless complete PhC. \label{fig:1DCavityLDOS_LDeviation_varyOmega}}
\end{subfigure}
\hfill
\begin{subfigure}[t]{0.48\textwidth}
    \centering
        \includegraphics[width=\linewidth]{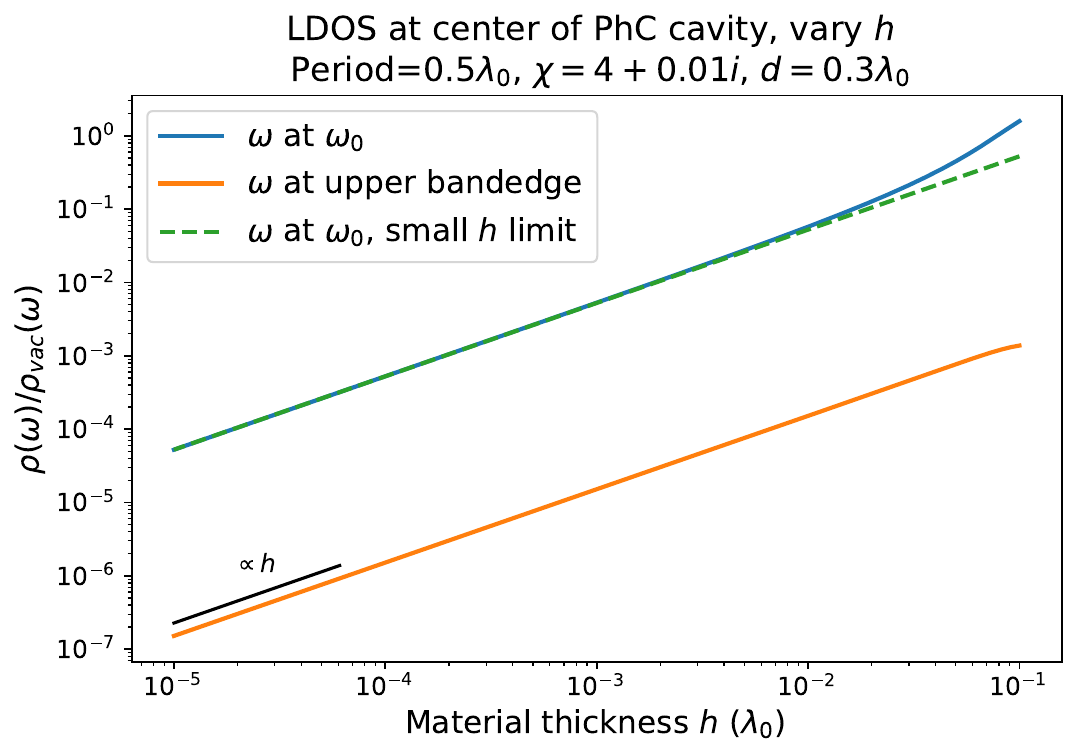}
        \caption{LDOS at and close to band edge for 1d PhC cavity with nonresonant $d$ and fixed $\chi=4 + 0.01i$. LDOS scales linearly with $h$ as $h/\lambda_{0} \to 0.$
        In particular, $\rho(\omega_{0})/\rho_{\textrm{vac}}(\omega_{0}) \approx \frac{\pi}{\sqrt{3}} \sec^{2}(\frac{\omega_{0}d}{2})\frac{h}{\lambda_{0}}$ as $h/a\to 0$ (dashed line).
        % In particular, $\rho(\omega_{0})/\rho_{\textrm{vac}}(\omega_{0}) \approx \Re[-4\pi i(1 + i/\sqrt{3}) e^{i\omega_{0} d}/(1 + e^{i\omega_{0} d})^{2}]\frac{h}{\lambda_{0}} = \frac{\pi}{\sqrt{3}} \sec^{2}(\frac{\omega_{0}d}{2})\frac{h}{\lambda_{0}}$ as $h/a\to 0$ (dashed line).
        \label{fig:1DCavityLDOS_Ldeviation_varyh}}
\end{subfigure}
\caption{LDOS characteristics of 1d PhC cavity with period $0.5\lambda_{0}$ and mismatched $d \neq 0.5\lambda_{0}$.}
\label{fig:1DCavityLDOS_Ldeviation}
\end{figure}

Suppose $d \not\approx (n + \frac{1}{2})\lambda_{0}$ where $n$ is any nonnegative integer. This is the case for which LDOS suppression is achieved at the band edge, see Fig.~\ref{fig:1DCavityLDOS_Ldeviation}. As $h/\lambda_{0} \to 0,$ the LDOS at $\omega_{0}$ approaches 0 linearly as a function of $h.$

\subsubsection{LDOS spectrum of photonic crystal cavity with central cavity matching periodicity}

\begin{figure}
    \centering
\begin{subfigure}[t]{0.48\textwidth}
    \centering
        \includegraphics[width=\linewidth]{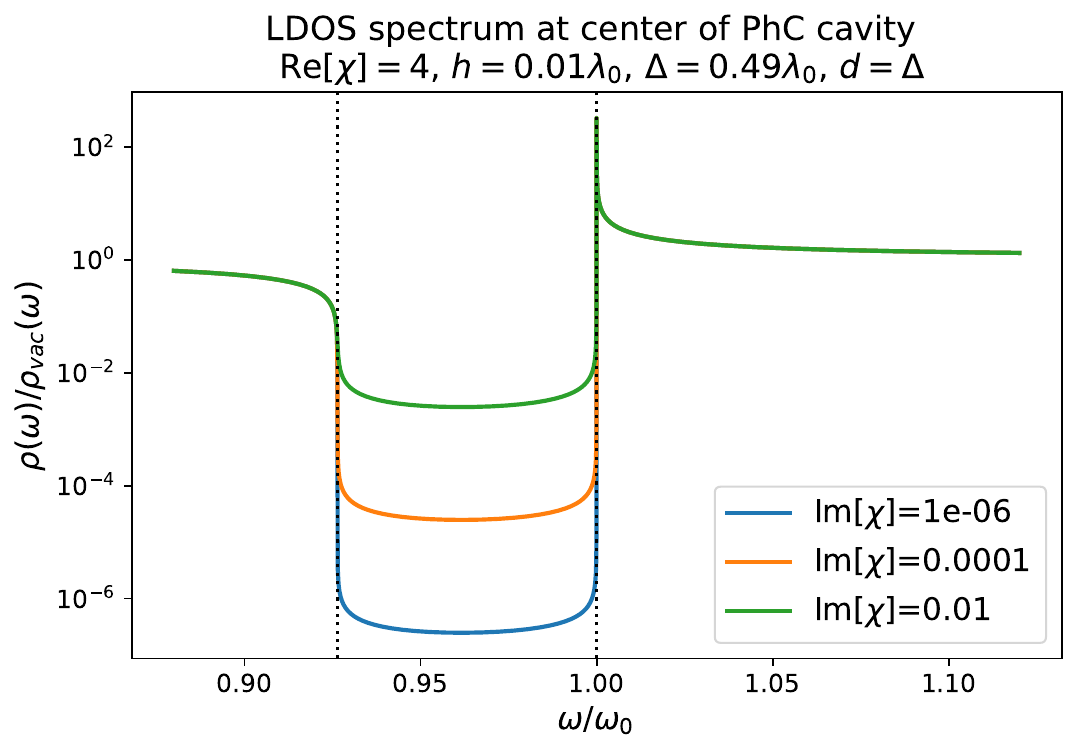}
        \caption{LDOS spectrum of a 1d PhC cavity with $d=\Delta \approx 0.5\lambda_{0}$ for $h=0.01\lambda_{0}$. Dotted vertical lines show the band gap of the hypothetical lossless complete PhC. \label{fig:1DCavityLDOS_Lmatch_varyOmega}}
\end{subfigure}
\hfill
\begin{subfigure}[t]{0.48\textwidth}
    \centering
        \includegraphics[width=\linewidth]{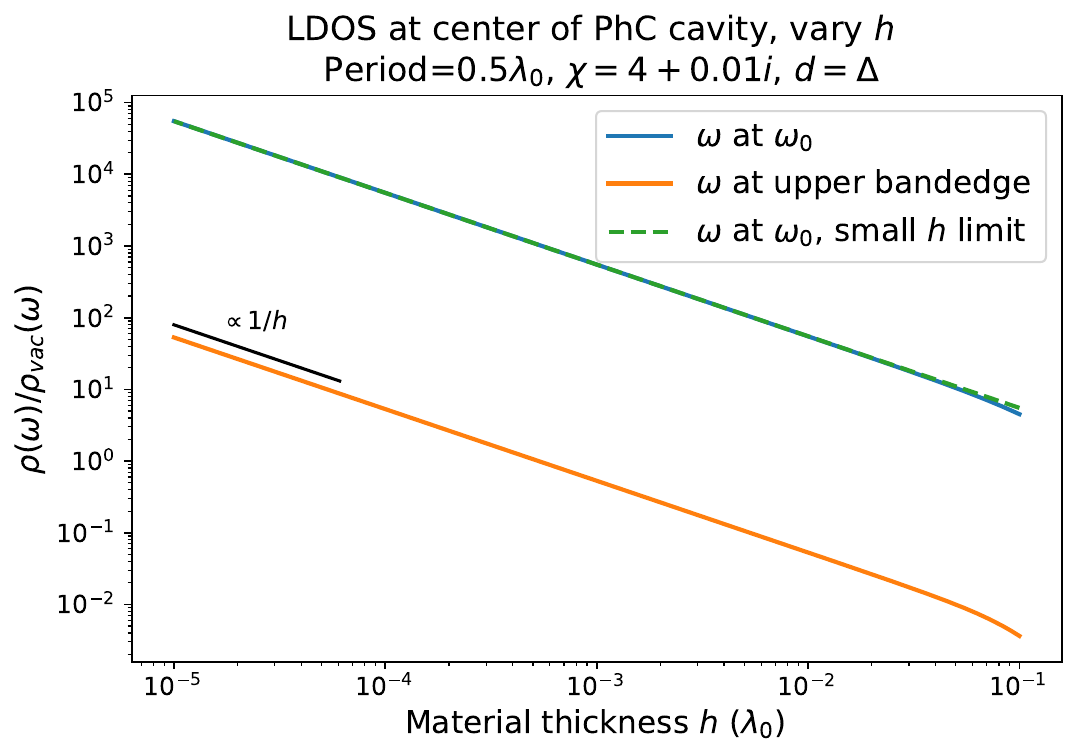}
        \caption{LDOS at and close to band edge for 1d PhC cavity with resonant $d$ and fixed $\chi=4 + 0.01i$. LDOS scales linearly with $1/h$ as $h/\lambda_{0} \to 0.$ In particular, for $d = a - h,$
        $\rho(\omega_{0})/\rho_{\textrm{vac}}(\omega_{0}) \approx \frac{\sqrt{3}}{\pi}\big(\frac{h}{\lambda_{0}}\big)^{-1}$ as $h/a\to 0$ (dashed line). If $d = (n + \frac{1}{2})\lambda_{0}$ for a nonnegative integer $n,$ then $\rho(\omega_{0})/\rho_{\textrm{vac}}(\omega_{0}) \approx \frac{\sqrt{3}}{4\pi}\big(\frac{h}{\lambda_{0}}\big)^{-1}$ as $h/a\to 0$.
        \label{fig:1DCavityLDOS_Lmatch_varyh}}
\end{subfigure}
\caption{LDOS characteristics of 1d PhC cavity with period $a=0.5\lambda_{0}$ and resonant $d \approx a$.}
\label{fig:1DCavityLDOS_Lmatching}
\end{figure}

Suppose $d \approx (n + \frac{1}{2})\lambda_{0}$ where $n$ is some nonnegative integer.
This is the case for which LDOS enhancement is achieved at the band edge, see Fig.~\ref{fig:1DCavityLDOS_Lmatching}. As $h/\lambda_{0} \to 0,$ the LDOS at $\omega_{0}$ diverges as linearly with $h^{-1}$.
This giant enhancement arises from the coupling of the dipole to a slow light edge mode, which approximately yields a well-known van Hove singularity with corresponding $\int_{\omega_{0}}^{\infty} W(\omega)\frac{1}{\sqrt{\omega-\omega_{0}}} \dd\omega \approx 1/\sqrt{\Delta\omega}$ scaling.
For any fixed material thickness $h > 0$, this divergence as $\Delta\omega \to 0$ saturates since $\rho(\omega_{0}) < \infty.$ 
By varying $h$ as $\Delta\omega \to 0,$ one can achieve persistent $1/\sqrt[3]{\Delta\omega}$ scaling as $\Delta\omega \to 0$ with no saturation (see below).

% \newpage
 
\subsubsection{Transition through cavity resonance with varying cavity size}

\begin{figure}
    \centering
    \includegraphics[width=0.5\linewidth]{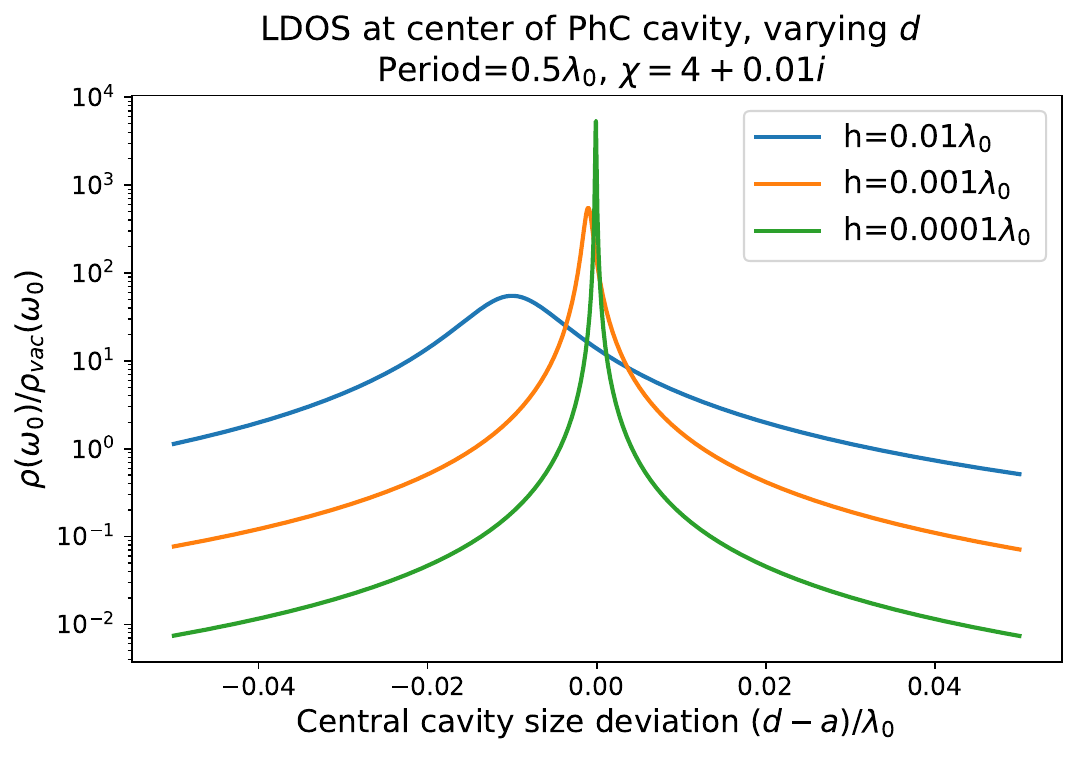}
    \caption{Plot of LDOS at the center of the cavity as a function of the cavity size $d$ around the resonant value $d=a=\lambda_{0}/2$.}
    \label{fig:phc1d_varyL}
\end{figure}

See Fig.~\ref{fig:phc1d_varyL}.

\subsubsection{Optimal thickness for minimization for a given bandwidth}

\begin{figure}
    \centering
    \includegraphics[width=0.5\linewidth]{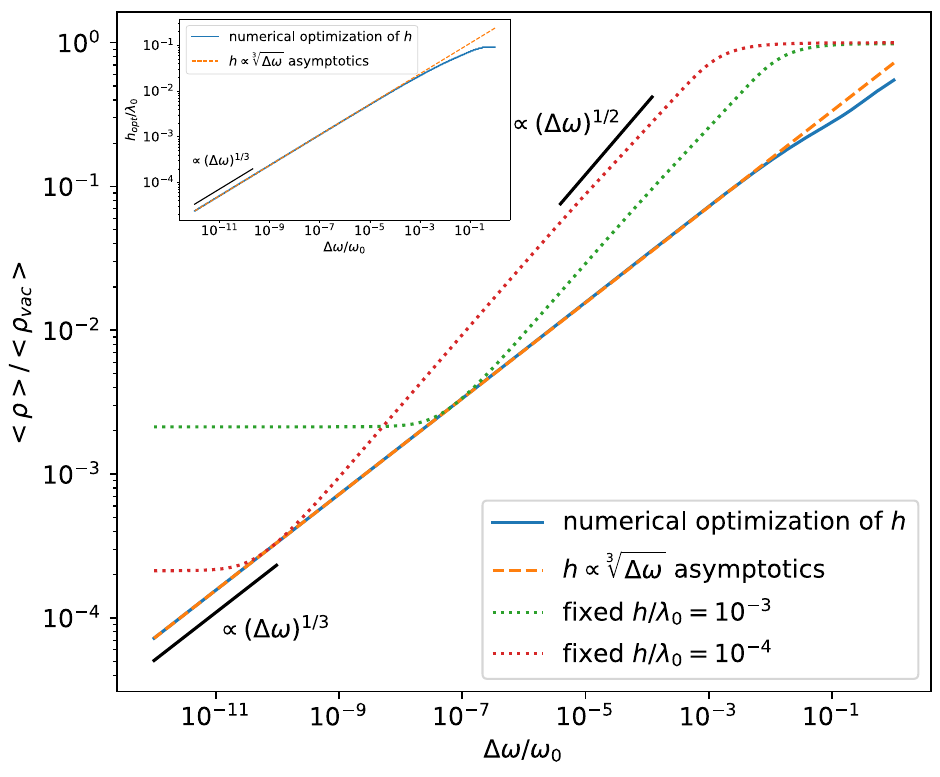}
    \caption{Spacing $a=0.5\lambda_{0}$, central cavity thickness $d = a/4$, and susceptibility $\chi = 4 + 0.1i.$ The dashed line is Eq.~\eqref{eq:appdwbetahcubed} with $\beta$ equal to Eq.~\eqref{eq:betaoptmin}.}
    \label{fig:phc1d_hoptandfixedh}
\end{figure}

The bandwidth-integrated LDOS is given by
\begin{align}
\frac{\langle\rho\rangle}{\langle\rho_{\textrm{vac}}\rangle}
    = \Re{\frac{1+r e^{i \tilde{\omega} d}}{1-r e^{i \tilde{\omega} d}}\frac{1}{\tilde{\omega}\mathcal{N}}}
\label{eq:PhCCavityavgLDOSRe}
\end{align}
where $\tilde{\omega} = \omega_{0} +i\Delta\omega.$
We find numerically that the optimal PhC thickness $h_{\textrm{opt}} \propto \sqrt[3]{\Delta \omega}$ as $\Delta\omega \to 0$, so assuming $d/\lambda_{0} \not\approx \frac{1}{2}$ modulo 1 and plugging $\Delta\omega~=~\beta\omega_{0}(h/\lambda_{0})^{3}$ for some $\beta>0$ into this equation and expanding leads to
\begin{equation}
    \frac{\langle\rho\rangle}{\langle\rho_{\textrm{vac}}\rangle} 
    \approx
    \Re\left[\frac{\pi}{\cos^{2}(\frac{\omega_{0}d}{2})}\sqrt{\frac{1}{3} + \frac{i\beta\chi^{*}}{2\pi^{2}|\chi|^{2}}}\right] \sqrt[3]{\frac{\Delta\omega}{\omega_{0}\beta}}.
    \label{eq:appdwbetahcubed}
\end{equation}
To optimize this for a fixed $0 < \Delta\omega/\omega_{0} \ll 1$ as a function of $\beta,$ we focus on
\begin{align}
    \sqrt{1 + \frac{3i\beta\chi^{*}}{2\pi^{2}|\chi|^{2}}}
    &=
    \sqrt[4]{
    \bigg(1 + \frac{3\beta\text{Im}[\chi]}{2\pi^{2}|\chi|^{2}}\bigg)^{2} + \bigg(\frac{3\beta\text{Re}[\chi]}{2\pi^{2}|\chi|^{2}}\bigg)^{2}
    }
    \exp{\frac{i}{2}\atan(\frac{\frac{3\beta\text{Re}[\chi]}{2\pi^{2}|\chi|^{2}}}{1 + \frac{3\beta\text{Im}[\chi]}{2\pi^{2}|\chi|^{2}}})}.
\end{align}
Using $\text{Re}[e^{\frac{i}{2}\atan(\phi)}] = \cos(\frac{1}{2}\atan(\phi)) = \frac{1}{\sqrt{2}}\sqrt{1 + \frac{1}{\sqrt{1 + \phi^{2}}}}$ leads to
\begin{align}
    \text{Re}
    \Bigg[
    \sqrt{1 + \frac{3i\beta\chi^{*}}{2\pi^{2}|\chi|^{2}}}
    \bigg]
    &=
    \frac{1}{\sqrt{2}}
    \sqrt{\sqrt{
    \bigg(1 + \frac{3\beta\text{Im}[\chi]}{2\pi^{2}|\chi|^{2}}\bigg)^{2} + \bigg(\frac{3\beta\text{Re}[\chi]}{2\pi^{2}|\chi|^{2}}\bigg)^{2}
    }
    +
    \bigg(1 + \frac{3\beta\text{Im}[\chi]}{2\pi^{2}|\chi|^{2}}\bigg)
    }.
\end{align}
Taking a derivative with respect to $\beta$ of the terms in the outer square root and setting it to 0 leads to a quadratic equation in terms of $\beta,$ with a positive root
\begin{align}
    \beta_{\textrm{opt}} \equiv \frac{2\pi^{2}\text{Im}[\chi]}{3}\bigg(-1 +\sqrt{9 + 8\bigg(\frac{\text{Re}[\chi]}{\text{Im}[\chi]}\bigg)^{2}}\bigg).
    \label{eq:betaoptmin}
\end{align}
See Fig.~\ref{fig:phc1d_hoptandfixedh}.
For a fixed material thickness $h,$ the structure exhibits $\sqrt{\Delta\omega}$ as $\Delta\omega \to 0$ before saturating due to $\rho(\omega_{0}) > 0$ for any fixed finite $h > 0.$
By optimizing over $h$ at each $\Delta\omega,$ one finds a group of structures which yields a curve scaling like $\sqrt[3]{\Delta\omega}$ as the bandwidth shrinks, with no saturation. 
For $\text{Im}[\chi] \ll |\text{Re}[\chi]|$, we find
\begin{align}
    \frac{\langle\rho\rangle}{\langle\rho_{\textrm{vac}}\rangle}
    &=
    \frac{\sqrt[3]{\frac{\pi}{2\sqrt{3}|\text{Re}[\chi]|}}}{\cos^{2}(\frac{\omega_{0}d}{2})} \bigg(
    1
    +
    \frac{\sqrt{2}}{3}\frac{\text{Im}[\chi]}{|\text{Re}[\chi]|}
    + \mathcal{O}(\text{Im}[\chi]^{2})
    \bigg)
    \sqrt[3]{\frac{\Delta\omega}{\omega_{0}}}.
\end{align}

\subsubsection{Optimal thickness for maximization for a given bandwidth}

\begin{figure}
    \centering
    \includegraphics[width=0.5\linewidth]{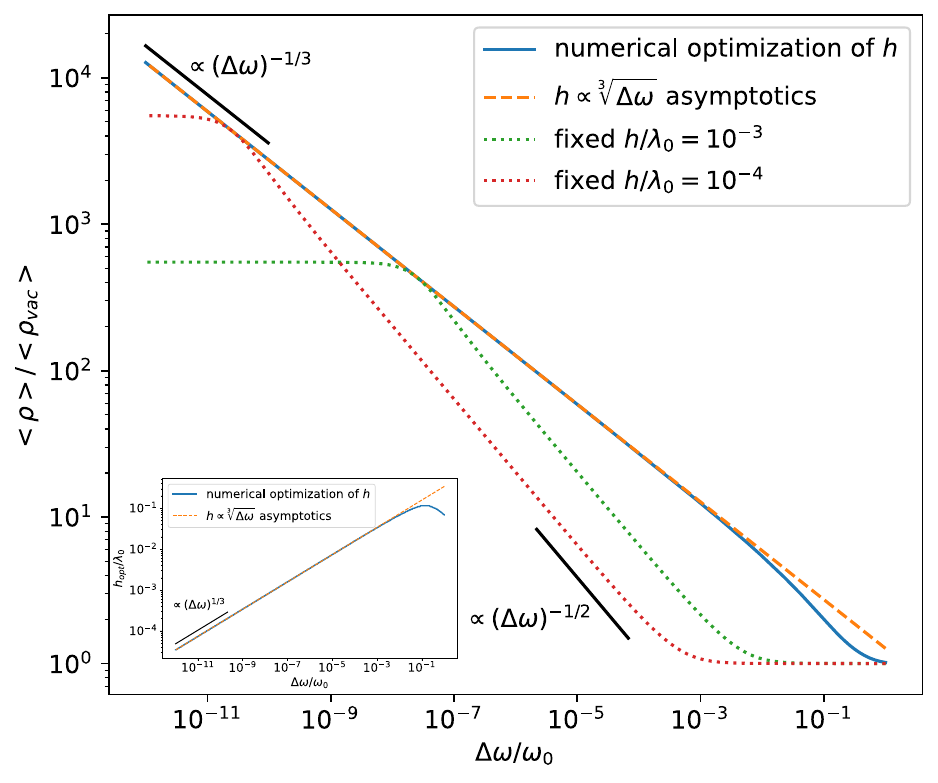}
    \caption{Spacing $a=0.5\lambda_{0}$, central cavity thickness $d = a-h$, and susceptibility $\chi = 4 + 0.1i.$ The dashed line is Eq.~\eqref{eq:appdwbetahcubedmax} with optimal $\beta$ solved for numerically.}
    \label{fig:phc1d_hoptandfixedh_max}
\end{figure}

For maximization, we set $d = a - h$. 
We find numerically that the optimal PhC thickness $h_{\textrm{opt}} \propto \sqrt[3]{\Delta \omega}$ as $\Delta\omega \to 0$, so assuming $d/\lambda_{0} \not\approx \frac{1}{2}$ modulo 1 and plugging $\Delta\omega = \beta \omega_{0}(h/\lambda_{0})^{3}$ for some $\beta>0$ into bandwidth-integrated equation and expanding leads to
\begin{equation}
    \frac{\langle\rho\rangle}{\langle\rho_{\textrm{vac}}\rangle} 
    \approx
    \Re\left[\frac{\sqrt{3}}{\pi}\frac{1}{\sqrt{1 + \frac{3i\beta\chi^{*}}{2\pi^{2}|\chi|^{2}}}}  \right] \sqrt[3]{\frac{\omega_{0}\beta}{\Delta\omega}}
    \label{eq:appdwbetahcubedmax}
\end{equation}
which proves the existence of lossy structures exhibiting $1/\sqrt[3]{\Delta\omega}$ scaling for the maximization scenario, with no saturation as $\Delta\omega \to 0.$ See Fig.~\ref{fig:phc1d_hoptandfixedh_max}.
For a fixed material thickness $h,$ the structure exhibits $1/\sqrt{\Delta\omega}$ as $\Delta\omega \to 0$ before saturating due to $\rho(\omega_{0}) < \infty$ for any fixed finite $h > 0.$
By varying $h$ at each $\Delta\omega,$ we find a group of structures which yields a curve scaling like $1/\sqrt[3]{\Delta\omega}$ as the bandwidth shrinks with no saturation.
Taking a derivative with respect to $\beta$ of the terms in the outer square root and setting it to 0, we found a quartic equation for $\beta,$ which in principle has a closed-form expression for the optimal $\beta.$ We do not write it out in this case.

\bibliography{refs}